\documentclass[a4paper,11pt]{article}
\pdfoutput=1
\usepackage{jheppub}

\usepackage{tikz}

\usepackage{enumitem}
\usepackage[T1]{fontenc} 
\usepackage{floatrow}
\usepackage{appendix}
\usepackage{slashed}
\usepackage{tabularx}
\usepackage[normalem]{ulem}

\allowdisplaybreaks

\usepackage{epsf}
\usepackage{amsmath}
\usepackage{empheq}
\usepackage[theorems,skins]{tcolorbox}
\usepackage{amsfonts}
\usepackage{amssymb}
\usepackage{psfrag,epsfig,graphicx,graphics}

\newcommand\numberthis[1][]{%
    \refstepcounter{equation}%
    \ifx#1\empty\else\label{eq:#1}\fi%
    \tag{\theequation}%
}

\usepackage{xargs} 
\usepackage[colorinlistoftodos,prependcaption,textsize=tiny]{todonotes}
\providecommand{\U}[1]{\protect\rule{.1in}{.1in}}

\def\k{{\mathbf k}}
\def\x{{\mathbf x}}  
\def\y{{\mathbf y}}
\def\r{{\mathbf r}}
\def\z{{\mathbf z}}

\def\slashchar#1{\setbox0=\hbox{$#1$}
   \dimen0=\wd0
   \setbox1=\hbox{/} \dimen1=\wd1
   \ifdim\dimen0>\dimen1
      \rlap{\hbox to \dimen0{\hfil/\hfil}}
      #1
   \else
      \rlap{\hbox to \dimen1{\hfil$#1$\hfil}}
      /
   \fi}

\def\nn{\nonumber}

\def\bei{\begin{itemize}}
\def\ei{\end{itemize}}

\def\beeq{\begin{eqnarray}} 
\def\beqa{\begin{eqnarray}}
\def\bea{\begin{eqnarray}}

\def\eea{\end{eqnarray}}
\def\eqa{\end{eqnarray}}
\def\eeeq{\end{eqnarray}}

\def\eqar{\end{array}}
\def\beqar{\begin{array}}

\def\beas{\begin{eqnarray*}}
\def\beqas{\begin{eqnarray*}}

\def\eqas{\end{eqnarray*}}
\def\eeas{\end{eqnarray*}}

\def\beq{\begin{equation}} 
\def\be{\begin{equation}}

\def\ee{\end{equation}}
\def\eq{\end{equation}}
\def\eeq{\end{equation}}

\def\beqd{\begin{displaymath}}
\def\eeqd{\end{displaymath}}
\def\eqd{\end{displaymath}}

\def\beeq{\begin{eqnarray}} \def\eeeq{\end{eqnarray}}

\newcommand{\fin}{\end{document}}

\newcommandx{\MF}[2][1=]{\todo[linecolor=blue,backgroundcolor=blue!25,bordercolor=blue,#1]{#2}}

\newcommandx{\JF}[2][1=]{\todo[linecolor=black,backgroundcolor=white!25,bordercolor=black,#1]{#2}}

\newcommandx{\TA}[2][1=]{\todo[linecolor=red,backgroundcolor=red!25,bordercolor=red,#1]{#2}}

\newcommandx{\GB}[2][1=]{\todo[linecolor=violet,backgroundcolor=violet!25,bordercolor=violet,#1]{#2}}

\title{\boldmath Next-to-Leading Order corrections to the Next-to-Eikonal DIS structure functions}

\author[a]{Tolga Altinoluk,}
\author[a]{Guillaume Beuf,}
\author[a]{Jules Favrel,}
\author[a]{Michael Fucilla,}

\affiliation[a]{National Centre for Nuclear Research (NCBJ), Pasteura 7, 02-093 Warsaw, Poland}

\emailAdd{tolga.altinoluk@ncbj.gov.pl}
\emailAdd{guillaume.beuf@ncbj.gov.pl}
\emailAdd{jules.favrel@ncbj.gov.pl}
\emailAdd{michael.fucilla@ncbj.gov.pl}

\abstract{
We compute next-to-leading order (NLO) corrections to next-to-eikonal (NEik) quark background contributions to DIS structure functions. 
Among NEik corrections,  $t$-channel quark exchanges provide the lowest order contributions in $\alpha_s$, and can be represented as insertions of the quark background field of the target.
At NLO, we compute NEik corrections induced by both quark and gluon background fields, and suppressed by an explicit factor of $\alpha_s$. We show that the NLO corrections to the NEik longitudinal structure function are finite, while those to the NEik transverse structure function exhibit rapidity and UV divergences. These divergences are analyzed, and the finite contributions are extracted.    
}

\begin{document} 
\maketitle
\flushbottom

\newpage
\section{Introduction}
\label{sec:intro}
%{\color{red} Generic introduction}

In the high-energy (small-$x$) regime, hadronic collisions are effectively described by the Color Glass Condensate (CGC) framework (see Refs. \cite{Gelis:2010nm,Albacete:2014fwa,Blaizot:2016qgz} for reviews). The CGC formalism is based on the phenomenon of gluon saturation, which arises in the Regge–Gribov limit as the decrease of Bjorken-$x$ drives a rapid growth of gluon densities. At sufficiently small $x$, nonlinear gluon interactions become important and tame the growth of the gluon densities, leading to the phenomenon known as {\it gluon saturation} that is  characterized by the emergent momentum scale $Q_s$. At small-$x$, the rapidity evolution which encodes the nonlinear effects is provided by the Balitsky–Kovchegov /Jalilian-Marian-Iancu-McLerran-Wiegert-Leonidov-Kovner (BK/JIMWLK) equation derived in  \cite{Balitsky:1995ub,Kovchegov:1999yj,Kovchegov:1999ua,Jalilian-Marian:1996mkd,Jalilian-Marian:1997qno,Jalilian-Marian:1997jhx,Jalilian-Marian:1997ubg,Kovner:2000pt,Weigert:2000gi,Iancu:2000hn,Iancu:2001ad,Ferreiro:2001qy}. 

Deep inelastic scattering (DIS) on dense targets provides a particularly clean probe of saturation dynamics. In the dipole factorization framework \cite{Bjorken:1970ah,Nikolaev:1990ja}, a virtual photon emitted by the incoming lepton fluctuates into a quark–antiquark dipole that subsequently scatters off the target. The interaction of the quark-antiquark pair with the dense target is encoded in Wilson lines within the CGC framework, while the splitting of the photon to the quark-antiquark pair is computed perturbatively. 

Hints of gluon saturation have been observed at Relativistic Heavy Ion Collider (RHIC), the Large Hadron Collider (LHC), and HERA, yet a definitive experimental confirmation has not been achieved.  The forthcoming Electron–Ion Collider (EIC) in the USA is expected to provide a significantly cleaner environment and higher luminosity than HERA. Moreover, the use of the nuclear targets is expected to enhance the saturation effects despite the lower center-of-mass energies. In order to fully exploit the future experimental data from EIC for understanding gluon saturation, the CGC framework has to be improved to provide more precise theoretical predictions. This improvement can be achieved either through higher-order corrections in strong coupling $\alpha_s$ or by refining the kinematic approximations adopted in the computation of the observables within the CGC framework. In recent years, substantial progress has been made in both directions.

The next-to-leading order (NLO) corrections in $\alpha_s$ to various DIS related observables have been computed within the CGC framework. These observables include inclusive DIS with massless quarks \cite{Balitsky:2010ze,Balitsky:2012bs,Beuf:2011xd,Beuf:2016wdz,Beuf:2017bpd,Ducloue:2017ftk,Hanninen:2017ddy} and corresponding fits to HERA data \cite{Beuf:2020dxl}, inclusive DIS with heavy-quarks \cite{Beuf:2021qqa,Beuf:2021srj,Beuf:2022ndu} and its phenomenological analyses \cite{Hanninen:2022gje,Casuga:2025etc}, diffractive structure functions \cite{Beuf:2022kyp,Beuf:2024msh}, single inclusive DIS (SIDIS) \cite{Caucal:2024cdq,Bergabo:2024ivx,Bergabo:2022zhe,Altinoluk:2024vgg,Altinoluk:2025dwd}, diffractive single hadron production \cite{Fucilla:2023mkl}, inclusive dijet/dihadron production \cite{Caucal:2021ent,Taels:2022tza,Caucal:2023fsf,Bergabo:2023wed,Bergabo:2022tcu,Iancu:2022gpw}, diffractive dijet/dihadron production \cite{Boussarie:2016ogo,Boussarie:2019ero,Fucilla:2022wcg}, inclusive photon+jet production\cite{Taels:2023czt}, inclusive photon+dijet production \cite{Roy:2019hwr}, exclusive light \cite{Boussarie:2016bkq,Mantysaari:2022bsp} and heavy \cite{Mantysaari:2021ryb,Mantysaari:2022kdm} vector meson production at NLO. These developments provide essential building blocks for high-precision saturation studies at the EIC. 

As noted earlier, a complementary approach to enhancing the precision of calculations within the saturation framework is to refine the kinematical approximations employed. The central approximation in the CGC framework is the eikonal approximation, which essentially retains only the contributions that are leading in collision energy, neglecting all energy-suppressed terms in the evaluation of observables. 
From the perspective of a target that is boosted along the $x^-$ direction with a large Lorentz factor $\gamma_t$ and is described by the background field ${A}_a^{\mu}(x)$, the eikonal approximation relies on three key assumptions: (i) the background field is localized in the longitudinal direction (around $x^+ = 0$); (ii) only the dominant component of the background field (in our setup, the $-$'' component) contributes to the interaction with the projectile parton, while other components (transverse and $+$'' components, suppressed by the Lorentz boost) are neglected; and (iii) the target dynamics is frozen, which is equivalent to assuming that the background field is independent of $x^-$ due to Lorentz time dilation. Collectively, these assumptions yield the well-known shockwave approximation, in which the target’s background field takes the form
\begin{equation}
{A}^{\mu}_a(x^-,x^+,\x)\approx \delta^{\mu -}\delta(x^+){A}^-_a(\x)\, . 
\end{equation}
While the eikonal approximation is robust at LHC energies and has successfully described experimental data, energy-suppressed corrections become increasingly relevant at RHIC energies and for the future EIC. Incorporating these corrections is therefore necessary to improve the precision of phenomenological studies. Relaxing any of the eikonal assumptions introduces corrections beyond the leading power in the high-energy limit. These effects are systematically accounted for by next-to-eikonal (NEik) corrections, which correspond to terms suppressed by powers of $1/\gamma_t$ in the boosted background field. NEik corrections encode finite-energy effects, extending the applicability of the CGC beyond the strict high-energy limit.  

Although relaxing any of the above listed eikonal assumptions would yield NEik effects  for a target described solely gluon background field, additional NEik contributions arise from the quark content of the target. In this case, projectile partons can interact via $t$-channel quark exchange, generating subeikonal effects beyond those associated with the gluon background. Under a boost of $\gamma_t$ along the $x^-$ direction, currents associated with the target scale as
\begin{align}
 \label{eq:Target_Current_Scaling_1}
J^-(x)\propto\gamma_t\, , \quad J^i(x)\propto (\gamma_t)^0\, \quad J^+(x)\propto (\gamma_t)^{-1} \, .    
\end{align}

%\end{eqaution}
%
The quark background field $\Psi(x)$ associated with the target can be decomposed into its so-called “good” and “bad” components, defined by
\begin{align}
\label{Eq:GoodComp}
\Psi^{(-)}(x)&=\frac{\gamma^+\gamma^-}{2}\Psi(x) \, ,\\
\label{Eq:BadComp}
\Psi^{(+)}(x)&=\frac{\gamma^-\gamma^+}{2}\Psi(x) \, .
\end{align}
The quark contribution to target currents are constructed as bilinears of $\Psi(x)$, and their components satisfy
\begin{align}
{\overline \Psi}(x)\gamma^-\Psi(x)&=\overline {\Psi^{(-)}}(x)\gamma^-\Psi^{(-)}(x) \, , \label{quark_current}\\
{\overline \Psi}(x)\gamma^j\Psi(x)&=\overline {\Psi^{(-)}}(x)\gamma^j\Psi^{(+)}(x)+\overline {\Psi^{(+)}}(x)\gamma^j\Psi^{(-)}(x) \, , \\
{\overline \Psi}(x)\gamma^+\Psi(x)&=\overline {\Psi^{(+)}}(x){\gamma^+}\Psi^{(+)}(x) \, .
\end{align}
To ensure that the target currents follow the scaling behavior introduced in Eq.~\eqref{eq:Target_Current_Scaling_1}, the good and bad components of the quark field scale with the boost parameter $\gamma_t$ as
\begin{align}
\label{eq:good-bad_scaling}
\Psi^{(-)}(x) \propto (\gamma_t)^{1/2} \, ,\\
\Psi^{(+)}(x) \propto (\gamma_t)^{-1/2} \, . 
\end{align}
Since the scaling of the quark field components is parametrically smaller than $\gamma_t$, each insertion of a single quark background field is a non-eikonal effect, due to the suppression of the effective range of integration along $x^+$ as $1/\gamma_t$, by Lorentz contraction.
The only possibility for the quark background field to contribute at eikonal accuracy is to have a composite insertion of two factors $\Psi^{(-)}(x)$ or $\overline{\Psi^{(-)}}(x)$ at the same $x^+$. By experience, such insertion  never happen in the quark propagator, but only double insertions of the more suppressed components $\Psi^{(+)}(x)$ (with an instantaneous Coulomb interaction in between, like in light-front perturbation theory).
In the case of the gluon propagator, the insertion of the background quark color current (analog to \eqref{quark_current} with a color generator $t^a$ included) do contribute at eikonal accuracy in general. However, in the light-cone gauge $A^+=0$ that we are using, this contribution to the gluon propagator decouples due to the structure of the vacuum gluon propagator. 
Consequently, the quark background field does not contribute at eikonal order. The enhanced component, $\Psi^{(-)}(x)$, contributes at next-to-eikonal (NEik) order, while the suppressed component, $\Psi^{(+)}(x)$, contributes only at next-to-next-to-eikonal (NNEik) order and beyond.

With the motivation of increasing the precision of the theoretical framework for future phenomenological studies, derivation of the NEik corrections in the CGC framework received significant effort over the last decade. 
Early studies addressed corrections arising from the finite longitudinal width of the target \cite{Altinoluk:2014oxa,Altinoluk:2015gia}, which were subsequently applied to particle production and correlation analyses in both dilute-dilute \cite{Altinoluk:2015xuy,Agostini:2019avp,Agostini:2019hkj} and dilute-dense \cite{Agostini:2022ctk,Agostini:2022oge} collisions. NEik corrections to quark and scalar propagators were computed in \cite{Altinoluk:2020oyd,Altinoluk:2021lvu,Agostini:2023cvc} and applied to DIS dijet production \cite{Altinoluk:2022jkk,Agostini:2024xqs}, while back-to-back quark-gluon dijet production in DIS, including $t$-channel quark exchange, was studied to probe quark TMDs \cite{Altinoluk:2023qfr}. Similar analyses for gluon TMDs, and the interplay between NEik and kinematic twist corrections, were performed in \cite{Altinoluk:2024zom}, and the gluon propagator incorporating all NEik corrections was revisited in \cite{Altinoluk:2024dba} for parton-nucleus scattering applications. NEik contributions from $t$-channel quark exchanges in back-to-back dijet production in proton-nucleus collisions were computed in \cite{Altinoluk:2024tyx}, allowing the study of various quark TMDs. Finally, NEik corrections to inclusive DIS and SIDIS cross sections were calculated in \cite{Altinoluk:2025ang}, focusing solely on the contributions arising from $t$-channel quark exchanges.  

Helicity-dependent observables, quark and gluon helicity evolutions, and single- and double-spin asymmetries have been explored extensively at NEik accuracy \cite{Kovchegov:2015pbl,Kovchegov:2016zex,Kovchegov:2016weo,Kovchegov:2017jxc,Kovchegov:2017lsr,Kovchegov:2018znm,Kovchegov:2018zeq,Kovchegov:2020kxg,Kovchegov:2020hgb,Adamiak:2021ppq,Kovchegov:2021lvz,Kovchegov:2021iyc,Cougoulic:2022gbk,Kovchegov:2022kyy,Borden:2023ugd,Kovchegov:2024aus,Borden:2024bxa,Adamiak:2025dpw,Kovchegov:2025gcg,Borden:2025ehe}, and helicity-dependent extensions of the CGC have been formulated in \cite{Cougoulic:2019aja,Cougoulic:2020tbc}. Rapidity evolution of gluon TMDs, interpolating between moderate and low $x$, was studied in \cite{Balitsky:2015qba,Balitsky:2016dgz,Balitsky:2017flc}, while similar interpolations for inclusive DIS and exclusive Compton scattering were investigated in \cite{Boussarie:2020fpb,Boussarie:2021wkn,Boussarie:2023xun}. NEik corrections to both quark and gluon propagators have also been formulated within the high-energy operator product expansion (OPE) \cite{Chirilli:2018kkw,Chirilli:2021lif}. Sub-eikonal corrections were further examined using an effective Hamiltonian approach \cite{Li:2023tlw,Li:2024fdb,Li:2024xra}, and alternative frameworks allowing longitudinal momentum exchange between projectile and target were developed in \cite{Jalilian-Marian:2017ttv,Jalilian-Marian:2018iui,Jalilian-Marian:2019kaf}. Finally, the impact of sub-eikonal corrections on orbital angular momentum was analyzed in \cite{Hatta:2016aoc,Kovchegov:2019rrz,Boussarie:2019icw,Kovchegov:2023yzd,Kovchegov:2024wjs}.

Although NLO corrections to certain NEik operators have been derived in the context of small-$x$ evolution, observables evaluated at NEik accuracy have so far been treated exclusively at leading order. In this work, we take the first step toward establishing a complete description of observables computed simultaneously at NLO and NEik accuracy within the CGC framework. Our analysis focuses on the inclusive DIS cross section at NEik order, previously obtained at leading order in the coupling constant in \cite{Altinoluk:2025ang}. Building on the expressions derived in \cite{Altinoluk:2025ang}, we compute the NLO corrections to the NEik contribution to the inclusive DIS cross section. The manuscript is organized as follows. In Section \ref{Sec:VariousProp}, we briefly review the parton propagators at NEik accuracy, derived in \cite{Altinoluk:2024dba}, which serve as the essential building blocks for computing the NLO corrections to the NEik cross section. Section \ref{Sec:LOCrossSec} outlines the main steps and results of the leading-order calculation of the NEik inclusive DIS cross section, as first presented in \cite{Altinoluk:2025ang}. Sections \ref{Sec:NLOCrossSecQuark} and \ref{Sec:NLOXsec_gb} are devoted to deriving the NLO corrections to the NEik inclusive DIS cross section, focusing on the contributions from the quark (and antiquark) background field and the gluon background field, respectively. In Section \ref{Sec:sum_out}, we summarize our findings and provide a brief outlook. Appendices \ref{Sec:AppUsefulFormulas}, \ref{Sec:AppDiracTrace} and \ref{Sec:AppInsideMedium} contain additional technical details and supplementary derivations.

%%%%%%%%%%%%%%%%%%%%%%%%%%%%%%%%%%%%%%%%%%%%%%%%%%%%%%%%%%%%%%%%%%%%%%%%%%%%%%%%%%%%%%%%%%%%
%%%%%%%%%%%%%%%%%%%%%%%%%%%%%%%%%%%%%%%%%%%%%%%%%%%%%%%%%%%%%%%%%%%%%%%%%%%%%%%%%%%%%%%%%%%%
\section{CGC beyond eikonal accuracy and various parton propagators}
\label{Sec:VariousProp}
%%%%%%%%%%%%%%%%%%%%%%%%%%%%%%%%%%%%%%%%%%%%%%%%%%%%%%%%%%%%%%%%%%%%%%%%%%%%%%%%%%%%%%%%%%%%
%%%%%%%%%%%%%%%%%%%%%%%%%%%%%%%%%%%%%%%%%%%%%%%%%%%%%%%%%%%%%%%%%%%%%%%%%%%%%%%%%%%%%%%%%%%%

Since our primary interest lies in the inclusive DIS cross section, we frame our discussion of NEik corrections around this observable. At leading order in the eikonal approximation, the inclusive DIS cross section can be expressed within the dipole factorization framework for longitudinally and transversely polarized virtual photon as
\begin{align}
\sigma^{\gamma^*}_L(x_{Bj},Q^2)=&\,\frac{4\, \alpha_{\text{em}}N_c}{(2\pi)^2}\sum_f e_f^2 \, \int_0^1 dz \, \int d^2\x_1 \, d^2\x_2 \Big[1-d(\x_1,\x_2)\Big] \nn \\
& %\hspace{3cm} 
\times
\, 4Q^2z^2(1-z)^2\, K_0^2(|\x_1-\x_2|\overline Q) \; , \\
\sigma^{\gamma^*}_T(x_{Bj},Q^2)=&\,
\frac{4\, \alpha_{\text{em}}N_c}{(2\pi)^2}\sum_f e_f^2 \, \int_0^1 dz \, \int d^2\x_1 \, d^2\x_2 \Big[1-d(\x_1,\x_2)\Big] \nn \\
& %\hspace{2.0cm} 
\times
\, \Big(\big[ z^2+(1-z)^2\big]\,
{\overline Q}^2
K_1^2(|\x_1-\x_2|\overline Q)+m^2K_0^2(|\x_1-\x_2|\overline Q)\Big) \; ,
\end{align}
where $z$ and $(1-z)$ corresponds to the longitudinal momentum fraction carried by the quark and antiquark relative to the longitudinal momentum of the incoming virtual photon, and $m$ is the quark mass (for which we keep the flavor index $f$ implicit to simplify notations). $K_{\alpha}(\cdots)$ is the modified Bessel function of the second type and $\overline{Q}^2=m^2+z(1-z)Q^2$ with $Q^2$ being the virtuality of the incoming photon. The inclusive DIS cross sections for the scattering of transverse or longitudinal photon on the target are related to the transverse or longitudinal structure functions via 
\begin{align}
\label{sigma2F}
\sigma^{\gamma^*}_{T,L}(x_{Bj},Q^2)=\frac{(2\pi)^2\, \alpha_{\text{em}}}{Q^2} F_{T,L}(x_{Bj},Q^2) \; .
\end{align}
The operator $d(\x_i,\x_j)$ that appears in the inclusive cross sections is the dipole operator that accounts for the multiple scattering of the partons off the dense target and it is defined as     
\begin{equation}
    d (\mathbf{x}_1,\mathbf{x}_2) = \left \langle \frac{1}{N_c} {\rm Tr_c}  \bigg[ \mathcal{U}_{F} (\mathbf{x}_1) \mathcal{U}^{\dagger}_{F} (\mathbf{x}_2) \bigg ] \right \rangle_A \; ,
\end{equation}
where ${\mathcal U}_F(\x_i)\equiv {\mathcal U}_F(+\infty,-\infty;\x_i)$ is the Wilson line in fundamental representation in a gluon background field $A^-(z^+,\z)$ which is given as 
\begin{equation}
    \mathcal{U}_{F} (x^+, y^+, \mathbf{z}) = \mathcal{P} \exp \left( - i g_s \int_{y^+}^{x^+} d z^+ \; t \cdot A^{-} (z^+, \mathbf{z})\right) \; , 
    \label{eq:WilsFun}
\end{equation}
with \( t \) being the \( SU(3) \) generator in the fundamental representation, and 
\( \mathcal{P} \) indicates ordering of those color generators along the path.  Moreover, the notation $\langle\cdots\rangle$ stands for target averaging in the spirit of CGC formalism. As mentioned earlier, at NEik order one relaxes the shockwave approximation and consider a target with a finite longitudinal width with a support of $[-L^+/2, L^+/2]$. In the rest of the manuscript, we use\footnote{\label{footnote_1}As discussed in detail in \cite{Altinoluk:2024zom}, since we are working in gauge where the background gauge fields vanish outside of the target such that 
\begin{equation}
    A^{\mu} \Big( \frac{L}{2}^+, \mathbf{z} \Big) = A^{\mu} \Big( -\frac{L}{2}^+, \mathbf{z} \Big) = 0 \; . \nn
\label{Eq:TargGauge}
\end{equation}
Therefore, if the target width $L^+$ is appearing on the limits of the longitudinal integrals of the Wilson line structures, one can safely extend all the background Wilson lines beyond the longitudinal extend of the target, from $-\infty$ to $+\infty$.}
\begin{align}
{\mathcal U}_F(\z)\equiv {\mathcal U}_F\Big(\frac{L}{2}^+,-\frac{L}{2}^+; \z\Big)= \mathcal{P}_+ \exp \left( - i g_s \int_{L^+} d z^+ t \cdot A^{-} (z^+, \mathbf{z}) \right) \; , 
\label{eq:WilsFun2}
\end{align}
as a shorthand notation. In a gluon background field, the NEik corrections arising from the finite longitudinal width and from interactions with the transverse components of the background field are encoded in the so-called decorated Wilson lines, which contain field-strength–tensor insertions along the longitudinal direction and take the form 
\begin{gather}
\label{eq:decW_1}
\hspace{-1.8cm}
    \mathcal{U}_{F,j}^{(1)} \left( \mathbf{z} \right)   = 2 i g_s \int_{-\frac{L}{2}^+}^{\frac{L}{2}^+} d z^{+} z^+ \mathcal{U}_F\left(\frac{L}{2}^+, z^+ ; \mathbf{z} \right) t \cdot \mathcal{F}_j^{\; -} (z^+, \mathbf{z} ) \; \mathcal{U}_F\left(z^{+},-\frac{L}{2}^+ ; \mathbf{z}\right) \; ,
\end{gather}
\begin{align}
\label{eq:decW_2}
 \mathcal{U}_{F}^{(2)} \left( \mathbf{z} \right)  &= (i g_s)^2 \int_{-\frac{L}{2}^+}^{\frac{L}{2}^+} d z^{+} \int_{-\frac{L}{2}^+}^{\frac{L}{2}^+} d v^{+} (z^+ - v^+) \theta (z^+ - v^+) \mathcal{U}_F\left(\frac{L}{2}^+, z^+ ; \mathbf{z} \right)
  \nonumber \\
  &  \hspace{2cm} \times t \cdot \mathcal{F}_j^{\; -} (z^+, \mathbf{z}) \; \mathcal{U}_F\left(z^{+}, v^{+} ; \mathbf{z} \right) t \cdot \mathcal{F}_j^{\; -} (v^+, \mathbf{z}) \mathcal{U}_F\left(v^{+},-\frac{L}{2}^+ ; \mathbf{z}\right) \;  ,
\end{align}
and
\begin{gather}
\label{eq:decW_3}
\hspace{-2.4cm}
    \mathcal{U}_{F,i j}^{(3)} \left( \mathbf{z}\right) =  \int_{-\frac{L}{2}^+}^{\frac{L}{2}^+} d z^{+} \mathcal{U}_F\left(\frac{L}{2}^+, z^{+} ; \mathbf{z} \right) g_s \; t \cdot \mathcal{F}_{i j}(z^+, \mathbf{z}) \mathcal{U}_F\left(z^{+},-\frac{L}{2}^+ ; \mathbf{z} \right) \; ,
\end{gather}
with following convention for the field strength tensor 
\begin{align}
{\mathcal F}_{\mu\nu}=\partial_{\mu}A_{\nu} - \partial_{\nu}A_{\mu}+ig\big[ A_{\mu}, A_{\nu}\big] \, . 
\end{align}
At NEik order, one constructs decorated operators by replacing one of the standard eikonal Wilson lines inside the color trace with one of the decorated Wilson lines listed above. For inclusive DIS cross section, only decorated dipole structures contribute at NEik order when evaluated in a gluon background field. These decorated dipole operators take the form
\begin{equation}
    d_j^{(1)} (\mathbf{x}_{1*},\mathbf{x}_2) = \left \langle \frac{1}{N_c} {\rm Tr_c}  \bigg[ \mathcal{U}_{F,j}^{(1)} (\mathbf{x}_{1}) \mathcal{U}^{\dagger}_{F} (\mathbf{x}_2) \bigg ] \right \rangle \; ,
\end{equation}
\begin{equation}
    d^{(2)} (\mathbf{x}_{1*},\mathbf{x}_2) = \left \langle \frac{1}{N_c} {\rm Tr_c}  \bigg[ \mathcal{U}_{F}^{(2)} (\mathbf{x}_{1}) \mathcal{U}^{\dagger}_{F} (\mathbf{x}_2) \bigg ] \right \rangle \; ,
\end{equation}
\begin{equation}
    d_{ij}^{(3)} (\mathbf{x}_{1*},\mathbf{x}_2) = \left \langle \frac{1}{N_c} {\rm Tr_c}  \bigg[ \mathcal{U}_{F,ij}^{(3)} (\mathbf{x}_{1}) \mathcal{U}^{\dagger}_{F} (\mathbf{x}_2) \bigg ] \right \rangle \; .
\end{equation}
The star in the transverse coordinate of the arguments of decorated dipole operators indicates the position of the corresponding decoration. 

Apart from the quark and gluon propagators that traverse the whole extend of the medium, to compute some of the observables at NEik accuracy, one requires explicit expressions for the quark and gluon propagators with their initial and/or final positions located inside the medium. For the computation of scattering processes at NEik accuracy, the propagators with these special configurations are only needed at eikonal order, we therefore restrict ourselves to that precision. All relevant propagators were derived in our earlier works \cite{Altinoluk:2022jkk,Altinoluk:2023qfr,Altinoluk:2024dba}. For convenience, we summarize their expressions below.  

Let us start with the Feynman quark propagator in the absence of a background field. In $D$-dimensions it reads 
\begin{align}
\hspace{-0.4cm}
    S_{F} (x,y) & = \int \frac{d^D k}{(2 \pi)^D} e^{-i(x-y) \cdot k} \frac{i (\slashed{k} + m)}{k^2 - m^2 + i \epsilon} \nonumber \\ 
    & = \theta(x^+-y^+) \int \frac{d^{D-1} \underline{k}}{(2 \pi)^{D-1}} \frac{ \theta(k^+)}{  2 k^+ } e^{ - i \check{k} \cdot (x-y) } ( \slashed{\check{k}} + m ) 
    \nonumber \\ 
    &
    - \theta(y^+ - x^+) \int \frac{d^{D-1} \underline{k}}{(2 \pi)^{D-1}} \frac{ \theta(-k^+)}{  2 k^+ } e^{ - i \check{k} \cdot (x-y) } ( \slashed{\check{k}} + m ) 
    \nonumber \\
    &+ i \gamma^+ \delta(x^+ - y^+) \delta^{(D-2)} (\mathbf{x}-\mathbf{y}) \int \frac{d k^+}{2 \pi} \frac{e^{-i k^+ (x^- - y^-) }}{2 k^+}  
    \nonumber \\ 
    &\equiv \theta(x^+-y^+) S_{F,q} (x,y) + \theta(y^+-x^+) S_{F,\bar{q}} (x,y)+ \delta(x^+ - y^+) S_{F,q}^{i.} (x,y) \; ,
    \label{Eq:FullFreePropQuark}
\end{align}
with the notation $\check{k}^{\mu}$ stands for the on-shell analog of a given momentum 4-vector $k^{\mu}$. More precisely, it is defined in such a way that their $+$ and transverse components coincide, $\check{k}^+=k^+$ and $\check{\k}=\k$, whereas the $-$ component of $\check{k}^{\mu}$ is given by its on-shell value, i.e. $\check{k}^-=(\k^2+m^2)/(2k^+)$.

The before-to-after quark propagator at NEik accuracy, for which quark traverses the whole medium, i.e.  from the position $y^+<-L^+/2$ before the medium to $x^+>L^+/2$ after the medium, reads  
\begin{align} 
\label{eq:quark_BA}
S_{F,q}^{b.a.} (x, y) &= \int \frac{d^{D-1} \underline{p}}{(2 \pi)^{D-1}} \int \frac{d^{D-1} \underline{k}}{(2 \pi)^{D-1}}\frac{ \theta(p^{+}) }{2p^+}\frac{\theta(k^{+}) }{2k^+}e^{-i x \cdot \check{p}} e^{i y \cdot \check{k}} \; (2 \pi) \delta (p^{+}-k^{+}) \nonumber \\  
& \hspace{3.5cm}
\times \int d^{D-2} \mathbf{z} \; e^{-i \mathbf{z} \cdot(\mathbf{p}-\mathbf{k})} 
 (\slashed{\check{p}}+m) \, \gamma^{+} \,  \mathcal{U}^{\star}_F \, \left(\mathbf{z}  \right)(\slashed{\check{k}}+m)  \; ,
\end{align}
where we have introduced the notation for momentum 3-vector $\underline{k}=(k^+, \k)$.  Moreover, we also defined the {\it fully decorated} Wilson line $ \mathcal{U}_F^{\star} \left( \mathbf{z} \right)$, in terms of the standard and previously introduced decorated Wilson, as 
\begin{gather}
\hspace{-0.5cm}
    \mathcal{U}_F^{\star} \left( \mathbf{z} \right) = \mathcal{U}_F\left( \mathbf{z} \right)  - \frac{\left(\mathbf{p}^j+\mathbf{k}^j\right)}{2\left(p^{+}+k^{+}\right)} \mathcal{U}_{F,j}^{(1)} \left( \mathbf{z}  \right)   -\frac{i}{\left(p^{+}+k^{+}\right)} \mathcal{U}_{F}^{(2)} \left( \mathbf{z}  \right) + \frac{\left[\gamma^i, \gamma^j\right]}{4\left(p^{+}+k^{+}\right)} \mathcal{U}_{F,i j}^{(3)} \left( \mathbf{z}   \right) \; ,
    \label{Eq:FullyDec}
\end{gather}
with the standard Wilson line defined in Eq. \eqref{eq:WilsFun2} while the decorated Wilson lines are defined in Eqs. \eqref{eq:decW_1}, \eqref{eq:decW_2} and \eqref{eq:decW_3}.  One should also consider the case in which the points $x$ and $y$ are interchanged, meaning that $y^+>L^+/2$ and $x^+<-L^+/2$, which corresponds to what is refereed to as the before-to-after antiquark propagator which reads 
\begin{align} 
\label{eq:antiquark_BA}
S_{F,\bar{q}}^{b.a.} (x, y)&= (-1)\int \frac{d^{D-1} \underline{p}}{(2 \pi)^{D-1}} \int \frac{d^{D-1} \underline{k}}{(2 \pi)^{D-1}} \frac{\theta( - p^{+})}{2p^+} \frac{\theta( - k^{+})}{2k^+} e^{-i x \cdot \check{p}} e^{i y \cdot \check{k}} \ (2 \pi) \delta(p^{+}-k^{+}) 
\nonumber \\ 
& \hspace{4cm}
\times 
\int d^{D-2} \mathbf{z} \; e^{-i \mathbf{z} \cdot(\mathbf{p}-\mathbf{k})}  (\slashed{\check{p}}+m) \gamma^{+} \mathcal{U}^{\star \dagger}_F\left(\mathbf{z}  \right) (\slashed{\check{k}}+m) \; , 
\end{align}
where the conjugate {\it fully decorated} Wilson line is defined as  
\begin{gather}
\hspace{-0.2cm}
    \mathcal{U}_F^{\star \dagger} \left( \mathbf{z} \right) = \mathcal{U}_F^{\dagger }\left( \mathbf{z} \right)  + \frac{\left(\mathbf{p}^j+\mathbf{k}^j\right)}{2\left(p^{+}+k^{+}\right)} \mathcal{U}_{F,j}^{(1) \dagger} \left( \mathbf{z} \right) + \frac{i}{\left(p^{+}+k^{+}\right)} \mathcal{U}_{F}^{(2) \dagger} \left( \mathbf{z} \right) - \frac{\left[\gamma^i, \gamma^j\right]}{4\left(p^{+}+k^{+}\right)} \mathcal{U}_{F,i j}^{(3) \dagger} \left( \mathbf{z} \right) \; .
\end{gather}

The before-to-after quark and antiquark propagators computed at NEik accuracy, given in Eqs. \eqref{eq:quark_BA} and \eqref{eq:antiquark_BA}, incorporate the NEik corrections arising from the finite longitudinal width of the target and from interactions with the transverse components of the background gluon field. However, as discussed previously, there exists an additional source of NEik corrections—those originating from going beyond the static approximation of the background field. In particular, these corrections stem from the non-vanishing $z^-$ dependence of the background gluon fields. As discussed in \cite{Altinoluk:2021lvu,Altinoluk:2022jkk}, when the computation of an observables is restricted to NEik accuracy, these non-static corrections decouple from the other type of NEik contributions, and the interplay among them need can be neglected. Therefore, the NEik corrections associated with the non-static background field can be incorporated into the following effective before-to-after quark and antiquark propagators. For a quark propagating from  $y^+ < -L^+/2$ to $x^+ > L^+/2$, one has 
\begin{align} 
S_{F,q}^{\rm n.s.} (x, y) &= \int \frac{d^{D-1} \underline{p}}{(2 \pi)^{D-1}} \int \frac{d^{D-1} \underline{k}}{(2 \pi)^{D-1}} \frac{ \theta(p^{+}) }{2p^+}\frac{\theta(k^{+}) }{2k^+}e^{-i x \cdot \check{p}} e^{i y \cdot \check{k}} \; \int d z^- e^{iz^- (p^+-k^+)} 
\nonumber \\  
& 
\times \int d^{D-2} \mathbf{z} \; e^{-i \mathbf{z} \cdot(\mathbf{p}-\mathbf{k})}  (\slashed{\check{p}}+m)\gamma^{+} \mathcal{U}_F\left(\mathbf{z}, z^-  \right)(\slashed{\check{k}}+m)  \; ,
\label{Eq:NonStatQuarkProp}
\end{align}
and for an antiquark  propagating from $x^+ < - L^+/2$ to $y^+ > L^+/2$, one has 
\begin{align} 
S_{F,\bar{q}}^{\rm n.s.} (x, y) & = (-1)\int \frac{d^{D-1} \underline{p}}{(2 \pi)^{D-1}} \int \frac{d^{D-1} \underline{k}}{(2 \pi)^{D-1}} \frac{\theta( - p^{+})}{2p^+} \frac{\theta( - k^{+})}{2k^+} e^{-i x \cdot \check{p}} e^{i y \cdot \check{k}} \;  \int d z^- e^{iz^- (p^+-k^+)}  
\nonumber \\ 
& \times 
\int d^{D-2} \mathbf{z} \; e^{-i \mathbf{z} \cdot(\mathbf{p}-\mathbf{k})}  (\slashed{\check{p}}+m) \gamma^{+} \mathcal{U}^{ \dagger}_F\left(\mathbf{z} , z^- \right) (\slashed{\check{k}}+m) \; ,
\label{Eq:NonStatAntiQuarkProp}
\end{align}
with the non-static Wilson line $ \mathcal{U}_F\left(\mathbf{z}, z^-  \right)$ is defined as\footnote{In the literature, these terms are referred to Generalized Eikonal contributions due to the fact that while keeping the explicit $z^-$ dependence in the background field, their form is the same as the eikonal contributions.}
\begin{equation}
    \mathcal{U}_{F} (\mathbf{z}, z^-) = \mathcal{P}_+ \exp \left( - i g_s \int_{-\frac{L^+}{2}}^{\frac{L^+}{2}} d z^+ t \cdot A^{-} (z^+, \mathbf{z}, z^-) \right) \; .
    \label{eq:WilsFun3}
\end{equation}

The quark propagator from inside to after the medium, with $ - L^+/2 < y^+ < L^+/2$ to $x^+ > L^+/2$,  is given by 
\begin{align} 
S_{F,q}^{i.a.} (x, y) &=\int \frac{d^{D-1} \underline{k}}{(2 \pi)^{D-1}}  \frac{\theta\left(k^{+}\right)}{2k^+}  e^{-i x \cdot \check{k}}  (\slashed{\check{k}}+m) 
\, \mathcal{U}_F\left( x^+,y^+, \mathbf{y} \right) 
\nn \\
& \hspace{6cm}
 \times 
\left[ 1 - \frac{\gamma^+ \gamma^i}{2 k^+} i \overleftarrow{D_{\mathbf{y}^i}^F} \right] e^{i y^- k^+ } e^{-i \mathbf{y} \cdot \mathbf{k} }  \; ,
\label{eq:quark_P_IA}
\end{align}
while the inside-to-after antiquark propagator, which can be obtained by interchanging $x$ and $y$, and therefore corresponds to the propagation of an antiquark from $ - L^+/2 < x^+ < L^+/2$ to $y^+ > L^+/2$, reads
\begin{align} 
\label{eq:In_aft_AntiQ}
S_{F,\bar{q}}^{i.a.} (x, y) & = (-1) \int \frac{d^{D-1} \underline{p}}{(2 \pi)^{D-1}}  \frac{\theta\left(-p^{+}\right)}{2 p^+}  e^{i y \cdot \check{p}} e^{-i x^- p^+ } e^{i \mathbf{x} \cdot \mathbf{p} }  
\nn \\
&  \hspace{5cm}
\times
\left[ 1 - \frac{\gamma^+ \gamma^i}{2 p^+} i \overrightarrow{D_{\mathbf{y}^i}^F} \right]  \mathcal{U}^{\dagger}_F\left( y^+,x^+, \mathbf{y} \right) (\slashed{\check{p}}+m)  .
\end{align}
In the expressions above, the forward and backward covariant derivatives are defined as $\overleftarrow{D_{\mathbf{y}^i}^F} = \overleftarrow{\partial}_{\mathbf{y}^i} - i g t \cdot A_i(y^+, \mathbf{y}) $ and $\overrightarrow{D_{\mathbf{y}^i}^F} = \overrightarrow{\partial}_{\mathbf{y}^i} + i g t \cdot A_i(y^+, \mathbf{y}) $. \\

The quark propagator from before the medium at point  $y^+ < -L^+/2$ to inside the medium to point $-L^+/2 < x^+ < L^+/2$, is given by 
\begin{align} 
S_{F,q}^{b.i.} (x, y)
& =\int \frac{d^{D-1} \underline{k}}{(2 \pi)^{D-1}}  \frac{\theta\left(k^{+}\right)}{2 k^+}  e^{i y \cdot \check{k}}  e^{-i x^- k^+ } e^{i \mathbf{x} \cdot \mathbf{k} } 
\nn  \\ & \hspace{4.5cm}
\times 
\left[ 1 - \frac{\gamma^+ \gamma^i}{2 k^+} i \overrightarrow{D_{\mathbf{x}^i}^F} \right] (\slashed{\check{k}}+m) \mathcal{U}_F\left( x^+,y^+, \mathbf{x} \right)   \; ,
\label{eq:quark_P_BI}
\end{align}
while before-to-inside antiquark propagator, corresponding to $ x^+ < - L^+/2$ to $- L^+/2 < y^+ < L^+/2$, reads
\begin{align} 
\label{eq:Bef_In-AntiQ}
S_{F,\bar{q}}^{b.i.} (x, y) &= (-1) \int \frac{d^{D-1} \underline{p}}{(2 \pi)^{D-1}}  \frac{\theta\left(-p^{+}\right)}{2 p^+}  e^{-i x \cdot \check{p}} (\slashed{\check{p}}+m)  
\nn \\
&  \hspace{4cm}
\times \mathcal{U}^{\dagger}_F\left( y^+, x^+, \mathbf{y} \right) \left[ 1 - \frac{\gamma^+ \gamma^i}{2 p^+} i \overleftarrow{D_{\mathbf{y}^i}^F} \right] e^{i y^- p^+ } e^{-i \mathbf{y} \cdot \mathbf{p} } .
\end{align}

The next propagator that one should consider is the inside-to-inside quark propagator with $ -L^+/2< x^+ < L^+/2$ to $- L^+/2 < y^+ < L^+/2$. In \cite{Altinoluk:2024dba}, the inside-to-inside quark and antiquark propagators have been computed separately assuming $x^+>y^+$ for the quark and $y^+>x^+$ for the antiquark propagators. However, for most of the inside-to-inside configurations both points are assumed to be inside the medium but no ordering between them is assumed. In that case, one should sum the separately computed quark and antiquark propagators. Hence, assuming only $ -L^+/2< x^+ < L^+/2$ to $- L^+/2 < y^+ < L^+/2$, the full inside-to-inside quark propagator can be written as 
\begin{align}
S_{F}^{\text { i.i. }}  (x, y) & =  \int \frac{d k^{+}}{2 \pi} \frac{1}{2 k^{+}} e^{-i k^{+}\left(x^{-}-y^{-}\right)} \bigg \{i \gamma^{+} \delta\left(x^{+}-y^{+}\right) \delta^{(D-2)} (\mathbf{x}-\mathbf{y}) 
\nonumber \\ 
&+ \left[k^{+} \gamma^{-}+m+i \gamma^i \overrightarrow{D}_{\mathbf{x}^i}^F\right] \frac{\gamma^{+}}{2 k^{+}} \int d^{D-2} \mathbf{z} \; \delta^{(D-2)}(\mathbf{x}-\mathbf{z}) \delta^{(D-2)} (\mathbf{z}-\mathbf{y}) \nonumber \\ 
& \times\left[\theta\left(x^{+}-y^{+}\right) \theta\left(k^{+}\right) \mathcal{U}_F\left(x^{+}, y^{+} , \mathbf{z} \right)-\theta\left(y^{+}-x^{+}\right) \theta\left(-k^{+}\right) \mathcal{U}_F^{\dagger}\left(y^{+}, x^{+}, \mathbf{z} \right)\right] 
\nonumber \\ 
& \left.\times\left[k^{+} \gamma^{-}+m-i \gamma^j \overleftarrow{D}_{\mathbf{y}^j}^F\right]\right\} \; ,
\label{Eq:Ins-Ins-Quark}
\end{align}
where the first term corresponds to the instantaneous term from the vacuum quark propagator. 

The last propagator that should be considered is the inside-to-inside gluon propagator, with $ -L^+/2< x^+ < L^+/2$ to $- L^+/2 < y^+ < L^+/2$. The detailed derivation of this propagator is presented in \cite{Altinoluk:2024dba}. As in the case of the inside-to-inside quark propagator, one should consider both orderings with $x^+>y^+$ and $y^+>x^+$. Taking into both orderings into account, full inside-to-inside gluon propagator reads 
\begin{align}
\label{eq:in-in_gluon}
G_F^{\mu \nu}(x, y)^{ \rm i.i.} 
&  = 
\int  \frac{d k^{+}}{2 \pi}  \frac{1}{2 k^{+}}  e^{-i k^{+}(x^{-}-y^{-})}
\bigg\{ \frac{2 i}{k^{+}} \, g^{\mu +} g^{\nu +} \, \delta\left(x^{+}-y^{+}\right) \,  \delta^{(D-2)} (\mathbf{x}-\mathbf{y})
 \nonumber \\ 
 &  + 
 \Big[g_i^\mu-\frac{i g^{\mu +}}{k^{+}} \overrightarrow{D}_{\mathbf{x}^i}^A\Big] 
 \int d^{D-2} \mathbf{z} \,  \delta^{(D-2)} (\mathbf{x}-\mathbf{z}) \,  \delta^{(D-2)} (\mathbf{z}-\mathbf{y})  
\nonumber \\ 
& \times 
\Big[\theta\left(x^{+}-y^{+}\right) \theta\left(k^{+}\right) \mathcal{U}_A\left(x^{+}, y^{+}, \mathbf{z} \right)  
- 
\theta\left(y^{+}-x^{+}\right) \theta\left(-k^{+}\right) \mathcal{U}_A^{\dagger}\left(y^{+}, x^{+}, \mathbf{z} \right)\Big]
\nonumber \\
& \times 
\Big[g_i^\nu+\frac{i g^{\nu +}}{k^{+}} \overleftarrow{D}_{\mathbf{y}^i}^A\big]\bigg\}  \; .
\end{align}

This concludes the discussion of the special propagator configurations required for computing the NEik corrections to the inclusive DIS cross section. 

%%%%%%%%%%%%%%%%%%%%%%%%%%%%%%%%%%%%%%%%%%%%%%%%%%%%%%%%%%%%%%%%%%%%%%%%%%%%%%%%%
%%%%%%%%%%%%%%%%%%%%%%%%%%%%%%%%%%%%%%%%%%%%%%%%%%%%%%%%%%%%%%%%%%%%%%%%%%%%%%%%%
\section{NEik DIS cross-section at the leading order}
\label{Sec:LOCrossSec}
%%%%%%%%%%%%%%%%%%%%%%%%%%%%%%%%%%%%%%%%%%%%%%%%%%%%%%%%%%%%%%%%%%%%%%%%%%%%%%%%%
%%%%%%%%%%%%%%%%%%%%%%%%%%%%%%%%%%%%%%%%%%%%%%%%%%%%%%%%%%%%%%%%%%%%%%%%%%%%%%%%%

The inclusive DIS cross section at NEik accuracy was computed recently in \cite{Altinoluk:2025ang}, where the analysis focused exclusively on NEik corrections arising from $t$-channel quark exchanges. It was shown that quark background field contributions to inclusive DIS cross section start at ${\cal O}({\alpha_s}^0)$ and can be expressed in terms of quark and antiquark collinear parton distribution functions (PDFs). Furthermore, it was argued that the NEik contributions induced by quark background field coincide with the low $x$ expansion of the parton model results for inclusive DIS. Although the primary aim of this manuscript is to compute quark background field contributions to inclusive DIS at ${\cal O}({\alpha_s})$, i.e. at NLO, in this section we summarize the main steps of the calculation of the LO contributions for completeness. 

Let us consider the elastic scattering of a photon on a dense target, with an incoming photon with momentum $q$ and polarization $\lambda_1$, and outgoing photon with momentum $q'$ and polarization $\lambda_2$. Both incoming and outgoing photons are virtual and can be longitudinal or transverse. In $D$ dimensions, the photon-to-photon transition matrix can be defined from the $S$-matrix as
\begin{equation}
    S_{\lambda_1, \lambda_2} = (2 \pi)^D \delta^{D} (q-q') \delta_{\lambda_1, \lambda_2} + 2 q^+ (2 \pi) \delta (q^+-q'^+) i \mathcal{M}_{\lambda_1, \lambda_2} (q,q') \; ,
    \label{Eq:S_M_matrix_relation}
\end{equation}
with the first term corresponding to the possibility of no scattering of the photon with the target. The forward elastic scattering amplitude is then given by the transition amplitude with same initial and final states: $q = q'$ and $\lambda_2 = \lambda_1$. 
Using the optical theorem, the total cross for longitudinal and transverse photons can be written as 
\begin{align}
    \sigma^{\gamma^*}_{L} & = 2 \; {\rm Im} \; \big\langle \mathcal{M}_{L, L} (q,q) \big\rangle = 2 \; {\rm Re} \big\langle (-i) \mathcal{M}_{L, L} (q,q) \big\rangle \; , 
    \label{Eq:LongCrossDef} \\
    \sigma^{\gamma^*}_{T} &= \frac{1}{D-2} \sum_{\lambda} 2 \; {\rm Im} \; \big\langle \mathcal{M}_{\lambda, \lambda} (q,q) \big\rangle = \frac{1}{D-2} \sum_{\lambda} 2 \; {\rm Re} \big\langle (-i) \mathcal{M}_{\lambda, \lambda} (q,q) \big\rangle \; ,
    \label{Eq:TransvCrossDef}
    \end{align}
where we averaged over the $(D-2)$ transverse polarizations $\lambda$ when writing the total cross section for transverse photon.  The averaging of the forward elastic scattering amplitudes in Eqs. \eqref{Eq:LongCrossDef} and \eqref{Eq:TransvCrossDef} are CGC-like target averages. For a given color operator ${\cal O}$, the CGC-like average $\langle{\cal O} \rangle$ and the quantum expectation value $\langle p_{t}|\mathcal{O}|p_{t} \rangle$ in the state $|p_t\rangle$ of the target are related to each other via \cite{Altinoluk:2023qfr}   
\begin{equation}
    \left\langle \mathcal{O}\right\rangle = \lim_{p'_{t}\rightarrow p_{t}}\frac{\langle p'_{t}|\mathcal{O}|p_{t} \rangle}{\langle p'_{t}|p_{t}\rangle} \; ,
   \label{Eq:DefForwMatrElem}
\end{equation}
where the target states are normalized as 
\begin{equation}
    {\langle p'_{t}|p_{t}\rangle} = 2 p^-_{t} (2\pi)^{D-1} \delta(p^-_{t}-p'^-_{t})\delta^{(D-2)}(\mathbf{p}_{t}-\mathbf{p}'_{t}) \; .
\end{equation}
\begin{figure}
    \centering
    \includegraphics[width=0.47 \linewidth]{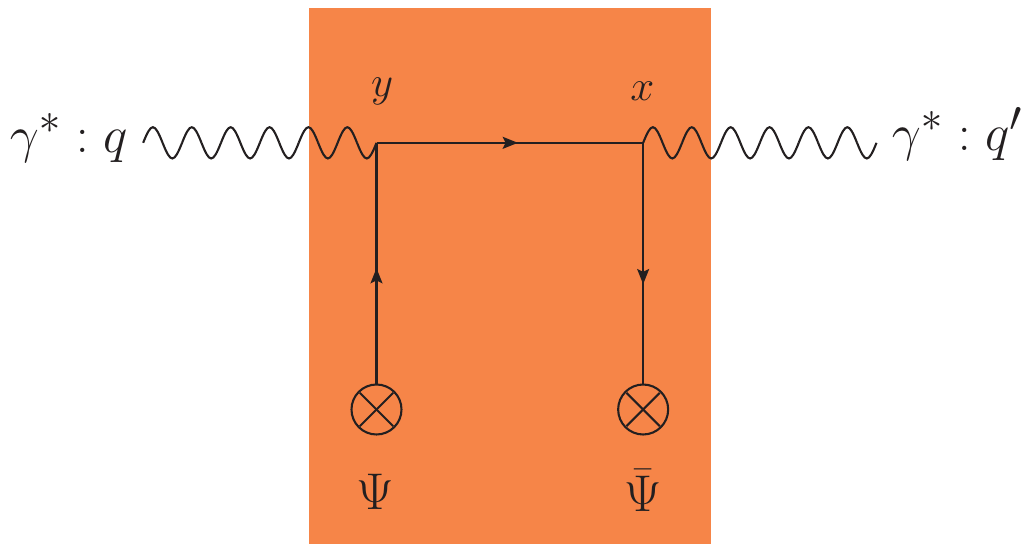} \hspace{0.5 cm}
    \includegraphics[width=0.47 \linewidth]{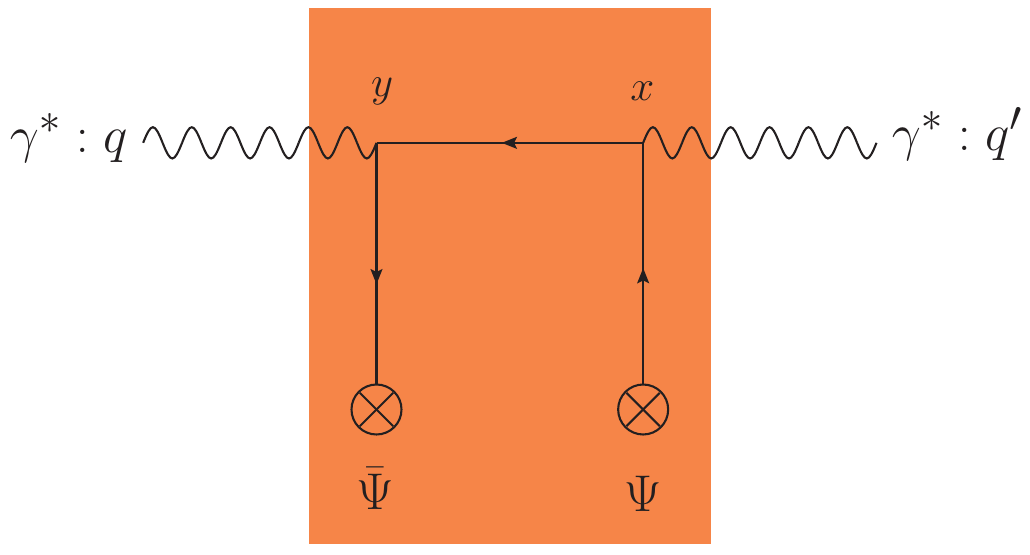}
    (a) \hspace{7.2 cm} (b)
    \caption{(a) Leading order contribution to DIS at the NEik accuracy from the quark background of the target. (b) Leading order contribution to DIS at the NEik accuracy from the antiquark background of the target.}
    \label{fig:LODISNEik}
\end{figure}

At LO, the photon-target scattering includes two contributions, one probing the quark and the second probing the antiquark background of the target (see Fig. \ref{fig:LODISNEik}). Let us first focus on the quark contribution of one flavor $f$ to the NEik inclusive DIS cross section at LO. To study this contribution, we consider the uncontracted $S$-matrix element, $S^{\mu\nu}$, which is related to the S-matrix element via 
\begin{align}
S = \varepsilon_{\gamma_{\lambda}, \mu} \, \varepsilon_{\gamma_{\lambda}, \nu}^{*}  \, S^{\mu \nu} \; ,
\end{align}
with $\varepsilon_{\gamma_{\lambda}, \mu}$ is the photon polarization vector. The uncontracted $S$-matrix element for the quark contribution, represented diagrammatically in (a) of Fig. \ref{fig:LODISNEik}, reads 
\begin{align}
\hspace{-0.4cm}
\label{eq:Smunu_LO}
    [S_{{\rm LO}, \Psi}]^{\mu \nu} & = - e_f^2 \; g^2 \int d^D y \int d^D x \; e^{i q' \cdot x -i q \cdot y} \; \mathcal{T}: \overline{\Psi}^{(-)} \!(\underline{x}) \, \gamma^{\mu}\,  S_{F} (x, y)^{\rm i.i. } \gamma^{\nu} \, \Psi^{(-)} (\underline{y}) \nonumber \\ 
     &=- e_f^2 \; g^2 \int d^D y \int d^D x \; e^{i q' \cdot x -i q \cdot y} \; \mathcal{T}: \overline{\Psi} (\underline{x}) \, \frac{\gamma^{-} \gamma^{+}}{2} \gamma^{\mu} S_{F} (x, y)^{
     \rm i.i. } \gamma^{\nu} \frac{\gamma^{+} \gamma^{-}}{2} \Psi(\underline{y}) \; ,
\end{align}
where ${\cal T}$ represents the light cone-time ordering along the $+$ direction and $S_{F} (x, y)^{\rm i.i. } $ is the total inside-to-inside quark propagator given in Eq. \eqref{Eq:Ins-Ins-Quark}. In the second equality, we used the explicit expressions for the good component of the quark background field given in Eq. \eqref{Eq:GoodComp}.  For simplicity of notation, the flavor index $f$ of the quark field $\Psi$ is kept implicit. The longitudinal photon contribution can be obtained by using the effective polarization vector
\begin{equation}
    \varepsilon_{\gamma_{L}, \mu} = \frac{Q}{q^+} {g_{\mu}}^+ \; ,
    \label{Eq:LonPolVec}
\end{equation}
leading to a result proportional to $[S_{{\rm LO}, \Psi}]^{+ +}$ which clearly vanishes due to the Dirac structure. Therefore, the NEik corrections stemming from $t$-channel quark exchange to the inclusive DIS cross section vanish for a longitudinal photon (or equivalently that correction to the $F_L$ structure function vanishes).

To compute the inclusive DIS cross section for a transverse photon, one must average over $(D-2)$ photon polarizations and then perform the sum over photon polarization, see Eq. \eqref{Eq:TransvCrossDef}, which yields 
\begin{align}
    \big[S_{{\rm LO}, \Psi}\big]_{\rm T} & = \frac{- g_{\perp}^{\mu \nu} }{D-2}\big[S_{{\rm LO}, \Psi}\big]_{\mu \nu} \nonumber \\
    & = \frac{e_f^2 \; g^2}{D-2} \int d^D y \int d^D x \; e^{i q' \cdot x -i q \cdot y} \; \mathcal{T}: \overline{\Psi} (\underline{x}) \frac{\gamma^{-} }{2} \gamma_{\perp \mu} \gamma^{+} S_{F} (x, y)^{
     \rm i.i. } \gamma^{+} \gamma^{\mu}_{\perp} \frac{ \gamma^{-}}{2} \Psi(\underline{y}) \; .
\label{eq:S_quark_LO_T}
\end{align}
Using the explicit expression for the inside-to-inside quark propagator given in Eq. \eqref{Eq:Ins-Ins-Quark}, in Eq. \eqref{eq:S_quark_LO_T}, one realizes that instantaneous contribution of the inside-to-inside quark propagator vanishes, and the $\gamma^{+} S_{F} (x, y)^{\rm i.i. } \gamma^{+}$ structure inside Eq. \eqref{eq:S_quark_LO_T} can be written as 
\begin{align}
&
   \gamma^{+} S_{F} (x, y)^{\text { i.i. }} \gamma^{+} = \delta^{(D-2)}(\mathbf{x}-\mathbf{y}) \gamma^{+} \int \frac{d k^{+}}{2 \pi}  e^{-i k^{+}\left(x^{-}-y^{-}\right)}  \nonumber \\ 
   &
   \times \bigg \{  \left[\theta\left(x^{+}-y^{+}\right) \theta\left(k^{+}\right) \mathcal{U}_F\left(x^{+}, y^{+} ; \mathbf{z} \right)-\theta\left(y^{+}-x^{+}\right) \theta\left(-k^{+}\right) \mathcal{U}_F^{\dagger}\left(y^{+}, x^{+}, \mathbf{z} \right)\right] \bigg \} \; .
   \label{Eq:LONEik1}
\end{align}
The two terms in Eq. \eqref{Eq:LONEik1} correspond to opposite light-cone time ordering along $+$ direction. After plugging Eq. \eqref{Eq:LONEik1} back into Eq. \eqref{eq:S_quark_LO_T}, integrations over $x^-$ and $y^-$ can be performed trivially. Upon integration, the first term in Eq. \eqref{Eq:LONEik1} becomes proportional to $\theta (q^+)$, while the second term becomes proportional to $\theta (-q^+)$. However, in our setup the photon moves with large $+$ momentum component. Therefore, only the term that is proportional to $\theta (q^+)$ contributes to the $S$-matrix for a transversely polarized photon and the result reads 
\begin{align}
\label{eq:Eq:LONEik2}
     [S_{{\rm LO}, \Psi}]_{\rm T} = - 2 q^{+} (2 \pi) \delta (q^+-q'^+)  \frac{e_f^2 \; g^2}{  4 q^+p_t^-} 2 p_t^{-} \int d^{D-2} \mathbf{y}  \int_{-\frac{L^+}{2}}^{\frac{L^+}{2}} d x^+ \int_{-\frac{L^+}{2}}^{\frac{L^+}{2}} d y^+ \theta (x^+ - y^+) \nonumber \\ \times e^{i q'^- x^+ -i q^- y^+} \; \mathcal{T}_+ : \overline{\Psi} (x^+, \mathbf{y}) \mathcal{U}_F (x^+, y^+, \mathbf{y}) \frac{\gamma^{-}}{2} \Psi(y^+, \mathbf{y}) \; ,
\end{align}
Since the ordering between $x^+$ and $y^+$ is explicit in Eq. \eqref{eq:Eq:LONEik2} is explicit thanks to the $\theta$-function, the action of the $\mathcal{T}_+$ is now trivial and can be neglected. We can now consider the imaginary part of the LO forward scattering amplitude for transversely polarized photon, which is written in terms of the $S$-matrix as 
\begin{align}
&
    2 \; {\rm Im} : [\mathcal{M}_{{\rm LO}, \Psi}]_{\rm T}  = 2 \; {\rm Im} : \frac{(-i) [S_{{\rm LO}, \Psi}]_{\rm T}}{2 q^{+} (2 \pi) \delta (q^+-q'^+) } 
    \nonumber \\ 
    & =  \frac{e_f^2 \; g^2}{ q^+ }  \int d^{D-2} \mathbf{y}  \int_{-\frac{L^+}{2}}^{\frac{L^+}{2}} d x^+ \int_{-\frac{L^+}{2}}^{\frac{L^+}{2}} d y^+ \theta (x^+ - y^+)  {\rm Re} : \overline{\Psi} (x^+, \mathbf{y}) \mathcal{U}_F (x^+, y^+, \mathbf{y}) \frac{\gamma^{-}}{2} \Psi(y^+, \mathbf{y}) \; .
    \label{Eq:LOPsiBeforeTrans}
\end{align}
Taking explicitly its real part and considering its quantum expectation value between the target states, one simply gets 
\begin{align}
\label{eq:matrix_elem_1}
 2 \langle p_t'| \; {\rm Im} : [\mathcal{M}_{{\rm LO}, \Psi}]_{\rm T} \; | p_t \rangle & = \frac{e_f^2 \; g^2}{  2 q^+ } \int d^{D-2} \mathbf{y}  \int_{-\frac{L^+}{2}}^{\frac{L^+}{2}} d x^+ \int_{-\frac{L^+}{2}}^{\frac{L^+}{2}} d y^+ 
 \nn \\
 & \hspace{2cm}
  \times
 \langle p_t'| \overline{\Psi} (x^+, \mathbf{y}) \mathcal{U}_F (x^+, y^+, \mathbf{y}) \frac{\gamma^{-}}{2} \Psi(y^+, \mathbf{y}) | p_t \rangle \; .
\end{align}
As discussed in \cite{Altinoluk:2023qfr}, the translation of a local operator is obtained by the action of momentum operator $\hat{p}_{\mu}$ as 
\begin{align}
\hat{\mathcal O}(x)= e^{i \hat{p} \cdot y }  \mathcal{O} (x-y) e^{-i \hat{p} \cdot y } \; ,
\end{align}
so that the matrix elements of non-local operators obey
\begin{align}
\langle p'_t| \hat{\mathcal O}_1(x_1)\cdots \hat{\mathcal O}_n(x_n)|p_t\rangle 
& = \langle p'_t| e^{i\hat{p}\cdot y}\hat{\mathcal O}_1(x_1-y) \cdots \hat{\mathcal O}_n(x_n-y)e^{-i\hat{p}\cdot y}|p_t\rangle \nn \\
&= e^{i(p'_t - p_t)\cdot y} \, \langle p'_t| \hat{\mathcal O}_1(x_1-y)\cdots \hat{\mathcal O}_n(x_n-y)|p_t\rangle \, .  
\label{eq:non_loc_trans}
\end{align}
Using Eq. \eqref{eq:non_loc_trans} for performing a shift, Eq. \eqref{eq:matrix_elem_1} can be rewritten as
\begin{align}
\label{eq:matrix_elem_shifted}
   2 \langle p_t'| \; {\rm Im} : [\mathcal{M}_{{\rm LO}, \Psi}]_{\rm T} \; | p_t \rangle & = 
   \frac{e_f^2 \; g^2 \pi }{   2 q^+ p_t^- } 
   \Big[2 p_t^{-} (2 \pi)^{D-1} \delta (p_t'^- - p_t^-) \delta^{(D-2)} (\mathbf{p}_t' - \mathbf{p}_t)\Big]
   \nonumber \\ 
   & \times 
   \int \frac{d z^+}{2 \pi} \langle p_t| \overline{\Psi} (z^+, \mathbf{0}) \mathcal{U}_F (z^+, 0^+, \mathbf{0}) \frac{\gamma^{-}}{2} \Psi(0^+, \mathbf{0}) | p_t \rangle \; , 
\end{align}
where we have performed the trivial integrations and taken $L^+\to\infty$ using the arguments presented in footnote \ref{footnote_1}. The inclusive DIS cross section for transverse photon now can be obtained trivially by substituting Eq. \eqref{eq:matrix_elem_shifted} into Eq. \eqref{Eq:TransvCrossDef} and one gets 
\begin{align}
\label{eq:Xsec_tr_LO}
   \sigma_{{\rm LO}, \Psi}^{\rm T} = \frac{e_f^2 \; g^2 \pi }{ 2 q^+ p_t^- } \int \frac{d z^+}{2 \pi} \langle p_t| \overline{\Psi} (z^+, \mathbf{0}) \mathcal{U}_F (z^+, 0^+, \mathbf{0}) \frac{\gamma^{-}}{2} \Psi(0^+, \mathbf{0}) | p_t \rangle \, . 
   %= \frac{e_f^2 \; g^2 \pi }{ W^2} f_q (x=0) \; ,
\end{align}
{By analogy with the standard quark PDF definition, we define the background quark PDF as\footnote{{Note that this background quark distribution is different from the usual quark distribution. In the usual distribution, each field contains both the background and the fluctuation contribution. In particular, the usual quark distribution contains the contribution from the correlator of two quark fluctuation fields expressed as a quark propagator in gluon background. Such contribution is eikonal, and has been studied for example in Refs.~\cite{Hatta:2017cte,Bhattacharya:2025fnz,Kovchegov:2025yyl}. Here instead, we separate eikonal contributions, NEik, and so on at the cross section level, and focus in the present study on the NEik contributions, that we express in terms of the background quark contribution. By contrast, the ${\rm x}\rightarrow 0$ limit of the eikonal contribution studied in Refs.~\cite{Hatta:2017cte,Bhattacharya:2025fnz,Kovchegov:2025yyl} is included in the dipole factorization contribution to the cross section in our setup. In particular, note that in our power counting for the high-energy limit, the quark background is strongly suppressed outside of the target, whereas the quark fluctuation field typically has power tails in $z^+$, due to the behavior of the free quark propagator.}}}
\begin{align}
    q_f({\rm x}) = \int \frac{d z^+}{2 \pi} e^{-i {\rm x} p_t^- z^+} \langle p_t'| \overline{\Psi} (z^+, \mathbf{0}) \mathcal{U}_F (z^+, 0^+, \mathbf{0}) \frac{\gamma^{-}}{2} \Psi(0^+, \mathbf{0}) | p_t \rangle \; .
    \label{Eq:QuarkPDFDef}
\end{align}
in terms of the quark background field $\Psi$ and gluon background field $A^{\mu}$.
Finally, using Eqs. \eqref{eq:Xsec_tr_LO} and \eqref{Eq:QuarkPDFDef},  the inclusive DIS cross section for transverse photon can be written in terms of the background quark PDF as 
%
%\begin{align}
%\label{eq:Xsec_tr_LO_fin}
%   \sigma_{{\rm LO}, \Psi}^{\rm T}  = \frac{e_f^2 \; g^2 \pi }{ W^2} f_q (x=0) \; . 
%\end{align}
% 
\begin{align} 
\label{eq:Xsec_tr_LO_fin}
   \sigma_{{\rm LO}, \Psi}^{\rm T}  = e_f^2\,  \frac{(2\pi)^2 \alpha_{\text{em}}}{   2 q^+ p_t^- }   q_f ({\rm x}=0) +{\rm NNEik} \; . 
\end{align}
where we have used $\alpha_{\text{em}}=g^2/(4\pi)$. By using Eq. \eqref{sigma2F}, that provides the relation between the DIS structure functions and the inclusive cross section and using the fact that $x_{Bj}\sim Q^2/{ 2 q^+ p_t^-}$, one obtains the quark contribution to the transverse  structure function as 
\begin{align}
\label{FTx=0}
F_T(x_{Bj}, Q^2)\Big|_{\rm LO, \, \Psi} = e_f^2\, \frac{Q^2}{  2 q^+ p_t^-}\,  q_f({\rm x}=0) + {\rm NNEik} 
 = e_f^2\, x_{Bj}\,  q_f({\rm x}=0) + {\rm NNEik} \; .
\end{align}

As discussed in detail in Ref. \cite{Altinoluk:2025ang}, the correct value of ${\rm x}$ entering the parton distribution functions can be determined by computing corrections to the eikonal approximation at higher power orders. For instance, in the  back-to-back DIS dijet production, the resummation of a certain subset of corrections to the eikonal approximation was shown to generate an ${\rm x}$-dependent phase in the definition of the transverse-momentum-dependent (TMD) gluon distribution. Consequently, this subset of corrections beyond the eikonal approximation can be interpreted as the Taylor expansion of ${\rm x}$ times the gluon TMD around ${\rm x}=0$~\cite{Altinoluk:2024zom}. A similar phase factor is incorporated into the unintegrated gluon distribution in \cite{Boussarie:2021wkn,Boussarie:2020fpb} in order to capture part of the subeikonal effects in the context of DIS. 

In the present study, since the quark contribution to the DIS structure functions already appears at NEik order, determining the correct value of ${\rm x}$ entering the ${\rm x}$-dependent phase factor in the quark PDF in Eq. \eqref{Eq:QuarkPDFDef} would require the computation of the NNEik corrections, which lies beyond the scope of this manuscript. Nevertheless, guided by the collinear factorization framework for DIS, one expects the momentum fraction ${\rm x}$ entering the quark PDF to be $x_{Bj}$. In the high-energy regime, where $x_{Bj}\sim Q^2/{  2 q^+ p_t^-} \ll 1$ and since the DIS structure function is already explicitly proportional to $x_{Bj}$ in Eq. \eqref{FTx=0}, replacing ${\rm x}\to x_{Bj}$ in the quark PDF would only induce corrections starting at NNEik order. Given that the accuracy of the present calculation is limited to NEik order, one may therefore safely write 
\begin{align}
\label{FT_q_LO_fin}
F_T(x_{Bj}, Q^2)\Big|_{\rm LO, \, \Psi} = e_f^2\, x_{Bj}\,  q_f(x_{Bj}) + {\rm NNEik} \; .
\end{align}

The computation of the antiquark contribution to the DIS structure functions at NEik accuracy at LO (see (b) of Fig. \ref{fig:LODISNEik}) closely follows the one performed for the quark contribution. Therefore, we outline only the main steps of the calculation and omit the repetitive details. The uncontracted $S$-matrix element associated with the antiquark contribution is given by
\begin{align}
\label{eq:S_munu_antiq}
    [S_{{\rm LO}, \overline{\Psi}}]^{\mu \nu} =- e_f^2 \; g^2 \int d^D y \int d^D x \; e^{i q' \cdot x -i q \cdot y} \; \mathcal{T}: \overline{\Psi} (\underline{y}) \frac{\gamma^{-} \gamma^{+}}{2} \gamma^{\mu} S_{F} (y, x)^{
     \rm i.i. } \gamma^{\nu} \frac{\gamma^{+} \gamma^{-}}{2} \Psi(\underline{x}) \; . 
\end{align}
Here $S_{F} (y, x)^{\rm i.i. }$ denotes the full inside-to-inside quark propagator defined in Eq. \eqref{Eq:Ins-Ins-Quark}. Owing to the close similarity between the Dirac structures in Eq. \eqref{eq:S_munu_antiq} and Eq. \eqref{eq:Smunu_LO}, it is straightforward realize that, just as in the quark case, the contribution from the longitudinally polarized photon vanishes. On the other hand, the contribution from the transversely polarized photon reads
\begin{align}
    2 \; {\rm Im} : [\mathcal{M}_{{\rm LO}, \overline{\Psi}}]_{\rm T} 
   &  = 
    2 \; {\rm Im} : \frac{(-i) [S_{{\rm LO}, \overline{\Psi}}]_{\rm T}}{2 q^{+} (2 \pi) \delta (q^+-q'^+) } \nonumber \\
    &  = - \frac{e_f^2 \; g^2}{   q^+ }  \; {\rm Re} : \int d^{D-2} \mathbf{x}  \int_{-\frac{L^+}{2}}^{\frac{L^+}{2}} d y^+ \int_{-\frac{L^+}{2}}^{\frac{L^+}{2}} d x^+ \theta (x^+ - y^+)
     \nonumber \\ 
     & \hspace{3.5cm}
     \times 
     \mathcal{T}_+ : \overline{\Psi} (y^+, \mathbf{x}) \mathcal{U}_F (y^+, x^+, \mathbf{x}) \frac{\gamma^{-}}{2} \Psi(x^+, \mathbf{x}) \; .
    \label{Eq:LOPsiBarBeforeT}
\end{align}
Note that a relative minus sign appears when comparing the corresponding quark expression in Eq. \eqref{Eq:LOPsiBeforeTrans} with the antiquark expression in Eq. \eqref{Eq:LOPsiBarBeforeT}. This difference arises because, for the antiquark contribution, only the second term in the non-instantaneous part of the inside-to-inside quark propagator (Eq. \eqref{Eq:Ins-Ins-Quark}) contributes, whereas in the quark case it is the first term that survives. Upon implementing light-cone time ordering enforced by the operator ${\cal T}_+$, taking the real part, and evaluating the resulting expression as the quantum expectation value in the target states $|p_t\rangle$, one gets 
\begin{align}
    2 \langle p_t'| \; {\rm Im} : [\mathcal{M}_{{\rm LO}, \overline{\Psi}}]_{\rm T} \; | p_t \rangle  & = \frac{e_f^2 \; g^2}{  2 q^+}  \int d^{D-2} \mathbf{x}  \int_{-\frac{L^+}{2}}^{\frac{L^+}{2}} d y^+ \int_{-\frac{L^+}{2}}^{\frac{L^+}{2}} d x^+
     \nonumber \\ 
     & \hspace{-2cm}
     \times \langle p_t'| {\rm Tr_{D}} \left[ \mathcal{U}_F (\infty^+, x^+, \mathbf{x}) \Psi (x^+, \mathbf{x}) \overline{\Psi}(y^+, \mathbf{x}) \frac{\gamma^{-}}{2} \mathcal{U}_F^{\dagger} (\infty^+, y^+, \mathbf{x}) \right] | p_t \rangle \; .
\end{align}
Employing Eq. \eqref{eq:non_loc_trans} to perform the non-local translation, analogously to the quark case, the antiquark contribution to the inclusive DIS cross section for transverse photon can be written as
\begin{align}
\label{antiQ_LO}
 \sigma_{{\rm LO}, \overline{\Psi}}^{\rm T} &  
   %= \frac{e_f^2 \; g^2 \pi }{ W^2} \int \frac{d z^+}{2 \pi} \langle p_t| {\rm Tr_{D}} \left[ \mathcal{U}_F (\infty^+, 0^+, \mathbf{0}) \Psi (0^+, \mathbf{0}) \overline{\Psi}(z^+, \mathbf{0}) \frac{\gamma^{-}}{2} \mathcal{U}_F^{\dagger} (\infty^+, z^+, \mathbf{0}) \right] | p_t \rangle \nonumber \\ & 
   = e_f^2 \, \frac{(2\pi)^2 \, \alpha_{\text{em}} }{  2 q^+ p_t^-} \overline{q}_f ({\rm x}=0) \; ,
 %\end{split}  
\end{align}
with the background antiquark PDF is defined as 
\begin{align}
    \overline{q}_f ({\rm x}) = \int \frac{d z^+}{2 \pi} e^{i {\rm x} p_t^- z^+} \langle p_t| {\rm Tr_{D,c}} \left[ \mathcal{U}_F (\infty^+, 0^+, \mathbf{0}) \Psi (0^+, \mathbf{0}) \overline{\Psi}(z^+, \mathbf{0}) \frac{\gamma^{-}}{2} \mathcal{U}_F^{\dagger} (\infty^+, z^+, \mathbf{0}) \right] | p_t \rangle \; .
\end{align}
Using the relation between the total cross sections and the DIS structure functions given in Eq. \eqref{sigma2F}, the antiquark background contribution to the transverse structure function can be obtained from Eq. \eqref{antiQ_LO} straightforwardly. Moreover, the power counting argument justifying the replacement  ${\rm x}\to x_{Bj}$ in the antiquark distribution applies in exactly the same manner as for the quark contribution. One therefore readily obtains  
\begin{align}
\label{FT_LO_qbar_fin}
F_T(x_{Bj}, Q^2)\Big|_{{\rm LO}, \overline{\Psi}}= e_f^2\, x_{Bj}\,  \overline{q}_f({x_{Bj}})+{\rm NNEik} \; .
\end{align}

All in all, combining Eq. \eqref{FT_q_LO_fin} and \eqref{FT_LO_qbar_fin} and summing over the quark flavors, one obtains the LO expressions for the quark and antiquark background contributions to the DIS structure functions at NEik which reads 
\begin{align}
\label{FT_LO_fin}
F_T(x_{Bj},Q^2)\Big|_{{\rm LO}, \Psi+\overline{\Psi}}&=\sum_{f} e_f^2\, x_{Bj}\Big[ q_f(x_{Bj})+\overline{q}_f(x_{Bj})\Big]+{\rm NNEik} \, , \\
F_L(x_{Bj},Q^2)\Big|_{{\rm LO}, \Psi+\overline{\Psi}}&= 0 +{\rm NNEik}  \, . 
\end{align}
%  

%%%%%%%%%%%%%%%%%%%%%%%%%%%%%%%%%%%%%%%%%%%%%%%%%%%%%%%%%%%%%%%%%%%%%%%%%%%%%%%%
%%%%%%%%%%%%%%%%%%%%%%%%%%%%%%%%%%%%%%%%%%%%%%%%%%%%%%%%%%%%%%%%%%%%%%%%%%%%%%%%
\section{NEik DIS cross-section at the next-to-leading order: Fermion background field contribution}
\label{Sec:NLOCrossSecQuark}
\begin{figure}
    \centering
    \includegraphics[width=0.47 \linewidth]{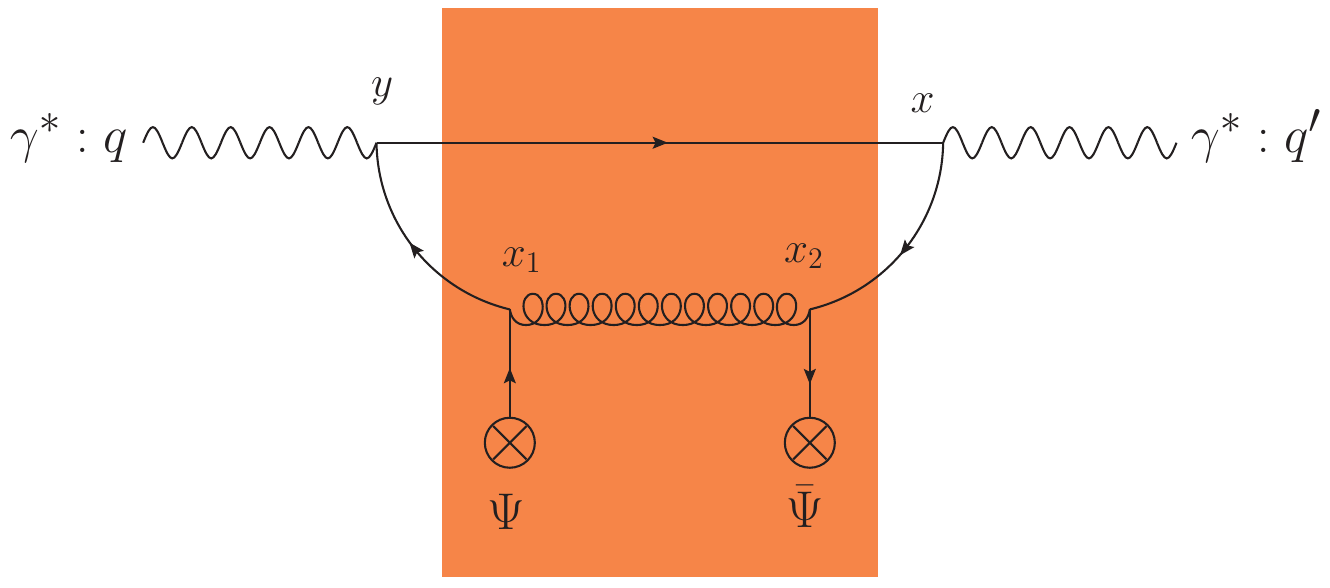} \hspace{0.5 cm}
    \includegraphics[width=0.47 \linewidth]{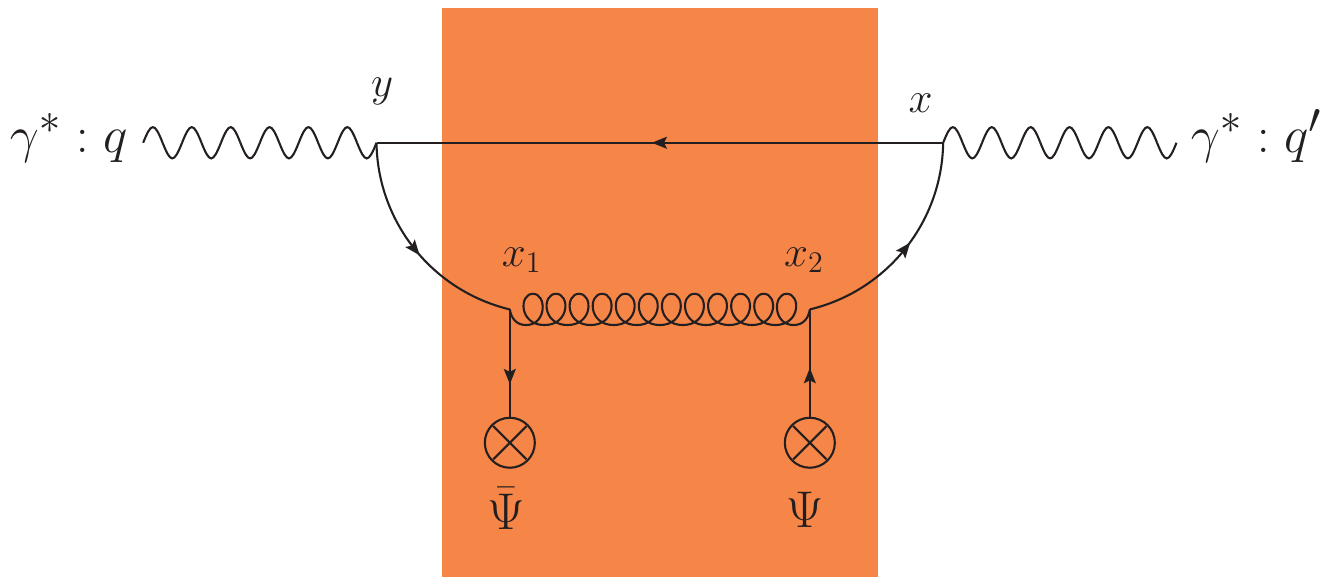}
    (a) \hspace{7.2 cm} (b)
    \caption{(a) Next-to-leading order contribution to DIS at the NEik accuracy from the quark background of the target. (b) Next-to-leading order contribution to DIS at the NEik accuracy from the antiquark background of the target.}
    \label{fig:NLODISNEik}
\end{figure}
%%%%%%%%%%%%%%%%%%%%%%%%%%%%%%%%%%%%%%%%%%%%%%%%%%%%%%%%%%%%%%%%%%%%%%%%%%%%%%%%
%%%%%%%%%%%%%%%%%%%%%%%%%%%%%%%%%%%%%%%%%%%%%%%%%%%%%%%%%%%%%%%%%%%%%%%%%%%%%%%%
We now turn to the computation of the NLO corrections to the inclusive NEik DIS cross section. In this section, we concentrate specifically on the NEik contributions generated by the quark background field of the target. As in the LO analysis presented in Sec. \ref{Sec:LOCrossSec}, the ${\mathcal O}(\alpha_s)$  corrections to the inclusive NEik cross section are obtained through the application of the optical theorem.

%%%%%%%%%%%%%%%%%%%%%%%%%%%%%%%%%%%%%%%%%%%%%%%%%%%%%%%%%%%%%%%%%%%%%%%%%%%%%%%%
\subsection{Quark background field contribution}
\label{Sec:QuarkCon}
%%%%%%%%%%%%%%%%%%%%%%%%%%%%%%%%%%%%%%%%%%%%%%%%%%%%%%%%%%%%%%%%%%%%%%%%%%%%%%%%
 Let us start our analysis by considering the ${\mathcal O}(\alpha_s)$ correction to the quark background field contribution to the inclusive NEik DIS cross section (see (a) of Fig. \ref{fig:NLODISNEik}). At NEik accuracy, this is provided by a configuration in which the incoming photon splits into a quark-antiquark pair before the medium.\footnote{ There is another configuration in which the incoming photon splits into quark-antiquark pair inside the medium. However, as discussed in \cite{Altinoluk:2022jkk}, this configuration is suppressed by one power of energy and therefore already counts as a NEik correction. Including interactions with quark background field of the target introduces a further suppression for this configuration and therefore pushes the resulting amplitude beyond NEik accuracy. Therefore such configurations can be safely neglected within the accuracy of our calculation.} For such a configuration, the uncontracted $S$-matrix element can be written as 
\begin{align}
    [S_{{\rm NLO}, \Psi}]^{\mu \nu} = e_f^2 \; g^2 \int_{-\infty^+}^{-\frac{L}{2}^+} d^D y \int_{ \frac{L}{2}^+}^{\infty^+} d^D x \; e^{i q' \cdot x -i q \cdot y} \; \mathcal{T}: {\rm Tr_{D,c}} \left[  \gamma^{\nu} S_{F,q}^{b.a.} (x, y) \gamma^{\mu}  S_{F, \bar{q}}^{\Psi} (y, x) \right] \, . 
    \label{Eq:QuarkConBeg}
\end{align}
with $S_{F,q}^{b.a.} (x, y) $ is the before-to-after quark propagator given in Eq. \eqref{eq:quark_BA}.   
Here, compared to its LO analog expression, we have introduced a relative sign together with trace over the Dirac indices, i.e. a factor of $(-1) {\rm Tr_D}$, in order to account for the ordering of the background quark fields $\Psi$ and $\overline{\Psi}$.  These quark background fields are now embedded inside the effective before-to-after antiquark propagator in a quark background which is denoted as $S_{F, \bar{q}}^{\Psi} (y,x)$ whose explicit expression reads 
\begin{align}
\label{eq:Eff_Anti_Q_1}
    S_{F, \bar{q}}^{\Psi} (y, x) & = - g_s^2  \int_{-\frac{L}{2}^+}^{\frac{L}{2}^+} d^D x_1 \int_{-\frac{L}{2}^+}^{\frac{L}{2}^+} d^D x_2 \; \nonumber \\ 
    &
    \times \, S_{F,\bar{q}}^{\rm b.i.} (y, x_1) \;t^b \gamma^{\rho} \, \Psi^{(-)} (\underline{x_1})\,  
    \big[G_{F, \rho \sigma}(x_1, x_2)^{ \rm i.i.}\big]^{ba} 
    \; \overline{\Psi}^{(-)} (\underline{x_2})\,  t^a \gamma^{\sigma} \, S_{F,\bar{q}}^{\rm i.a.} (x_2, x) \; ,
\end{align}
with the before-to-inside ($S_{F,\bar{q}}^{\rm b.i.}$) and inside-to-after ($S_{F,\bar{q}}^{\rm i.a.}$) antiquark propagators given in Eqs. \eqref{eq:Bef_In-AntiQ} and \eqref{eq:In_aft_AntiQ} respectively, while the inside-to-inside gluon propagator is provided in Eq. \eqref{eq:in-in_gluon}. 
Since $S_{F,\bar{q}}^{\rm b.i.}$ and $S_{F,\bar{q}}^{\rm i.a.}$
contain only contributions of negative light-cone momentum, and that the quark background field is taken in the static approximation at this order, only the contribution of negative light-cone momentum in the inside-to-inside gluon Feynman propagator can contribute to Eq.~\eqref{eq:Eff_Anti_Q_1} due to light-cone momentum conservation. Thus, only the second term of the non-instantaneous part of the inside-to-inside gluon propagator in Eq.~\eqref{eq:in-in_gluon} contributes, which enforces the ordering $x_1^+<x_2^+$ of the quark background field insertions, as represented on Fig. \ref{fig:NLODISNEik}. Upon substituting the explicit expressions of the propagators, the effective  before-to-after antiquark propagator in a quark background can be written as 
\begin{align}
\label{eq:Eff_Anti_Q_2}
 S_{F, \bar{q}}^{\Psi} (y, x) 
 & = g_s^2  
 \int_{-\frac{L}{2}^+}^{\frac{L}{2}^+} \hspace{-0.1 cm} d^D x_1 
 \int_{x_1^+}^{\frac{L}{2}^+} \hspace{-0.1 cm}  d^D x_2 
 \int \frac{d^{D-1} \underline{p_i}}{(2 \pi)^{D-1}}  \frac{\theta\left(-p_i^{+}\right)}{2 p_i^+}
  \int \frac{d^{D-1} \underline{p_f}}{(2 \pi)^{D-1}}  \frac{\theta(-p_f^{+})}{2 p_f^+} 
  \int \frac{d p^{+}}{2 \pi} \frac{\theta(-p^{+} )}{2 p^{+}} 
  \nonumber \\ 
  & \times 
  e^{- i p^+ (x_1^{-} - x_2^{-} ) -i y \cdot \check{p}_f + i x_1^- p_f^+ -i \mathbf{x}_1 \cdot \mathbf{p}_f +i x \cdot \check{p}_i - i x_2^- p_i^+ - i \mathbf{x}_2 \cdot \mathbf{p}_i  }
  \nn \\
  &\times 
   (\slashed{\check{p}}_f + m ) \, \bigg\{ \mathcal{U}_F \left( y^+, x_1^+, \mathbf{x}_1 \right)
    \bigg[ 1 - \frac{\gamma^+ \gamma^k}{2 p_f^+} i \overleftarrow{D}_{\mathbf{x}_1^k}^F \bigg] 
    t^b \gamma^{\rho} \Psi^{(-)} (\underline{x_1}) 
    \bigg[g_{i \rho}-\frac{i {g_{\rho}}^{+}}{p^{+}} \overrightarrow{D}_{\mathbf{x}_1^i}^A\bigg]
  \nonumber \\ 
  & \times 
  \int d^{D-2} \mathbf{z} \; \delta^{(D-2)} (\mathbf{x}_1-\mathbf{z}) \;  \delta^{(D-2)} (\mathbf{z}-\mathbf{x}_2) 
  \big[ \mathcal{U}_A \left(x_1^{+}, x_2^{+}, \mathbf{z} \right) \big]^{ba}
  \nonumber \\ 
   &\times 
 \left[g_{i \sigma} +\frac{i {g_{\sigma}}^+}{k^{+}} \overleftarrow{D}_{\mathbf{x}_2^i}^A\right] \overline{\Psi}^{(-)} (\underline{x_2}) t^a \gamma^{\sigma} \left[ 1 - \frac{\gamma^+ \gamma^j}{2 p^+} i \overrightarrow{D_{\mathbf{x}_2^j}^F} \right]  \mathcal{U}_F\left( x_2^+,x^+, \mathbf{x}_2 \right) \bigg\}(\slashed{\check{p}}_i+m) \; , 
\end{align}
where the covariant derivatives act only inside the curly brackets an not on the overall phase factors. Upon expressing the good components of the quark and antiquark fields using Eq. \eqref{Eq:GoodComp}, the Dirac structure of the effective propagator simplifies substantially, allowing  Eq. \eqref{eq:Eff_Anti_Q_2} to be recast into  
\begin{align}
\label{eq:Eff_Anti_Q_3}
    S_{F, \bar{q}}^{\Psi} (y, x) 
    & = 
    \frac{g_s^2}{4} 
    \int_{-\frac{L}{2}^+}^{\frac{L}{2}^+} \hspace{-0.1 cm} d x_1^+ 
    \int_{x_1^+}^{\frac{L}{2}^+} \hspace{-0.1 cm}  d x_2^+ 
    \int d^{D-2} \mathbf{x}_1  
    \int \frac{d p^{+}}{2 \pi} \frac{\theta(-p^{+} )}{2 p^{+}} 
     \int \frac{d^{D-2} \mathbf{p}_i}{(2 \pi)^{D-2}} 
      \int \frac{d^{D-2} \mathbf{p}_f}{(2 \pi)^{D-2}}  
    \nonumber \\ 
    & \hspace{-1cm}
    \times 
    %\int \frac{d^{D-2} \mathbf{p}_i}{(2 \pi)^{D-2}} 
    %\int \frac{d^{D-2} \mathbf{p}_f}{(2 \pi)^{D-2}}  
    e^{ -i y \cdot \check{p}_f -i \mathbf{x}_1 \cdot \mathbf{p}_f + i x \cdot \check{p}_i - i \mathbf{x}_1 \cdot \mathbf{p}_i  } 
    \frac{(\slashed{\check{p}}_f + m )}{2 p^+ } 
    \mathcal{U}_F \left( y^+, x_1^+, \mathbf{x}_1 \right)  
     t^b \gamma^{i} \gamma^+ \gamma^-  
      \Psi (x_1^+, \mathbf{x}_1) 
     \nonumber \\ 
     & \hspace{-1cm}
     \times 
         \big[ \mathcal{U}_A \left(x_1^{+}, x_2^{+}, \mathbf{x}_1 \right) \big]^{ba}  \; 
     \overline{\Psi} (x_2^+, \mathbf{x}_1)  
     \gamma^- \gamma^+ t^a \gamma^{i}  
     \mathcal{U}_F\left( x_2^+,x^+, \mathbf{x}_1 \right) 
     \frac{(\slashed{\check{p}}_i+m)}{2 p^+ } \bigg|_{p_i^+=p_f^+=p^+} \; ,
\end{align}
where we have used the fact that at NEik accuracy quark fields are treated within the static approximation and performed the integrations over $x_1^-$ and $x_2^-$.\footnote{ At this point, we note that this integration has implied $\theta (-p_i^+) = \theta (-p_f^+) = \theta (-p^+) $, which are identical to the one appearing in the term we selected in the inside-to-inside gluon propagator. Choosing the other term would have led to Heaviside $\theta$-functions with arguments of opposite sign.}  Finally, using Eq. \eqref{eq:Eff_Anti_Q_3}  for the effective before-to-after antiquark propagator and  Eq. \eqref{eq:quark_BA} for the before-to-after quark propagator, the uncontracted $S$-matrix element given in Eq. \eqref{Eq:QuarkConBeg} can be written as 
\begin{align}
\label{eq:S_NLO_mu_nu_1}
 &
 [S_{{\rm NLO}, \Psi}]^{\mu \nu} 
  = 
 \frac{e_f^2 \; g^2 \; g_s^2}{4} 
 \int_{-\infty^+}^{-\frac{L}{2}^+} d^D y 
 \int_{ \frac{L}{2}^+}^{\infty^+} d^D x 
 \int_{-\frac{L}{2}^+}^{\frac{L}{2}^+} \hspace{-0.1 cm} d x_1^+ 
 \int_{x_1^+}^{\frac{L}{2}^+} d x_2^+ 
 \int d^{D-2} \mathbf{x}_1 
 \int d^{D-2} \mathbf{x}_2 
 \nonumber \\ 
 & \times 
 \int \frac{d p^{+}}{2 \pi} \frac{\theta(-p^{+} )}{2 p^{+}} 
 \int \frac{d k^{+}}{2 \pi} \theta (k^+) 
 \int \frac{d^{D-2} \mathbf{p}_i}{(2 \pi)^{D-2}} 
 \int \frac{d^{D-2} \mathbf{p}_f}{(2 \pi)^{D-2}} 
  \int \frac{d^{D-2} \mathbf{k}_i}{(2 \pi)^{D-2}} 
  \int \frac{d^{D-2} \mathbf{k}_f}{(2 \pi)^{D-2}}  
  \nonumber \\ 
  & \times 
  e^{ i q' \cdot x -i q \cdot y -i y \cdot \check{p}_f + i x \cdot \check{p}_i + i y \cdot \check{k}_i - i x \cdot \check{k}_f } e^{ -i \mathbf{x}_1 \cdot (\mathbf{p}_f - \mathbf{p}_i) + i \mathbf{x}_2 \cdot (\mathbf{k}_i - \mathbf{k}_f)} 
  \nonumber \\ 
  & \times 
  \mathcal{T}_+ : {\rm Tr_{D,c}} 
  \bigg[ 
  \gamma^{\nu} \, \frac{(\slashed{\check{k}}_f+m)}{2 k^+} \, \gamma^{+} \, \mathcal{U}_F\left(\mathbf{x}_2  \right) 
  \frac{(\slashed{\check{k}}_i+m)}{2 k^+} \,  \gamma^{\mu} \, \frac{(\slashed{\check{p}}_f + m )}{2 p^+ } 
  \mathcal{U}_F \Big( -\frac{L}{2}^+, x_1^+, \mathbf{x}_1 \Big)
   t^b \gamma^{i} \gamma^+ \gamma^-  
  \nonumber \\ 
  &  \times
    \Psi (x_1^+ , \mathbf{x}_1) \big[ \mathcal{U}_A \left(x_1^{+}, x_2^{+}, \mathbf{x}_1 \right) \big]^{ba}  
   \overline{\Psi} (x_2^+, \mathbf{x}_1) 
  %\nn\\ 
   %& \times
   \gamma^- \gamma^+ t^a \gamma^{i}  \mathcal{U}_F\left( x_2^+, \frac{L}{2}^+, \mathbf{x}_1 \right) \frac{(\slashed{\check{p}}_i+m)}{2 p^+ } \bigg] \Bigg|_{\substack{p_i^+ = p_f^+ = p^+ \\ k_i^+ = k_f^+ = k^+}}  .
\end{align}
One can immediately realize that in Eq. \eqref{eq:S_NLO_mu_nu_1}, the dependence on the positions $x$ and $y$ appears only on the phase factors and factorizes from the rest. Therefore, one can perform the integrals over $x^-$, $y^-$, $\x$ and $\y$ trivially which yields to $\delta$-functions of initial and final transverse and $+$ momenta.  The integrations over $y^+$ and $x^+$ can also be performed in a trivial manner yielding to the initial and final energy denominators:
\begin{align}
    \int_{-\infty^+}^{- \frac{L}{2}^+} d y^+ e^{-i y^+ (q^- + \check{p}_f - \check{k}_i ) } 
    & \simeq 
    \frac{i}{(q^- + \check{p}_f - \check{k}_i )} 
    = 
    i \frac{2 k^+ (k^+-q^+)}{q^+\big[\mathbf{k}_i^2 + {\bar Q}^2\big]} \; , 
    \nonumber \\
    \int^{\infty^+}_{\frac{L}{2}^+} d x^+ e^{i x^+ (q^- + \check{p}_i - \check{k}_f ) } 
    & \simeq \frac{i}{(q^- + \check{p}_i - \check{k}_f )} = i \frac{2 k^+ (k^+-q^+)}{q^+\big[\mathbf{k}_f^2 + {\bar Q}^2\big]} \; ,
    \label{Eq:Energy denominator}
\end{align}
where we have approximated the longitudinal width of the target to be infinitely small for the $x^+$ and $y^+$ integrals.\footnote{
The integration domain of, for example, $x^+$ can be decomposed as
\begin{gather*}
 \int_{\frac{L}{2}^+}^{\infty^+} d x^+
 \;=\;
 \int_{0}^{\infty^+} d x^+ \,\cdots
 \;-\;
 \int_{0}^{\frac{L}{2}^+} d x^+ \,\cdots \, .
\end{gather*}
The second term already contributes only at NEik order, i.e.\ it is of $\mathcal{O}(L^+)$ in the $L^+ \to 0$ expansion. Consequently, in the quark background contribution it yields a next-to-NEik (NNEik) correction.
}
Moreover, we have defined 
\begin{equation}
    \bar{Q}=\sqrt{\frac{k^+ (q^+ - k^+)}{(q^+)^2} Q^2 + m^2} \; .
\end{equation}

All in all, after performing the integrations over $x$ and $y$ in Eq. \eqref{eq:S_NLO_mu_nu_1}, the uncontracted $S$-matrix element for the quark background contribution can be written as 
\begin{align}
[S_{{\rm NLO}, \Psi}]^{\mu \nu} 
&= 
- e_f^2 \; g^2 \; g_s^2 (2 \pi) \delta (q^+-q'^+) 
\int d^{D-2} \mathbf{x}_1 
\int d^{D-2} \mathbf{x}_2   
\int_0^{q^+} \frac{d k^{+}}{2 \pi} \frac{(k^+)^2 (k^+-q^+)}{2 (q^+)^2} \nonumber
\\ 
& \hspace{-1cm}
 \times 
 \int \frac{d^{D-2} \mathbf{k}_i}{(2 \pi)^{D-2}} 
 \int \frac{d^{D-2} \mathbf{k}_f}{(2 \pi)^{D-2}}  
 \frac{e^{ - i \mathbf{x}_{12} \cdot (\mathbf{k}_f - \mathbf{k}_i)}}{[\mathbf{k}_i^2 + \bar{Q}^2][\mathbf{k}_f^2 + \bar{Q}^2]} 
 \int_{-\frac{L}{2}^+}^{\frac{L}{2}^+} \hspace{-0.1 cm} d x_1^+ 
 \int_{x_1^+}^{\frac{L}{2}^+} d x_2^+  
 \nonumber \\ 
 &  \hspace{-1cm}
 \times 
 \mathcal{T}_+ : {\rm Tr_{D,c}} \bigg\{ \gamma^{\nu} \frac{(\slashed{\check{k}}_f+m)}{2 k^+} \gamma^{+} \mathcal{U}_F\left(\mathbf{x}_2  \right) \frac{(\slashed{\check{k}}_i+m)}{2 k^+}  \gamma^{\mu} \frac{(\slashed{\check{p}}_f + m )}{2 (k^+-q^+) }
 \nn \\
 & \hspace{1cm}
 \times
  \mathcal{U}_F \left( -\frac{L}{2}^+ , x_1^+, \mathbf{x}_1 \right) t^b \gamma^{i} \gamma^+ \gamma^-   
   \Psi (x_1^+ , \mathbf{x}_1) 
   \big[ \mathcal{U}_A \left(x_1^{+}, x_2^{+}, \mathbf{x}_1 \right) \big]^{ba}  
  \nn \\
 & \hspace{1cm} 
 \times
  \overline{\Psi} (x_2^+, \mathbf{x}_1) 
   \gamma^- \gamma^+ t^a \gamma^{i}  \mathcal{U}_F\left( x_2^+, \frac{L}{2}^+ , \mathbf{x}_1 \right) \frac{(\slashed{\check{p}}_i+m)}{2 (k^+-q^+) } \bigg\} \bigg|_{\substack{p_i^+ = p_f^+ = k^+ - q^+ \\ k_i^+ = k_f^+ = k^+ \\ \mathbf{k}_i = \mathbf{p}_f \; , \; \mathbf{p}_i = \mathbf{k}_f}}  \; .
   \label{Eq:QuarkBackConFullyGeneral}
   \end{align}

Employing Eq. \eqref{Eq:LonPolVec} for the polarization vector of a longitudinally polarized photon, the NLO corrections to the quark contribution to the S-matrix element can be obtained straightforwardly from Eq. \eqref{Eq:QuarkBackConFullyGeneral}. For this case, the Dirac structure simplifies substantially,  and Eq. \eqref{Eq:QuarkBackConFullyGeneral} reduces to  
\begin{align}
[S_{{\rm NLO}, \Psi}]_{\rm L} 
 &
= 
- e_f^2 \; g^2 \; g_s^2 (2 \pi) \delta (q^+-q'^+) 
\int d^{D-2} \mathbf{x}_1 
\int d^{D-2} \mathbf{x}_2   
\int_0^{q^+} \frac{d k^{+}}{2 \pi} \frac{(k^+)^2 (k^+-q^+)}{2 (q^+)^2} \nonumber
\\ 
& \hspace{-1.5cm}
 \times 
 \int \frac{d^{D-2} \mathbf{k}_i}{(2 \pi)^{D-2}} 
 \int \frac{d^{D-2} \mathbf{k}_f}{(2 \pi)^{D-2}}  
 \frac{e^{ - i \mathbf{x}_{12} \cdot (\mathbf{k}_f - \mathbf{k}_i)}}{[\mathbf{k}_i^2 + \bar{Q}^2][\mathbf{k}_f^2 + \bar{Q}^2]} 
 \int_{-\frac{L}{2}^+}^{\frac{L}{2}^+} \hspace{-0.1 cm} d x_1^+ 
 \int_{x_1^+}^{\frac{L}{2}^+} d x_2^+  
 \Big\{ 2(D-2)  
  \overline{\Psi} (x_2^+, \mathbf{x}_1)  \gamma^-
 \nonumber \\ 
 &  \hspace{-1.5cm}
 \times 
 t^a 
 \mathcal{U}_F\Big( x_2^+, \frac{L}{2}^+ , \mathbf{x}_1 \Big) 
 \mathcal{U}_F(\mathbf{x}_2) 
 \mathcal{U}_F \Big( -\frac{L}{2}^+, x_1^+, \mathbf{x}_1 \Big)
 t^b \Psi (x_1^+ , \mathbf{x}_1) \big[ \mathcal{U}_A \left(x_1^{+}, x_2^{+}, \mathbf{x}_1 \right) \big]^{ba}\Big\} \; .
\label{Eq:QuarkBackCon_Long_1}
\end{align}
The transverse momentum integrals over  $\k_i$ and $\k_f$ factorize from the rest of the expression and can be performed by means of the integral in Eq. \eqref{Eq:AppTransvMomInt1}. The result reads 
\begin{align}
\label{eq:trans_int}
    \int \frac{d^{2-2 \epsilon} \mathbf{k}_i}{(2 \pi)^{2-2 \epsilon}}  \frac{e^{- i \mathbf{k}_i \cdot \mathbf{x}_{12} }}{\mathbf{k}_i^2 + \bar{Q}^2}  \int \frac{d^{2-2 \epsilon} \mathbf{k}_f}{(2 \pi)^{2-2 \epsilon}}  \frac{e^{- i \mathbf{k}_f \cdot \mathbf{x}_{12} }}{\mathbf{k}_f^2 + \bar{Q}^2} = \frac{K_{\epsilon}^2 (\bar{Q} |\mathbf{x}_{12}|) }{(2 \pi)^{2- 2\epsilon}} \left( \frac{\bar{Q}}{|\mathbf{x}_{12}|} \right)^{-2\epsilon} \; . 
\end{align}
where we have used the fact that $K_{\alpha} (x) = K_{-\alpha} (x)$. Substituting Eq. \eqref{eq:trans_int} into Eq.\eqref{Eq:QuarkBackCon_Long_1}, using the relation between S-matrix and the scattering amplitude given in Eq. \eqref{Eq:S_M_matrix_relation}, and considering its quantum expectation value in the target states, we obtain 
\begin{align}
&
 2 \langle p_t'| \; {\rm Im} : [\mathcal{M}_{{\rm NLO}, \Psi}]_{\rm L} \; | p_t \rangle 
 = 
 \frac{ e_f^2 g^2 g_s^2 Q^2 }{ \pi q^+} {\rm Re} : \int \frac{d^{2} \mathbf{x}_1 d^{2} \mathbf{x}_2}{(2 \pi)^2} \int_0^1 dz \; z^2 (1-z) K_{0}^2 (\bar{Q} |\mathbf{x}_{12}|) 
 \nonumber \\ 
 & 
 \times 
 \int_{-\frac{L}{2}^+}^{\frac{L}{2}^+} \hspace{-0.1 cm} d x_1^+ \int_{x_1^+}^{\frac{L}{2}^+} d x_2^+  \; 
 \Big\langle p_t'\Big|
 \overline{\Psi} (x_2^+, \mathbf{x}_1) \gamma^- t^a 
 \mathcal{U}_F\Big( x_2^+, \frac{L}{2}^+ , \mathbf{x}_1 \Big) 
 \mathcal{U}_F\left(\mathbf{x}_2  \right) 
 \nonumber \\ 
 & \hspace{3.7cm}
 \times 
 \mathcal{U}_F \Big( - \frac{L}{2}^+ , x_1^+, \mathbf{x}_1 \Big)
 t^b \Psi (x_1^+ , \mathbf{x}_1) \big[ \mathcal{U}_A \left(x_1^{+}, x_2^{+}, \mathbf{x}_1 \right) \big]^{ba} \Big| p_t \Big\rangle \; ,
    \label{Eq:QuarkBackConMatrElemLongM}
\end{align}
where we have defined the longitudinal momentum fraction $z=k^+/q^+$ and set $D=4$, since all integrals are finite. After performing the translation  
\begin{equation}
    \mathcal{O} (x_2^+, x_1^+ , \mathbf{x}_1, \mathbf{x}_2 ) = e^{i \hat{p} \cdot x_1 }  \mathcal{O} (x_2^+-x_1^+,0^+, \mathbf{0}, \mathbf{x}_{21}) e^{-i \hat{p} \cdot x_1 } \; ,
    \label{Eq:Transl_Quark_NLO}
\end{equation}
in the operator in Eq. \eqref{Eq:QuarkBackConMatrElemLongM}, and renaming the variables $ z^+ = x_2^+ - x_1^+$ and   $\mathbf{r} = \mathbf{x}_{21}$, we obtain 
\begin{align}
2 \langle p_t'| \; {\rm Im} : [\mathcal{M}_{{\rm NLO}, \Psi}]_{\rm L} \; | p_t \rangle 
& = \frac{e_f^2\, g^2}{ 2 q^+ p_t^-} 
\frac{ g_s^2 Q^2 }{ 2 \pi^2} 
\Big[ 2 p_t^- (2 \pi)^{3}  \delta (p_t'^- - p_t^-) \delta^{2} (\mathbf{p}_t' - \mathbf{p}_t)\Big] 
 \nonumber \\ 
 & \hspace{1.5cm}
 \times  \int d^{2} \mathbf{r} \int_0^1 dz \; z^2 (1-z) K_{0}^2 (\bar{Q} |\mathbf{r}|) \; {\rm Re} : f_q (\mathbf{r})  \; ,
    \label{Eq:QuarkBackConMatrElemLongM2}
\end{align}
where we introduced the target-averaged color operator $f_q (\mathbf{r})$ which is defined as 
\begin{align}
\label{eq:def_F_q_main}
    f_q (\mathbf{r}) & =  \int \frac{d z^+}{2 \pi} \; \theta (z^+)\; \big\langle p_t\big| \overline{\Psi} (z^+, \mathbf{0}) \gamma^- t^a \mathcal{U}_F\left( z^+,  \infty^+, \mathbf{0} \right) \mathcal{U}_F\left(\mathbf{r}  \right) 
    \nonumber \\ 
    & \hspace{3.5cm}
    \times \mathcal{U}_F \left( - \infty^+ , 0^+ , \mathbf{0} \right)t^b \Psi (0^+ , \mathbf{0}) \left[ \mathcal{U}_A \left(0^{+}, z^{+}, \mathbf{0} \right) \right]^{ba}  \big| p_t \big\rangle \; .
\end{align}
Here, the limit $L^+ \to \infty$ is taken in the Wilson lines, as the background fields are assumed to vanish outside the medium in our gauge.
Then, the quark background contribution to the NLO inclusive DIS cross section for a longitudinally polarized photon is obtained from Eq.~(\ref{Eq:QuarkBackConMatrElemLongM2}) by introducing the normalization of the target states as in Eq.~(\ref{Eq:DefForwMatrElem}). The final result reads 
\begin{align}
    \sigma_{{\rm NLO}, \Psi}^{\rm L}  = 8\, e_f^2\, \alpha_{\text{em}}\, \alpha_{s}\, x_{Bj}
     \int d^{2} \mathbf{r} \int_0^1 dz \; z^2 (1-z) K_{0}^2 (\bar{Q} |\mathbf{r}|) \; {\rm Re} : f_q (\mathbf{r}) \, . 
    \label{Eq:QuarkBackConCrossLong}
\end{align}
Note that the NLO correction to the quark contribution to the inclusive DIS cross section for a longitudinal photon, presented in Eq. \eqref{Eq:QuarkBackConCrossLong}, exhibits an explicit $x_{Bj}$ dependence and therefore is of NEik order. Motivated by the power counting discussion introduced in Sec. \ref{Sec:LOCrossSec}, one may introduce an ${\rm x}$-dependent phase in the definition of the color structure $f_q$ and modifying  it as 
\begin{align}
\label{eq:def_F_q_main-xbj}
f_q({\rm x}, \r)  & =  \int \frac{d z^+}{2 \pi} e^{-i{\rm x}p_t^-z^+} \; \theta (z^+)\; \big\langle p_t\big| \overline{\Psi} (z^+, \mathbf{0}) \gamma^- t^a \mathcal{U}_F\left( z^+,  \infty^+, \mathbf{0} \right) \mathcal{U}_F\left(\mathbf{r}  \right) 
    \nonumber \\ 
    & \hspace{3.5cm}
    \times \mathcal{U}_F \left( - \infty^+ , 0^+ , \mathbf{0} \right)t^b \Psi (0^+ , \mathbf{0}) \left[ \mathcal{U}_A \left(0^{+}, z^{+}, \mathbf{0} \right) \right]^{ba}  \big| p_t \big\rangle \; .
\end{align}
such that the leading term in the Taylor expansion of $f_q({\rm x},\r)$ around $\rm{x}=0$ reproduces the original $f_q(\r)$ defined in Eq. \eqref{eq:def_F_q_main}. As discussed in more detail in \cite{Altinoluk:2025ang} and in Sec. \ref{Sec:LOCrossSec}, since the cross section carries an explicit $x_{Bj}$ dependence, which makes it of order NEik. Consequently, replacing ${\rm x}\to x_{Bj}$ in $f_q({\rm x}, \r)$ in the distribution function appearing in the cross section leads to corrections only at NNEik order. Therefore, up to NNEik corrections one can write the cross section for longitudinal photon as 
\begin{align}
    \sigma_{{\rm NLO}, \Psi}^{\rm L}  = 8\, e_f^2\, \alpha_{\text{em}}\, \alpha_{s}\, x_{Bj}
    %\frac{e_f^2 g^2}{W^2} \frac{ g_s^2 Q^2 }{ 2 \pi^2} 
    \int d^{2} \mathbf{r} \int_0^1 dz \; z^2 (1-z) K_{0}^2 (\bar{Q} |\mathbf{r}|) \; {\rm Re} : f_q (x_{Bj},\mathbf{r}) + {\rm NNEik} \; .
\end{align} 

Before proceeding with the analysis of the transversely polarized photon, a comment is in order. As will be shown later in this section, the target-averaged color structure $f_q(\mathbf{r})$ or equivalently $f_q(x_{Bj},\mathbf{r})$, defined in Eqs. \eqref{eq:def_F_q_main} and \eqref{eq:def_F_q_main-xbj} , is also the only nonperturbative target matrix element entering the quark-background contribution to the NLO inclusive DIS cross section for a transversely polarized photon. Using the relation in Eq. \eqref{Eq:AppenAdjFun} together with the Fierz identity in Eq. \eqref{Eq:AppenColorFierz}, $f_q(x_{Bj},\mathbf{r})$ can be written as
\begin{align}
\label{eq:def_F_q_expl}
 f_q (x_{Bj}, \mathbf{r}) &= \int \frac{d z^+}{2 \pi} \; \theta (z^+)\;  e^{-ix_{Bj}p^-_tz^+}
 \big\langle p_t\big| 
 \Big[   
 \overline{\Psi} (z^+, \mathbf{0}) \, \frac{\gamma^-}{2} \, \mathcal{U}_F\left( z^+, 0^+ , \mathbf{0} \right) 
 \Psi (0^+ , \mathbf{0}) 
 {\rm Tr_c} \big[\mathcal{U}_F (\mathbf{r}) \mathcal{U}_F^{\dagger} (\mathbf{0}) \big]
  \nonumber \\ 
  & - \frac{1}{N_c}   \overline{\Psi} (z^+, \mathbf{0}) \, \frac{\gamma^-}{2}\,  
  \mathcal{U}_F\left( z^+, \infty^+ , \mathbf{0} \right) \,  
  \mathcal{U}_F (\mathbf{r}) \, 
  \mathcal{U}_F\left( - \infty^+, 0^+ ,\mathbf{0} \right) \Psi (0^+ , \mathbf{0}) \Big] \big| p_t \big\rangle \; .
\end{align}
From its explicit form in Eq. \eqref{eq:def_F_q_expl}, in the limit $\r\to0$, the target-averaged color structure $f_q(x_{Bj},\r)$ is directly related to the collinear quark PDF via 
 \begin{equation}
    {\rm Re} : f_q (x_{Bj},\mathbf{0}) = C_F q_f (x_{Bj}) \; .
    \label{Eq:FqToPDF}
\end{equation}

In order to compute the quark background contributions to the NLO inclusive DIS cross section for transversely polarized photon, we average over $(D-2)$ photon polarizations and then perform the summation over the photon polarization as in Eq. \eqref{Eq:TransvCrossDef}. Then from Eq. \eqref{Eq:QuarkBackConFullyGeneral}, we obtain 
\begin{align}
\big[S_{{\rm NLO}, \Psi}\big]_{\rm T} &= 
 \frac{-g_{\perp \mu \nu}}{(D-2)} [S_{{\rm NLO}, \Psi}]^{\mu \nu} 
 \nn \\
 &= \frac{e_f^2 g^2 g_s^2}{(4 \pi) 2 q^+} \frac{[2 q^+ (2 \pi) \delta (q^+-q'^+)]}{(D-2)} \int d^{D-2} \mathbf{x}_1 \int d^{D-2} \mathbf{x}_2  
 \int_0^{q^+} \frac{d k^{+}}{q^+} \frac{(k^+)^2 (k^+-q^+)}{(q^+)}
 \nonumber \\ 
 & \times  \int \frac{d^{D-2} \mathbf{k}_i}{(2 \pi)^{D-2}} \int \frac{d^{D-2} \mathbf{k}_f}{(2 \pi)^{D-2}}  \frac{e^{ - i \mathbf{x}_{12} \cdot (\mathbf{k}_f - \mathbf{k}_i)}}{[\mathbf{k}_i^2 + \bar{Q}^2][\mathbf{k}_f^2 + \bar{Q}^2]} \int_{-\frac{L}{2}^+}^{\frac{L}{2}^+} \hspace{-0.1 cm} d x_1^+ \int_{x_1^+}^{\frac{L}{2}^+} d x_2^+  
 \nonumber \\ 
 & \times \mathcal{T}_+ : {\rm Tr_{D,c}} 
 \bigg[ 
 \gamma_{\perp}^{\mu} \frac{(\slashed{\check{k}}_f+m)}{2 k^+} 
 \gamma^{+} \mathcal{U}_F\left(\mathbf{x}_2  \right) 
 \frac{(\slashed{\check{k}}_i+m)}{2 k^+}  \gamma_{\perp \mu} 
 \frac{(\slashed{\check{p}}_f + m )}{2 (k^+ - q^+ ) } 
 \mathcal{U}_F \Big( -\frac{L}{2}^+ , x_1^+, \mathbf{x}_1 \Big) 
   \nonumber \\ 
 & \hspace{2.5cm}
 \times  
  t^b \gamma^{i} \gamma^+ \gamma^- 
 \Psi (x_1^+ , \mathbf{x}_1) 
 \left[ \mathcal{U}_A \left(x_1^{+}, x_2^{+}, \mathbf{x}_1 \right) \right]^{ba} 
  \overline{\Psi} (x_2^+, \mathbf{x}_1) 
  \nonumber\\
  & \hspace{2.5cm}
  \times
  \gamma^- \gamma^+ t^a \gamma^{i}  
  \mathcal{U}_F\Big( x_2^+, \frac{L}{2}^+ , \mathbf{x}_1 \Big) 
  \frac{(\slashed{\check{p}}_i+m)}{2 (k^+ - q^+ ) } \bigg] \bigg|_{\substack{p_i^+ = p_f^+ = k^+ - q^+ \\ k_i^+ = k_f^+ = k^+ \\ \mathbf{k}_i = \mathbf{p}_f \; , \; \mathbf{p}_i = \mathbf{k}_f}} ,
    \label{Eq:QuarkBackConTransverse}
\end{align}
The Dirac trace in Eq. \eqref{Eq:QuarkBackConTransverse} can be computed in a straightforward, albeit tedious, manner. Details of the calculation are given in Appendix \ref{Sec:AppDiracTrace}, and the final result is
\begin{align}
 &\mathcal{T}_+ : {\rm Tr_{D,c}} 
 \bigg[ 
 \gamma_{\perp}^{\mu} \frac{(\slashed{\check{k}}_f+m)}{2 k^+} \gamma^{+} 
 \mathcal{U}_F\left(\mathbf{x}_2  \right) 
 \frac{(\slashed{\check{k}}_i+m)}{2 k^+}  
 \gamma_{\perp \mu} \frac{(\slashed{\check{p}}_f + m )}{2 (k^+ - q^+)  } 
 \mathcal{U}_F \Big( -\frac{L}{2}^+ , x_1^+, \mathbf{x}_1 \Big) 
 t^b \gamma^{i} \gamma^+ \gamma^-  
 \nonumber \\ 
 & \times 
  \Psi (x_1^+ , \mathbf{x}_1) 
  \left[ \mathcal{U}_A \left(x_1^{+}, x_2^{+}, \mathbf{x}_1 \right) \right]^{ba}  
  \overline{\Psi} (x_2^+, \mathbf{x}_1) \gamma^- \gamma^+ t^a \gamma^{i}  
  \mathcal{U}_F\Big( x_2^+, \frac{L}{2}^+ , \mathbf{x}_1 \Big) 
  \frac{(\slashed{\check{p}}_i+m)}{2 (k^+ - q^+ ) } 
  \bigg] \bigg| 
  {\substack{p_i^+ = p_f^+ = k^+ - q^+ \\ k_i^+ = k_f^+ = k^+ \\ \mathbf{k}_i = \mathbf{p}_f \; , \; \mathbf{p}_i = \mathbf{k}_f}} 
  \nonumber \\ 
  & = - \frac{(D-2) (q^+)^2 }{ (k^+)^2 (q^+ - k^+)^2}  
  \Big[ 
  \mathbf{k}_i \cdot \mathbf{k}_f (z^2 + (1-z)^2) 
  + m^2 
  + \frac{(D-4)}{2} (\mathbf{k}_i \cdot \mathbf{k}_f + m^2) \Big] 
 \Big( \mathcal{T}_+ : { \rm Tr_{D,c} } 
  \big[ \mathcal{O}_{q} \gamma^- \big]\Big)
  \nonumber \\ 
  & \hspace{0.4cm}
  - m \,  \frac{ (q^+)^2 (D-4)}{{ (k^+)^2 (q^+ - k^+)^2}} 
  \Big( \frac{k^+}{q^+} + \frac{D-4}{2} \Big) 
  (\mathbf{k}_{f}^{j}-\mathbf{k}_{i}^{j}) 
 \Big( \mathcal{T}_+ : { \rm Tr_{D,c} } 
  \big[ \mathcal{O}_{q} \gamma^{j} \gamma^- \big]  \Big)
  \nonumber \\ 
  & \hspace{0.4cm}
  + \frac{(D-6)}{4 (k^+)^2 (q^+ - k^+)^2} 
  \Big[ (q^+)^2 (D-6) + 4 k^+ q^+ \Big] \; 
  \mathbf{k}_{f}^{l} \; \mathbf{k}_{i }^{j} \; 
  \Big(\mathcal{T}_+ : { \rm Tr_{D,c} } \big[ \mathcal{O}_{q} (\gamma^{l} \gamma^{j} - \gamma^{j} \gamma^{l}) \gamma^- \big] \Big)
    \label{Eq:BigTraceQuarkFin}
\end{align}
We define the shorthand ${\mathcal O}_q$ to denote the operator appearing inside the Dirac trace in all three terms on the right-hand side of Eq. \eqref{Eq:BigTraceQuarkFin}, which is given by
\begin{align}
\mathcal{O}_{q} = \mathcal{O}_{q} (x_2^+, x_1^+, \mathbf{x}_1,  \mathbf{x}_2)  
 & \equiv 
\, t^a  
\mathcal{U}_F\Big( x_2^+, \frac{L}{2}^+ , \mathbf{x}_1 \Big) 
\mathcal{U}_F\left(\mathbf{x}_2  \right) 
\mathcal{U}_F \Big( -\frac{L}{2}^+ , x_1^+, \mathbf{x}_1 \Big) 
\nonumber \\
& \times 
t^b \, \Psi (x_1^+ , \mathbf{x}_1) \left[ \mathcal{U}_A \left(x_1^{+}, x_2^{+}, \mathbf{x}_1 \right) \right]^{ba}  \overline{\Psi} (x_2^+, \mathbf{x}_1) \; .
\label{Eq:Oope}
\end{align}
Let us consider the three terms on the right-hand side of Eq. \eqref{Eq:BigTraceQuarkFin} separately. First, it is straightforward to observe that the third term vanishes identically after integration over $\mathbf{k}_i$ and $\mathbf{k}_f$ in Eq. \eqref{Eq:QuarkBackConTransverse}, due to the contraction of an antisymmetric tensor, $(\gamma^{l}\gamma^{j} - \gamma^{j}\gamma^{l})$, with a symmetric one, $(\x_{12}^{l}, \x_{12}^{j})$. The second term in Eq. \eqref{Eq:BigTraceQuarkFin} vanishes trivially in the massless case because of its explicit mass dependence. Although the vanishing of this term is less trivial in the massive case, it can be demonstrated as follows. It is proportional to $\epsilon$, but divergences from the integrals could in principle compensate this suppression, in particular UV divergences from the transverse integrals, or divergences for $k^+\to q^+$.
After performing the translation defined in Eq. \eqref{Eq:Transl_Quark_NLO} and introducing the variable $\mathbf{r}=\mathbf{x}_{21}$, the UV contribution to this second term can be rewritten in a form proportional to
\begin{equation}
    \int d^{2-2 \epsilon} \mathbf{r} \; \frac{r^{j}}{|\boldsymbol{r}|} \bar{Q} K_{1-\epsilon} (\bar{Q} |\mathbf{r}|) K_{-\epsilon} (\bar{Q} |\mathbf{r}|) = 0 \; ,
\end{equation}
times the operator \eqref{Eq:Oope} taken at $\mathbf{x}_2=\mathbf{x}_1$.
This structure vanishes identically by symmetry for any value of $\epsilon$. Hence, in the massive case, the transverse integrations are finite and cannot compensate the suppression by the factor $\epsilon$. 
Moreover, in the limit $z=k^+/q^+ \to 1$, one has $\bar{Q}\to m$, so that in this case, the transverse integrations cannot generate factors of the type $(1-z)^{-\epsilon}$, which could have compensated the $\epsilon$ factor when integrating over $z$. For these reasons, the  second term in Eq. \eqref{Eq:BigTraceQuarkFin} stays proportional to $\epsilon$ when the integrations are performed, and thus vanishes at $D=4$.

%{  This structure vanishes identically
%in four dimensions. At first sight, one might worry that the subsequent integration over the longitudinal momentum fraction $z$ could still produce a singular contribution, as indeed happens for the first term in Eq. \eqref{Eq:BigTraceQuarkFin}. However, in the present massive case, no such divergence arises. In particular, there is no mechanism by which dimensional regularization could generate a compensating pole in $\epsilon$, for example through factors of the form $(1-z)^{\epsilon}$. Consequently, the entire contribution remains proportional to $\epsilon$ and therefore vanishes in the limit $D \to 4$. 
%}

It follows that the only nonvanishing contribution originates from the first term in Eq. \eqref{Eq:BigTraceQuarkFin}. Substituting this term into Eq. \eqref{Eq:QuarkBackConTransverse}, performing the integrations over the transverse momenta $\mathbf{k}_i$ and $\mathbf{k}_f$ using Eqs. \eqref{Eq:AppTransvMomInt1} and \eqref{Eq:AppTransvMomInt2}, and finally taking the quantum expectation value of the resulting expression in the target states, one obtains
\begin{align}
    2 \langle p_t'| \; {\rm Im} : [\mathcal{M}_{{\rm NLO}, \Psi}]_{\rm T} \; | p_t \rangle & = 
     -\frac{e_f^2 g^2 g_s^2}{(4 \pi) q^+}  \; {\rm Re} : 
     \int \frac{d^{2-2 \epsilon} \mathbf{x}_1 d^{2-2 \epsilon} \mathbf{x}_2}{(2 \pi)^{2- 2\epsilon}}  
     \int_0^{1} \frac{ d z }{(1 - z)} 
     \left( \frac{\bar{Q}^2}{\mathbf{x}_{12}^2} \right)^{-\epsilon} 
      \nonumber \\
     &  \hspace{-2cm}\times 
     \int_{-\frac{L}{2}^+}^{\frac{L}{2}^+} \hspace{-0.1 cm} d x_1^+ 
     \int_{x_1^+}^{\frac{L}{2}^+} d x_2^+ 
          \Big\{  \big[z^2 + (1-z)^2\big] 
     \bar{Q}^2 K_{1-\epsilon}^2 
     \big(\bar{Q} |\mathbf{x}_{12}|\big) 
     + 
     m^2 K_{\epsilon}^2 \big(\bar{Q} |\mathbf{x}_{12}|\big) 
      \nonumber \\ 
      & \hspace{3.7cm}
      - \epsilon \Big [ \bar{Q}^2 K_{1-\epsilon}^2 \big(\bar{Q} |\mathbf{x}_{12}|\big) + m^2 K_{\epsilon}^2 \big(\bar{Q} |\mathbf{x}_{12}|\big) \Big] \Big\} 
      \nonumber \\
      & \hspace{1cm}
      \times 
      \big\langle p_t'\big|\Big( \mathcal{T}_+ : { \rm Tr_{D,c} } \left[ \mathcal{O}_{q} (x_2^+, x_1^+, \mathbf{x}_1,  \mathbf{x}_2) \gamma^- \right] \Big)  \big|p_t \big\rangle \, . 
\end{align}
Performing the translation as in Eq. \eqref{Eq:Transl_Quark_NLO}, changing the variables $z^+=x_2^+-x_1^+$ and $\r=\x_{21}$, and introducing the normalization of the target states as in Eq. \eqref{Eq:DefForwMatrElem}, the quark contribution to the NLO inclusive DIS cross section for a transversely polarized photon can be obtained as 
\begin{align}
\hspace{-0.3cm}
    \sigma_{{\rm NLO}, \Psi}^{\rm T}  & =  
    \frac{e_f^2 g^2 g_s^2}{2} \frac{x_{Bj}}{Q^2} 
    {\rm Re} : \int_0^1 \frac{dz}{1-z}  \int \frac{ d^{2-2 \epsilon} \mathbf{r} }{(2 \pi)^{2-2 \epsilon}} 
    \Big( \frac{\bar{Q}^2 }{\mathbf{r}^2} \Big)^{-\epsilon} 
    \Big\{ 
    \big[ z^2 + (1-z)^2)\big] \bar{Q}^2 K_{1-\epsilon}^2 \big(\bar{Q} | \mathbf{r} |\big)
     \nonumber \\   
     & \hspace{3cm}
     + m^2 K_{\epsilon}^2 \big(\bar{Q} | \mathbf{r} |\big)  
     - \epsilon \big[
     \bar{Q}^2 K_{1-\epsilon}^2 \big(\bar{Q} | \mathbf{r} |\big) 
     + m^2 K_{\epsilon}^2 \big(\bar{Q} | \mathbf{r} |\big) \big] 
     \Big\} f_q (x_{Bj},\mathbf{r}) \; . 
    \label{Eq:Sigma_NLO_Psi_T}
\end{align}
with $f_q (x_{Bj},\mathbf{r})$ being the same target-averaged color structure that is defined in Eq. \eqref{eq:def_F_q_expl}. 

%%%%%%%%%%%%%%%%%%%%%%%%%%%%%%%%%%%%%%%%%%%%
%%%%%%%%%%%%%%%%%%%%%%%%%%%%%%%%%%%%%%%%%%%%
\subsection{Analysis of the divergences and extraction of the finite contributions}
%%%%%%%%%%%%%%%%%%%%%%%%%%%%%%%%%%%%%%%%%%%%
%%%%%%%%%%%%%%%%%%%%%%%%%%%%%%%%%%%%%%%%%%%%

The quark background contributions to the NLO inclusive DIS cross sections for longitudinally and transversely polarized photons are given in Eqs. \eqref{Eq:QuarkBackConCrossLong} and \eqref{Eq:Sigma_NLO_Psi_T}, respectively. While the cross section for a longitudinally polarized photon is free of divergences and entirely finite, the cross section for a transversely polarized photon exhibits singularities.

The first divergence in the NLO inclusive DIS cross section at next-to-eikonal (NEik) accuracy for a transversely polarized photon, given in Eq. \eqref{Eq:Sigma_NLO_Psi_T}, arises in the limit $(1-z)\to0$ for a generic, nonvanishing transverse separation $\r$. This limit corresponds to the kinematic regime in which the longitudinal momentum fraction of the incoming virtual photon carried by the antiquark line in diagram (a) of Fig. \ref{fig:NLODISNEik} becomes vanishingly small. 
At eikonal level and at LO, the interaction with the dense target is encoded entirely in Wilson lines constructed from the target’s gluon field, and the projectile consists solely of a color dipole without any additional partonic radiation. As a result, there is no available phase space for emissions carrying parametrically small longitudinal momentum, nor is any rapidity separation between projectile and target degrees of freedom required. Consequently, rapidity divergences cannot arise in the LO dipole contribution at the eikonal level. 
Rapidity divergences emerge only beyond LO, when an additional gluon carrying a small longitudinal momentum fraction $z_g$ is radiated or absorbed. In this case, the longitudinal phase space of the emitted gluon extends to arbitrarily small $z_g$, giving rise to logarithmic divergences associated with large rapidity separations. Since the small-$z_g$ dynamics is already resummed through the BK/B-JIMWLK evolution of the dipole amplitude appearing in the LO contribution, the corresponding divergence must be subtracted from the NLO impact factor in order to avoid double counting.

However, at NEik accuracy, quark background field are incorporated into the description of the target alongside the gluon background. The LO contribution to the inclusive DIS cross section for a transversely photon, originally computed in \cite{Altinoluk:2025ang} and rederived in Section \ref{Sec:LOCrossSec}, { is proportional to the background quark distribution, without further $\alpha_s$ suppression}. The expression given in Eq. \eqref{Eq:Sigma_NLO_Psi_T} then represents the corresponding NLO correction. Consequently, the contribution arising in the limit $(1-z)\to 0$ should be interpreted as part of the small-$x$ evolution of the operator defined in Eq.~(\ref{Eq:QuarkPDFDef}).

In addition to the rapidity divergences discussed above, the NLO contribution to the inclusive DIS cross section for a transversely polarized photon, given in Eq. \eqref{Eq:Sigma_NLO_Psi_T}, also exhibits ultraviolet (UV) divergences, which arise in the limit  $\r\to0$ for $0<z<1$. Furthermore, there exists a region where the two types of singularities overlap, i.e., the simultaneous $\r\to0$ and $(1-z)\to 0$ regime, which suggests the emergence of a double-logarithmic evolution. An interesting direction would be to derive this double-logarithmic evolution of the { background quark operator $q_f(x_{Bj})$ at $x_{Bj}\rightarrow 0$} directly from its operator definition; however, this is beyond the scope of the present manuscript and is left for future work. { In particular, it would be important to check the consistency of such double log resummation with the ones already formulated in the literature for other NEik distributions~\cite{Kirschner:1983di,Bartels:1995iu,Bartels:1996wc,Kovchegov:2018znm,Cougoulic:2022gbk,Borden:2023ugd,Borden:2024bxa,Borden:2025ehe}.} 

In the remainder of this section, we present a detailed analysis of the extraction of divergences from the NLO quark background contribution to the inclusive DIS cross section for a transversely polarized photon, treating massive and massless quarks separately.

%%%%%%%%%%%%%%%%%%%%%%%%%%%%%%%%%%%%%%%%%
\subsubsection{Isolating divergent and finite contributions: massless case}
\label{Sec:DivAnalysis_Massless}
%%%%%%%%%%%%%%%%%%%%%%%%%%%%%%%%%%%%%%%%%%%%

We begin our analysis of divergences and the extraction of the finite contributions by considering the massless quarks. In this limit, NLO quark contribution to the inclusive DIS cross section for transversely polarized photon, given in Eq. \eqref{Eq:Sigma_NLO_Psi_T}, reduces to 
\begin{align}
    \sigma_{{\rm NLO}, \Psi}^{\rm T, \; m=0}  & 
    =  \frac{e_f^2 g^2 g_s^2  }{2 } \frac{x_{Bj}}{Q^2}  {\rm Re} : 
    \int_0^1 \frac{dz}{z}  \int \frac{ d^{2-2 \epsilon} \mathbf{r} }{(2 \pi)^{2-2 \epsilon}} 
    \frac{1}{(\mathbf{r}^2)^{-\epsilon}} 
    \Big[ \big[z^2 + (1-z)^2\big] - \epsilon \;  \Big]
    \nonumber \\
    & \hspace{7cm}
    \times \, 
    \big(\bar{Q}^2\big)^{1-\epsilon}\,  K_{1-\epsilon}^2 \big(\bar{Q} | \mathbf{r} |\big) \,  f_q (x_{Bj}, \mathbf{r}) \; ,
    \label{Eq:Sigma_NLO_Psi_T_Massless}
\end{align}
% F
with now $\bar{Q}^2\mapsto z(1-z)Q^2$, and where we have interchanged $z \leftrightarrow (1-z)$ so that rapidity divergence now appears in the limit $z\to0$. Our first goal is to isolate the rapidity-singular part of the expression in Eq. \eqref{Eq:Sigma_NLO_Psi_T_Massless}. To this end, we note that in the ${\bar Q}\to0$ limit, the Bessel function of the first kind behaves as 
\begin{align}
\lim_{\bar Q\to 0}\, \Big[  \bar{Q}^{\alpha} K_{\alpha} \big(\bar{Q}|\mathbf{r}|\big)\Big] =   \frac{2^{{\alpha}-1}}{|\mathbf{r}|^{\alpha}} \Gamma ({\alpha}) + \mathcal{O} (\bar{Q}) \, ,
\label{Bessel_Q_to_0}
\end{align}
with $\Gamma(\cdots)$ being the gamma function. Using Eq. \eqref{Bessel_Q_to_0} and taking the $z\to0$ limit in Eq. \eqref{Eq:Sigma_NLO_Psi_T_Massless}, the rapidity-singular part of the NLO quark contribution to the inclusive DIS cross section for transversely polarized photon can be written as 
\begin{align}
\sigma_{{\rm NLO}, \Psi}^{{\rm T}, \; m=0} \bigg |_{z \rightarrow 0}  
&=  
\frac{e_f^2 g^2 g_s^2  }{ 2 } \frac{x_{Bj}}{Q^2} \, 
2^{-2\epsilon}\,  \big(1-\epsilon\big)\,  \Gamma^2 ( 1 -\epsilon  )  \, 
{\rm Re} : \int_0^1 \frac{dz}{z}  
\int \frac{ d^{2-2 \epsilon} \mathbf{r} }{(2 \pi)^{2-2 \epsilon}} 
\frac{1}{ (\mathbf{r}^2)^{1 - 2 \epsilon}} \, f_q (x_{Bj}, \mathbf{r}) \; .
 \label{Eq:Quark_Con_NLO_zTend0_1}
\end{align}
However, the rapidity-singular expression in Eq. \eqref{Eq:Quark_Con_NLO_zTend0_1} also contains a UV singularity in the $\r\to0$ limit, which can be isolated by subtracting and then re-adding the corresponding UV-divergent term. With this procedure, Eq. \eqref{Eq:Quark_Con_NLO_zTend0_1} can be rewritten as
\begin{align}
    \sigma_{{\rm NLO}, \Psi}^{{\rm T}, \; m=0} \bigg |_{z \rightarrow 0} 
    &=  
   \frac{e_f^2 g^2 g_s^2  }{2  } \frac{ x_{Bj} }{Q^2} \, 
   2^{-2\epsilon} \, \big(1-\epsilon\big) \,  \Gamma^2 ( 1 -\epsilon  ) \, 
   {\rm Re} : \int_0^1 \frac{dz}{z}  \int \frac{ d^{2-2 \epsilon} \mathbf{r} }{(2 \pi)^{2-2 \epsilon}} \frac{1}{ (\mathbf{r}^2)^{1 - 2 \epsilon}} 
     \nonumber \\ 
     & \times 
     \bigg\{ \bigg[ 
     f_q (x_{Bj}, \mathbf{r}) - \theta \bigg( \frac{1}{\Lambda^2} - \mathbf{r}^2 \bigg) 
     f_q (x_{Bj},\mathbf{0}) 
     \bigg] 
     + \theta \bigg( \frac{1}{\Lambda^2} - \mathbf{r}^2 \bigg) f_q (x_{Bj},\mathbf{0})  \bigg\} \; ,
    \label{Eq:Quark_Con_NLO_zTend0}
\end{align}
where the term inside the square bracket is UV-finite and the last term in the curly brackets represents the isolated UV-singular contribution. Therefore, this term corresponds to the both rapidity and UV-divergent contribution which reads
\begin{align}
 \sigma_{{\rm NLO}, \Psi}^{{\rm T}, \; m=0} \bigg |_{\rm rap. \;  \& \; UV \; div.} & =
 \frac{e_f^2 g^2 g_s^2  }{2  } \frac{x_{Bj}}{Q^2} \, 
   2^{-2\epsilon} \, \big(1-\epsilon\big) \,  \Gamma^2 ( 1 -\epsilon  ) \, 
   {\rm Re} : \int_0^1 \frac{dz}{z}  
     \nonumber \\ 
     & \hspace{3cm} 
     \times 
     \int \frac{ d^{2-2 \epsilon} \mathbf{r} }{(2 \pi)^{2-2 \epsilon}}\,   \theta \bigg( \frac{1}{\Lambda^2} - \mathbf{r}^2 \bigg)\,  \frac{1}{ (\mathbf{r}^2)^{1 - 2 \epsilon}} \, 
      f_q (x_{Bj},\mathbf{0})  \, . 
\end{align}
The transverse integration over $\r$ can be performed using Eq. \eqref{Eq:AppTransvCoordInt1}. Furthermore, by employing the relation between the target-averaged color structure and the background quark PDF given in Eq. \eqref{Eq:FqToPDF}, the rapidity- and UV-divergent term in the NLO massless quark contribution to the inclusive DIS cross section for a transversely polarized photon becomes
\begin{align}
 \sigma_{{\rm NLO}, \Psi}^{{\rm T}, \; m=0} \bigg |_{\rm rap. \;  \& \; UV \; div.} & = 
 e_f^2 \, g^2 \, \pi\, \frac{x_{Bj}}{Q^2}  \frac{\alpha_s}{2 \pi}\,  C_F \,q_f(x_{Bj})\;
   \frac{\Gamma ( 2 -\epsilon  )}{ \epsilon}
   \left( \frac{\Lambda^{2}}{ \pi \mu^2 } \right)^{-\epsilon}
   \int_0^1 \frac{dz}{z} \; , 
   \label{eq:Rap_UV_div_result}
\end{align}
where we have also used $\alpha_s = g_s^2 \mu^{-2 \epsilon}/4 \pi$. In Eq. \eqref{eq:Rap_UV_div_result}, the integration over $z$ has not been performed explicitly, since different regularization schemes can be employed for the rapidity divergence. For instance, in a cut-off scheme
the integral can be evaluated as 
\begin{equation}
\label{eq:cut_off}
     \int_{\alpha}^1 \frac{dz}{z} = - \ln \alpha  \; ,
\end{equation}
or alternatively, in the so-called $\eta^+$-scheme  \cite{Altinoluk:2025tms}, it can be expressed as 
\begin{equation}
\label{eq:eta_plus}
  \left[ \frac{q^+}{\nu^+} \right]^{\eta}  \int_{0}^1 dz \; z^{\eta-1} = \left[ \frac{q^+}{\nu^+} \right]^{\eta} \frac{1}{\eta} \; ,
\end{equation}
%s
with $\eta$ a dimensionless regularization parameter analog to $\epsilon$ in dimensional regularization, and $\nu^+$ an arbitrary light-cone momentum scale of reference, analog to $\mu$. 
Other rapidity regularization schemes could also be used.

As noted above, the remaining part of Eq. \eqref{Eq:Quark_Con_NLO_zTend0} corresponds to the rapidity-divergent but UV-finite contribution to the  NLO massless quark contribution to the inclusive DIS cross section for a transversely polarized photon, and it is given by  
\begin{align}
\label{eq:Rap_div_UV_fin}
 \sigma_{{\rm NLO}, \Psi}^{{\rm T}, \; m=0} \bigg |_{\rm rap. \; div.}
 &   =
     e_f^2 g^2 \pi  \, \frac{x_{Bj}}{Q^2} 
    \frac{\alpha_s}{2 \pi} \, 
    \frac{2^{-2\epsilon}\, \big(1-\epsilon\big)\, \Gamma^2(1-\epsilon)}{\mu^{-2\epsilon}}
    \int_0^1 \frac{dz}{z}
    \nonumber \\
    & \hspace{1.5cm}
    \times \, 
    {\rm Re} : \int \frac{ d^{2-2\epsilon} \mathbf{r} }{(2 \pi)^{2-2\epsilon}}
    \frac{1}{(\mathbf{r}^2)^{1-2\epsilon}}
    \bigg[
        f_q(x_{Bj},\mathbf{r})
        -
        \theta\!\left( \frac{1}{\Lambda^2} - \mathbf{r}^2 \right)
        f_q(x_{Bj},\mathbf{0})
    \bigg] \; .
\end{align}

Having isolated the rapidity divergences, we now subtract Eq. \eqref{Eq:Quark_Con_NLO_zTend0_1} from the exact expression of  NLO massless quark contribution to the inclusive DIS cross section for a transversely polarized photon given in Eq. \eqref{Eq:Sigma_NLO_Psi_T_Massless}. The resulting expression reads 
\begin{align}
    \sigma_{{\rm NLO}, \Psi}^{\rm T, \; m=0} -  \sigma_{{\rm NLO}, \Psi}^{{\rm T}, \; m=0} \bigg |_{z \rightarrow 0}  & =  
    \frac{e_f^2 g^2 g_s^2  }{2  } \frac{x_{Bj}}{Q^2} {\rm Re} : \int_0^1 \frac{dz}{z}  \int \frac{ d^{2-2 \epsilon} \mathbf{r} }{(2 \pi)^{2-2 \epsilon}} \frac{1}{(\mathbf{r}^2)^{-\epsilon}} 
    \nonumber \\ 
    & \times \bigg \{ \big ( 1 - \epsilon \big) 
    \Big[ 
    \big(\bar{Q}^2\big)^{1-\epsilon}\,  K_{1-\epsilon}^2 \big(\bar{Q} | \mathbf{r} |\big) 
    - 
    \frac{2^{-2\epsilon} \,  \Gamma^2 ( 1 -\epsilon  )}{ (\mathbf{r}^2)^{1-\epsilon} } \Big]
     f_q (x_{Bj},\mathbf{r}) 
     \nonumber \\
     & \hspace{3.5cm} 
     - 2 z (1-z) (\bar{Q}^2)^{1-\epsilon} K_{1-\epsilon}^2 (\bar{Q} | \mathbf{r} |) f_q (x_{Bj},\mathbf{r}) \bigg \} \; .  
     \label{eq:rap_subtraction}
\end{align}
One can immediately see that the term in the second line of Eq. \eqref{eq:rap_subtraction} is purely finite, whereas the term in the last line of Eq. \eqref{eq:rap_subtraction} contains a UV-divergence. This divergence can be isolated by applying the subtraction method once more. More precisely,  the target-averaged color structure {   $f_q(\r)$} in the last line of Eq. \eqref{eq:rap_subtraction} should be replaced by
\begin{align}
\label{eq:subtraction_procedure}
f_q(x_{Bj},\r)= \Big[ f_q (x_{Bj},\mathbf{r})-f_q (x_{Bj},\mathbf{0})\Big ] + f_q (x_{Bj},\mathbf{0}) \, . 
\end{align}
This procedure isolates the UV-divergent yet rapidity-finite contribution, which can be written as 
 \begin{align}
    \sigma_{{\rm NLO}, \Psi}^{{\rm T}, \; m=0} \bigg |_{\rm UV \; div.}  
   & = 
    -  e_f^2 \, g^2\,  g_s^2  \, \frac{x_{Bj}}{Q^2}  C_F \, q_f (x_{Bj}) 
    \int_0^1 dz   (1-z)  
    \nonumber\\
    & \hspace{4.2cm} 
    \times 
     \int \frac{ d^{2-2 \epsilon} \mathbf{r} }{(2 \pi)^{2-2 \epsilon}} 
    \frac{ \big(\bar{Q}^2\big)^{1-\epsilon} \, K_{1-\epsilon}^2 \big(\bar{Q} | \mathbf{r} |\big) }{(\mathbf{r}^2)^{-\epsilon}} \; , 
\end{align}
where we have used the relation between the target-averaged color structure and the quark PDF in the $\r\to0$ limit, as given in Eq. \eqref{Eq:FqToPDF}. The transverse integration over $\r$ can then be performed using Eq. \eqref{Eq:AppTransvCoordInt2}, which yields to 
\begin{align}
    \sigma_{{\rm NLO}, \Psi}^{{\rm T}, \; m=0} \bigg |_{\rm UV \; div.}  \!\!\!\!
   & = 
     - e_f^2 \, g^2  \, \frac{x_{Bj}}{Q^2} \frac{g_s^2\,  \mu^{-2 \epsilon} }{4 \pi} \,  C_F \, q_f (x_{Bj}) 
     \bigg( \frac{Q^2}{ 4 \pi \mu^2} \bigg)^{-\epsilon}  
     \nonumber \\
     & \hspace{4.8cm} 
     \times 
     \big(1-\epsilon\big) \,  \Gamma (\epsilon) \int_0^1 dz \; z^{-\epsilon} (1-z)^{1 - \epsilon} \; .
\end{align}
Integration over $z$ can be easily performed and the UV-divergent yet rapidity-finite term in the NLO massless quark contribution to the inclusive DIS cross section reads 
\begin{align}
\label{eq:Rap_fin_UV_div}
\sigma_{{\rm NLO}, \Psi}^{{\rm T}, \; m=0} \bigg |_{\rm UV \; div.}  
= - e_f^2\,  g^2\,  \pi \, \frac{x_{Bj}}{Q^2}\frac{\alpha_s}{2 \pi}\, C_F\,  q_f (x_{Bj})   \,  
\frac{ \Gamma \big(1+ \epsilon\big) \Gamma^{2} \big(1-\epsilon\big) }{ \Gamma \big(1-2 \epsilon\big)} \left( \frac{Q^2}{ 4 \pi \mu^2} \right)^{-\epsilon} 
\frac{1}{\epsilon}\, \frac{(1-\epsilon)}{(1- 2 \epsilon)} \; . 
\end{align}
The remainder of the UV-subtraction procedure in Eq. \eqref{eq:rap_subtraction} corresponds to the fully finite contribution to the massless quark contribution to the NLO DIS cross section for transversely photon, and it is given by 
\begin{align}
\label{eq:Rap_fin_UV_fin}
    \sigma_{{\rm NLO}, \Psi}^{{\rm T}, \; m=0} \bigg |_{\rm fin. }
    &=
    e_f^2 g^2 \pi \frac{x_{Bj}}{Q^2} \frac{\alpha_s}{2 \pi}
    \,{\rm Re} :
    \int_0^1 \frac{dz}{z}
    \int \frac{ d^{2} \mathbf{r} }{ \pi }
    \bigg\{
  \Big[ \bar{Q}^2 K_{1}^2\big(\bar{Q}|\mathbf{r}|\big) - \frac{1}{\mathbf{r}^2} \Big]
        f_q(x_{Bj},\mathbf{r}) 
        \nonumber \\
    & \hspace{4 cm}
        - 2 z(1-z)\,\bar{Q}^2 K_{1}^2\big(\bar{Q}|\mathbf{r}|\big)
        \Big[ f_q(x_{Bj},\mathbf{r}) - f_q(x_{Bj},\mathbf{0}) \Big]
    \bigg\} \; .
\end{align}
This concludes the analysis of the divergences in the massless quark contribution to the NLO inclusive DIS cross section for a transversely polarized photon. To summarize, the rapidity- and UV-divergent contribution is given in Eq. \eqref{eq:Rap_UV_div_result}, the rapidity-divergent but UV-finite contribution in Eq. \eqref{eq:Rap_div_UV_fin}, the rapidity-finite but UV-divergent contribution in Eq. \eqref{eq:Rap_fin_UV_div}, and the contribution that is finite with respect to both rapidity and UV divergences in Eq. \eqref{eq:Rap_fin_UV_fin}.

%%%%%%%%%%%%%%%%%%%%%%%%%%%%%%%%%%%%%%%%%%%%%%%%%%%%%%%%%%%%%%%
%%%%%%%%%%%%%%%%%%%%%%%%%%%%%%%%%%%%%%%%%%%%%%%%%%%%%%%%%%%%%%%
\subsubsection{Isolating divergent and finite contributions: massive case}
\label{Sec:DivAnalysis_Massive}
%%%%%%%%%%%%%%%%%%%%%%%%%%%%%%%%%%%%%%%%%%%%%%%%%%%%%%%%%%%%%%%
%%%%%%%%%%%%%%%%%%%%%%%%%%%%%%%%%%%%%%%%%%%%%%%%%%%%%%%%%%%%%%%

We now turn to the analysis of the massive quark contribution to the NLO inclusive DIS cross section for a transversely polarized photon. 
%The presence of a nonvanishing quark mass modifies the structure of the divergences encountered in the massless case. In this section, we systematically repeat the analysis carried out for massless quarks, isolating the rapidity-divergent, UV-divergent, and finite contributions, and highlighting the differences induced by the finite quark mass in the structure of the NLO correction.
%
%The quark contribution to the NLO inclusive DIS cross section for a transversely polarized photon, including both mass-indpendent terms and terms explicitly proportional to quark mass, is given in Eq. \eqref{Eq:Sigma_NLO_Psi_T}. The analysis of the mass-independent terms in the cross section was carried out in the previous subsection. In the present subsection, we focus exclusively on the terms proportional to the quark mass in Eq. \eqref{Eq:Sigma_NLO_Psi_T}.
Upon exchanging $z\leftrightarrow(1-z)$ in Eq. \eqref{Eq:Sigma_NLO_Psi_T}, the rapidity-singular contribution is isolated by taking the $z\to0$ limit, and it is given by 
\begin{align}
\sigma_{{\rm NLO}, \Psi}^{{\rm T}, \, m} \Big |_{z \rightarrow 0} 
& =  
\frac{e_f ^2\, g^2 \, g_s^2 }{2 } \frac{x_{Bj}}{Q^2} \, 
\mu^{-2\epsilon}\, \left( \frac{m^{2}}{\mu^2} \right)^{-\epsilon} 
(1-\epsilon) \,  m^2 \; {\rm Re} : 
\int_0^1 \frac{dz}{z}  
\nonumber\\
& \hspace{2cm}
\times 
\int \frac{ d^{2-2 \epsilon} \mathbf{r} }{(2 \pi)^{2-2 \epsilon}} \frac{1}{(\mathbf{r}^2)^{-\epsilon}} 
\Big[   K_{1-\epsilon}^2 \big( m | \mathbf{r} |\big) + K_{\epsilon}^2 \big(m| \mathbf{r} |\big)  \Big] \,  f_q (x_{Bj},\mathbf{r}) \; .
    \label{Eq:Quark_Con_NLO_zTend0_Mass}
\end{align}
The first term in the square brackets of Eq. \eqref{Eq:Quark_Con_NLO_zTend0_Mass} contains a UV divergence. This divergence can be isolated by applying the subtraction procedure defined in Eq. \eqref{eq:subtraction_procedure}. After implementing this procedure, the contribution that is simultaneously rapidity- and UV-divergent is given by
\begin{align}
\sigma_{{\rm NLO}, \Psi}^{{\rm T}, \, m}  \Big |_{\rm rap. \;  \& \; UV \; div.}  & =  
\frac{e_f^2 \,  g^2 \,  g_s^2 }{2  } \frac{x_{Bj}}{Q^2} \, \mu^{-2\epsilon} \,  
C_F \, q_f (x_{Bj})  
\left( \frac{m^{2}}{\mu^2} \right)^{-\epsilon} 
(1-\epsilon) \int_0^1 \frac{dz}{z} 
 \nonumber \\ 
 & \hspace{5cm} 
 \times 
  \int \frac{ d^{2-2 \epsilon} \mathbf{r} }{(2 \pi)^{2-2 \epsilon}} \frac{ m^2  K_{1-\epsilon}^2 ( m | \mathbf{r} |) }{(\mathbf{r}^2)^{-\epsilon}}  \; , 
\end{align}
where again we have used the relation between the target-averaged color structure $f_q(x_{Bj},\r)$ and the quark PDF in the $\r\to0$ limit given in Eq. \eqref{Eq:FqToPDF}. The transverse integration over $\r$ in the resulting expression can be carried out using  Eq. \eqref{Eq:AppTransvCoordInt2}, yielding 
\begin{align}
\label{eq:mass-rap_div-UV_div}
 \sigma_{{\rm NLO}, \Psi}^{{\rm T}, \, m} \Big |_{\rm rap. \;  \& \; UV \; div.} 
 = e_f^2 \,  g^2 \, \pi \, \frac{x_{Bj}}{Q^2}  \frac{\alpha_s}{2 \pi} \, C_F \, q_f (x_{Bj})  \frac{(1-\epsilon) \Gamma(1+\epsilon)}{\epsilon}  \left( \frac{m^{2}}{4 \pi \mu^2} \right)^{-\epsilon} \int_0^1 \frac{dz}{z}    \; . 
\end{align}
Here, as in the case of massless quarks, we have not performed the integration over $z$ explicitly. This integral can be evaluated either as  in Eq. \eqref{eq:cut_off} or as in Eq. \eqref{eq:eta_plus}, corresponding to the  cut-off and $\eta^+$ rapidity regulator schemes, respectively. 

The remainder of the UV subtraction procedure in Eq. \eqref{Eq:Quark_Con_NLO_zTend0_Mass}, corresponds to the rapidity-divergent but UV-finite contribution to the mass-dependent terms in the quark contribution to the NLO inclusive DIS cross section for transversely polarized photon, and it is given by  
 \begin{align}
 \label{eq:mass-rap_div-UV_fin}
   \sigma_{{\rm NLO}, \Psi}^{{\rm T}, \, m} \Big |_{\rm rap. \; div.}
    &=
    e_f^2 g^2 \pi \frac{x_{Bj}}{Q^2}
    \frac{\alpha_s}{\pi}
    \left( \frac{m^{2}}{\mu^2} \right)^{-\epsilon}
    (1-\epsilon) \, m^2\;
    {\rm Re} :
    \int_0^1 \frac{dz}{z}
    \int \frac{ d^{2-2\epsilon} \mathbf{r} }{(2\pi)^{1-2\epsilon}}
    \frac{1}{(\mathbf{r}^2)^{-\epsilon}} \nonumber \\
   & \times  
    \Big\{
        K_{1-\epsilon}^2(m|\mathbf{r}|)\, 
        \Big[ f_q(x_{Bj},\mathbf{r}) - f_q(x_{Bj},\mathbf{0}) \Big]
        +
        K_\epsilon^2(m|\mathbf{r}|)\, 
        f_q(x_{Bj},\mathbf{r})
    \Big\} \; .
 \end{align}
Since we have isolated the rapidity divergences, we can now subtract rapidity-singular contribution given in Eq. \eqref{Eq:Quark_Con_NLO_zTend0_Mass} from the mass-dependent terms in Eq. \eqref{Eq:Sigma_NLO_Psi_T} which yields 
\begin{align}
\label{eq:massive:exact-rapidity}
\sigma_{{\rm NLO}, \Psi}^{{\rm T}, \, m } -  \sigma_{{\rm NLO}, \Psi}^{{{\rm T}, \, m}} \big |_{z \rightarrow 0} 
&=
\frac{e_f^2 \,  g^2 \,  g_s^2}{2  } \frac{x_{Bj}}{Q^2} \,  {\rm Re} : 
\int_0^1 \frac{dz}{z}  
\int \frac{ d^{2-2 \epsilon} \mathbf{r} }{(2 \pi)^{2-2 \epsilon}} \frac{1}{ \left( \mathbf{r}^2 \right)^{-\epsilon} }
 \nonumber \\ 
 &
 \times 
 \Big \{  (1-\epsilon) 
 \Big[ \big(\bar{Q}^2\big)^{1-\epsilon}\, 
  K_{1-\epsilon}^2 \big(\bar{Q} | \mathbf{r} |\big) 
  - 
  (m^2)^{1-\epsilon} K_{1-\epsilon}^2 \big( m | \mathbf{r} |\big)
  \nonumber\\ 
  &  \hspace{2cm}
  + 
  m^2 \Big( \big(\bar{Q}^2\big)^{-\epsilon} K_{\epsilon}^2 \big(\bar{Q} | \mathbf{r}| \big)  
  - (m^2)^{-\epsilon} K_{\epsilon}^2 \big( m | \mathbf{r}| \big) \Big)
   \Big]
  \nonumber \\
   & \hspace{0.8cm}
    - 2 z (1-z) (\bar{Q}^2)^{1-\epsilon} K_{1-\epsilon}^2 (\bar{Q} | \mathbf{r} |)  \Big \} f_q (x_{Bj},\mathbf{r}) \; .
\end{align}
The last term in Eq. \eqref{eq:massive:exact-rapidity} contains a UV singularity, which can be isolated using the UV subtraction procedure described in Eq. \eqref{eq:subtraction_procedure}. Upon applying this procedure, the resulting UV-divergent but rapidity-finite contribution is given by 
\begin{align}
\label{eq:mass-rap_fin-UV_div}
    \sigma_{{\rm NLO}, \Psi}^{{\rm T}, \, m } \Big |_{\rm UV \; div.} &=
    - e_f^2 \, g^2 \, \pi  \frac{x_{Bj}}{Q^2} \, \frac{\alpha_s}{2 \pi} \,  
       C_F \, q_f (x_{Bj} ) \, 
        \frac{(1-\epsilon) \Gamma(1+\epsilon)}{\epsilon}  \left( \frac{m^{2}}{4 \pi \mu^2} \right)^{-\epsilon} 
       {\cal I} \left( \frac{Q^2}{m^2},\epsilon \right) \; .
\end{align}        
Here, we have used the relation between $f_q(x_{Bj},\r)$ and quark PDF in the $\r\to0$ limit. Moreover, the transverse integration over $\r$ has been computed using Eq. \eqref{Eq:AppTransvCoordInt2}. For convenience, the resulting expression is written in terms of a function ${\cal I}$ which is defined as 
\begin{align}
   {\cal I} \left( \frac{Q^2}{m^2},\epsilon \right) \equiv 2 \int_0^1 dz  \; z \; \left( 1 + \frac{z(1-z) Q^2}{m^2} \right)^{-\epsilon} 
 \end{align} 
Upon integration over $z$, for finite $\epsilon$, the function ${\cal I}$  is given by the Appell series and it reads 
\begin{align}
   {\cal I} \left( \frac{Q^2}{m^2},\epsilon \right) 
   = F_1 \bigg( 2, \epsilon, \epsilon, 3, \frac{2 }{ 1 + \sqrt{1 + {4 m^2}/{Q^2} } }, \frac{2 }{ 1 - \sqrt{1 + {4 m^2}/{Q^2} } } \bigg)  \; .
\end{align} 
Since this expression is required up to $ \mathcal{O} (\epsilon^2)$ in the $\epsilon \to 0$ limit, the function ${\cal I }$ can be written in a more practical form as 
\begin{align}
    {\cal I} \left( \frac{Q^2}{m^2},\epsilon \right) & = 1 - 2 \epsilon \int_0^1 dz  \; z \; \ln \left( 1 + \frac{z(1-z) Q^2}{m^2} \right) + \mathcal{O} (\epsilon^2) \nonumber \\ 
    &= 1 + 2 \epsilon \bigg[1 -  \sqrt{1 +  \frac{4 m^2}{Q^2}} \tanh ^{-1}\bigg(\frac{1}{\sqrt{1 +  {4 m^2}/{Q^2}}}\bigg) \bigg] + \mathcal{O} (\epsilon^2) 
    \end{align}
The remainder of the UV-subtraction procedure in Eq. \eqref{eq:massive:exact-rapidity} corresponds to the fully finite contribution to the mass-dependent terms in the quark background contribution to the NLO inclusive DIS cross section for transversely polarized photon and it is given by 
\begin{align}
\label{eq:mass-rap_fin-UV_fin}
    \sigma_{{\rm NLO}, \Psi}^{{\rm T}, \, m} \Big |_{\rm fin. }
    &=
    e_f^2 \, g^2 \, \pi \frac{x_{Bj}}{Q^2}
    \frac{\alpha_s}{2\pi}
    \;{\rm Re} :
    \int_0^1 \frac{dz}{z}
    \int \frac{ d^{2} \mathbf{r} }{\pi} 
    \nonumber
\\
   & \times 
    \Big\{ \Big[ \bar{Q}^2 K_1^2(\bar{Q}|\mathbf{r}|)- m^2 K_1^2(m|\mathbf{r}|)
            + m^2\Big( K_0^2(\bar{Q}|\mathbf{r}|) - K_0^2(m|\mathbf{r}|) \Big)\Big] f_q(x_{Bj},\mathbf{r})
            \nonumber\\ 
            & \hspace{1cm}
        -\,
        2 z(1-z)\,\bar{Q}^2
        K_1^2(\bar{Q}|\mathbf{r}|)\, 
        \Big[ f_q(x_{Bj},\mathbf{r}) - f_q(x_{Bj},\mathbf{0}) \Big]
    \Big\}
    \; . 
\end{align}
To sum up, NLO corrections to the NEik quark background contribution to  inclusive DIS cross section for transversely polarized photon for massive quarks gets the following contributions: the rapidity- and UV-divergent contribution given in Eq. \eqref{eq:mass-rap_div-UV_div}, the rapidity-divergent but UV-finite contribution given in Eq. \eqref{eq:mass-rap_div-UV_fin}, the rapidity-finite but UV-divergent contribution given in Eq. \eqref{eq:mass-rap_fin-UV_div} and the rapidity- and UV-finite contribution given in Eq. \eqref{eq:mass-rap_fin-UV_fin}.  
%

%%%%%%%%%%%%%%%%%%%%%%%%%%%%%%%%%%%%%%%%%%%%%%%%%%%%%%%%%%%%%%%%%%
%%%%%%%%%%%%%%%%%%%%%%%%%%%%%%%%%%%%%%%%%%%%%%%%%%%%%%%%%%%%%%%%%%
\subsection{Antiquark background field contribution}
%%%%%%%%%%%%%%%%%%%%%%%%%%%%%%%%%%%%%%%%%%%%%%%%%%%%%%%%%%%%%%%%%%
%%%%%%%%%%%%%%%%%%%%%%%%%%%%%%%%%%%%%%%%%%%%%%%%%%%%%%%%%%%%%%%%%%

The calculation of the antiquark contribution follows exactly the same strategy as for the quark contribution. We therefore outline only the main steps and then present the final result. The diagram corresponding to this contribution is shown in panel (b) of Fig. \ref{fig:NLODISNEik}. As in the quark case, to obtain a NEik contribution, the splitting of the quark-antiquark pair must occur outside the medium. Accordingly, the uncontracted $S$-matrix element for this configuration is given by 
\begin{align}
    [S_{{\rm NLO}, \overline{\Psi}}]^{\mu \nu} = e_f^2 \; g^2 \int_{-\infty^+}^{-\frac{L}{2}^+} d^D y \int_{ \frac{L}{2}^+}^{\infty^+} d^D x \; e^{i q' \cdot x -i q \cdot y} \; \mathcal{T}: {\rm Tr_D} \left[  \gamma^{\nu} S_{F, q}^{\Psi} (x, y) \gamma^{\mu} S_{F, \bar{q}}^{b.a.} (y,x) \right] \; ,
    \label{Eq:AntiQuarkConBeg}
\end{align}
with before-to-after antiquark propagator $S_{F, \bar{q}}^{b.a.} (y,x)$ defined in Eq. \eqref{eq:antiquark_BA}. Moreover, $S_{F, q}^{\Psi} (x, y) $ is the effective before-to-after quark propagator in a quark background field and it is defined as 
\begin{align}
& \hspace{-0.5cm}
S_{F, q}^{\Psi} (x, y)  = 
- g_s^2  
\int_{-\frac{L}{2}^+}^{\frac{L}{2}^+} d^D x_1 
\int_{x_1^+}^{\frac{L}{2}^+} d^D x_2 
\Big\{ S_{F,q}^{i.a.} (x, x_2) \;t^a \gamma^{\sigma} \Psi^{(-)} (\underline{x_2})
\nonumber \\ 
& \hspace{5cm}
\times 
 \big[G_{F, \rho \sigma}(x_2, x_1)^{ \rm i.i.}\big]^{ba} \; \overline{\Psi}^{(-)} (\underline{x_1}) \, t^b \, \gamma^{\rho} \, S_{F,q}^{b.i.} (x_1, y)\Big\} \; , 
 \label{eq:effective-BtoA}
\end{align}
with inside-to-after $S_{F,q}^{i.a.} $ and before-to-inside $S_{F,q}^{b.i.}$ quark propagators given in Eqs. \eqref{eq:quark_P_IA} and \eqref{eq:quark_P_BI} respectively, whereas the inside-to-inside gluon propagator $G_{F, \rho \sigma}^{ \rm i.i.}$ is given in Eq. \eqref{eq:in-in_gluon}. Using these equations for the propagators in Eq. \eqref{eq:effective-BtoA}, the effective before-to-after quark propagator in a quark background field can be written as  
\begin{align}
\hspace{-0.6cm}
S_{F, q}^{\Psi} (x, y) & = 
- \frac{g_s^2}{4}  
\int_{-\frac{L}{2}^+}^{\frac{L}{2}^+} \hspace{-0.1 cm} d x_1^+ 
\int_{x_1^+}^{\frac{L}{2}^+} \hspace{-0.1 cm}  d x_2^+ 
\int d^{D-2} \mathbf{x}_1  
\int \frac{d k^{+}}{2 \pi} \frac{\theta( k^{+} )}{2 k^{+}} 
\int \frac{d^{D-2} \mathbf{k}_i}{(2 \pi)^{D-2}}
\int \frac{d^{D-2} \mathbf{k}_f}{(2 \pi)^{D-2}} 
\nonumber \\ 
& \hspace{-1cm} 
\times 
e^{ -i x \cdot \check{k}_f - i \mathbf{x}_1 \cdot \mathbf{k}_f + i y \cdot \check{k}_i + i \mathbf{x}_1 \cdot \mathbf{k}_i  } \, 
\frac{(\slashed{\check{k}}_f + m )}{2 k^+ } \mathcal{U}_F \left( x^+, x_2^+, \mathbf{x}_1 \right) 
t^a \gamma^{k} \gamma^+ \gamma^-  
\Psi (x_2^+, \mathbf{x}_1) 
\nonumber \\ 
& \hspace{-1cm}
\times 
\left[ \mathcal{U}_A \left(x_2^{+}, x_1^{+}, \mathbf{x}_1 \right) \right]^{ab} 
 \overline{\Psi} (x_1^+, \mathbf{x}_1) 
 \, \gamma^- \gamma^+ t^b \gamma^{k} \, 
  \mathcal{U}_F\left( x_1^+,y^+, \mathbf{x}_1 \right) \frac{(\slashed{\check{k}}_i+m)}{2 k^+ } 
  \bigg|_{k_i^+=k_f^+=k^+} \, . 
\end{align}
Substituting this expression into Eq. \eqref{Eq:AntiQuarkConBeg} and using the explicit form of the before-to-after antiquark propagator, the uncontracted $S$-matrix element is obtained as 
\begin{align}
&
 [S_{{\rm NLO}, \overline{\Psi}}]^{\mu \nu} 
 = - e_f^2 \; g^2 \; g_s^2 (2 \pi) \delta (q^+-q'^+) 
 \int d^{D-2} \mathbf{x}_1 
 \int d^{D-2} \mathbf{x}_2   
 \int_0^{q^+} \frac{d k^{+}}{2 \pi} 
 \frac{k^+ (q^+-k^+)^2}{2 (q^+)^2} 
 \nonumber \\ 
 & \times  
 \int \frac{d^{D-2} \mathbf{k}_i}{(2 \pi)^{D-2}}
  \int \frac{d^{D-2} \mathbf{k}_f}{(2 \pi)^{D-2}}  
  \frac{e^{ - i \mathbf{x}_{12} \cdot (\mathbf{k}_f - \mathbf{k}_i)}}{[\mathbf{k}_i^2 + \bar{Q}^2][\mathbf{k}_f^2 + \bar{Q}^2]} 
  \int_{-\frac{L}{2}^+}^{\frac{L}{2}^+} \hspace{-0.1 cm} d x_1^+ 
  \int_{x_1^+}^{\frac{L}{2}^+} d x_2^+  
  \nonumber \\ 
  & 
  \times  {\rm Tr_{D,c}} \left[ \gamma^{\nu} \frac{(\slashed{\check{k}}_f+m)}{2 k^+} \mathcal{U}_F\left( x^+, x_2^+ , \mathbf{x}_1 \right) t^a \gamma^{k} \gamma^+ \gamma^- \Psi (x_2^+, \mathbf{x}_1) \left[ \mathcal{U}_A \left(x_2^{+}, x_1^{+}, \mathbf{x}_1 \right) \right]^{ab} \overline{\Psi} (x_1^+, \mathbf{x}_1)  \right. 
  \nonumber \\ 
  & \times 
  \gamma^- \gamma^+ t^b \gamma^k \frac{(\slashed{\check{k}}_i +m)}{2 k^+}  \left. \mathcal{U}_F\left( x_1^+, y^+ , \mathbf{x}_1 \right) \gamma^{\mu} \frac{(\slashed{\check{p}}_f+m)}{2 (k^+-q^+)} \gamma^+ \mathcal{U}_F^{\dagger} (\mathbf{x}_2) \frac{(\slashed{\check{p}}_i+m)}{2 (k^+-q^+)}  \right] \bigg|_{\substack{p_i^+ = p_f^+ = k^+ - q^+ \\ k_i^+ = k_f^+ = k^+ \\ \mathbf{k}_i = \mathbf{p}_f \; , \; \mathbf{p}_i = \mathbf{k}_f}} ,
    \label{Eq:AntiQuarkBackConFullyGeneral}
\end{align}
Since the fermionic fields are already ordered along the longitudinal $+$-coordinate,  the $\mathcal{T}_+$ operator does not appear explicitly in Eq. \eqref{Eq:AntiQuarkBackConFullyGeneral}.  Using Eq. \eqref{Eq:AntiQuarkBackConFullyGeneral} for the uncontracted $S$-matrix, the computation of the antiquark contribution to the NLO inclusive DIS cross section proceeds exactly the same steps as in the calculation of the quark contribution. After carrying out these steps, the cross sections for longitudinally and transversely polarized photons can be obtained as
\begin{align}
 \label{Eq:AntiQuarkBackConCrossLong}
    \sigma_{ {\rm NLO}, \overline{\Psi} }^{\rm L}  & = 
    \frac{ e_f^2 \, g^2 \, g_s^2\,   }{ 2\pi^2 } x_{Bj} \int d^{2} \mathbf{r} \int_0^1 dz \; z (1-z)^2 K_{0}^2 (\bar{Q} |\mathbf{r}|) \; 
    {\rm Re} : f_{\bar q} (x_{Bj},\mathbf{r}) \; , \\
    \label{Eq:AntiQuarkBackConCrossTrans}
  \sigma_{{\rm NLO}, \overline\Psi}^{\rm T} & =  \frac{e_f^2 \, g^2 \,  g_s^2}{2  } \frac{x_{Bj}}{Q^2}  {\rm Re} : 
    \int_0^1 \frac{dz}{z}  
   \int \frac{ d^{2-2 \epsilon} \mathbf{r} }{(2 \pi)^{2-2 \epsilon}} 
    \left( \frac{\bar{Q}^2 }{\mathbf{r}^2} \right)^{-\epsilon} 
    \Big\{ \big[z^2 + (1-z)^2\big] \bar{Q}^2 K_{1-\epsilon}^2 \big(\bar{Q} | \mathbf{r} |\big) 
    \nonumber \\  
    &  \hspace{2.8cm}
   + m^2 K_{\epsilon}^2 \big(\bar{Q} | \mathbf{r} |\big)  - \epsilon \big[ \bar{Q}^2 K_{1-\epsilon}^2 \big(\bar{Q} | \mathbf{r} |\big) + m^2 K_{\epsilon}^2 
   \big(\bar{Q} | \mathbf{r} |\big) \big] \Big\}  f_{\bar q} (x_{Bj},\mathbf{r}) \; ,
   \end{align}
with the target-averaged color structure related to the antiquark PDF is given by     
\begin{align}
f_{\bar q}(x_{Bj},\mathbf{r})
&=  
\int \frac{d z^+}{2 \pi} \; \theta (- z^+)\; e^{ix_{Bj}{p_t}^-z^+}
\big\langle p_t\big| {\rm Tr_{D,c}} 
\Big[ 
\mathcal{U}_F \left( \infty^+ , 0^+ , \mathbf{0} \right) t^a \Psi (0^+ , \mathbf{0}) 
\nonumber \\
& \hspace{1.2cm}
\times 
\left[ \mathcal{U}_A \left(0^{+}, z^{+}, \mathbf{0} \right) \right]^{ab} \overline{\Psi} (z^+, \mathbf{0})   t^b \gamma^- \mathcal{U}_F\left( z^+, - \infty^+, \mathbf{0} \right) \mathcal{U}_F^{\dagger} \left(\mathbf{r}  \right)  \Big] \big| p_t \big\rangle \; .
\end{align}
By comparing Eqs. \eqref{Eq:AntiQuarkBackConCrossLong} and \eqref{Eq:AntiQuarkBackConCrossTrans} with their corresponding expressions for the quark contribution to the cross section given in Eqs. \eqref{Eq:QuarkBackConCrossLong} and \eqref{Eq:Sigma_NLO_Psi_T}, it is straightforward to see that the two sets of expressions are identical up to the replacement $f_q(x_{Bj},\r)\leftrightarrow f_{\bar q}(x_{Bj},\r)$.  Consequently, the analysis of divergences and the extraction of finite contributions for the antiquark contribution proceeds in exactly the same manner as for the quark contribution, upon making this replacement.

%%%%%%%%%%%%%%%%%%%%%%%%%%%%%%%%%%%%%%%%%%%%%%%%%%%%%%%%%%%%%%%%% 
\subsection{Total fermionic contribution}
\label{sec:total_fermion}
%%%%%%%%%%%%%%%%%%%%%%%%%%%%%%%%%%%%%%%%%%%%%%%%%%%%%%%%%%%%%%%%% 
We can now combine the quark and anti-quark contributions to obtain the DIS structure functions at NEik and NLO. In the longitudinal case, we have
\begin{align}
F_L(x_{Bj},Q^2)\Big|_{{\rm NLO}, \Psi+\overline{\Psi}}&= 4\, x_{Bj} Q^2 \sum_{f} e_f^2 \, \, \frac{\alpha_{s}}{2 \pi} \, 
  \nonumber \\ & \times \int \frac{d^{2} \mathbf{r}}{\pi} \int_0^1 dz \; z^2 (1-z) K_{0}^2 (\bar{Q} |\mathbf{r}|) \; {\rm Re} : f(x_{Bj},\mathbf{r}) +{\rm NNEik}  \, , 
\end{align}
where we introduced the total target-averaged color structure $f (x_{Bj},\mathbf{r}) = f_q (x_{Bj},\mathbf{r}) + f_{\bar{q}} (x_{Bj},\mathbf{r}) $.
The transverse case is more delicate. Although we have isolated a purely finite contribution, the singular contributions shown must be appropriately subtracted once the evolution of the operator appearing at the leading order is determined. This cancellation will leave finite terms, dependent on the characteristic scale(s) of the evolution. The form of the structure function is thus
\begin{align}
    F_T(x_{Bj} &,Q^2)\Big|_{{\rm NLO}, \Psi+\overline{\Psi}}^{m=0}
    =
    x_{Bj} \sum_{f}
    e_f^2 \frac{\alpha_s}{2 \pi}
    \,{\rm Re} :
    \int_0^1 \frac{dz}{z}
    \int \frac{ d^{2} \mathbf{r} }{ \pi }
    \bigg\{
  \Big[ \bar{Q}^2 K_{1}^2\big(\bar{Q}|\mathbf{r}|\big) - \frac{1}{\mathbf{r}^2} \Big]
        f(x_{Bj}, \mathbf{r}) 
        \nonumber \\
    & 
        - 2 z(1-z)\,\bar{Q}^2 K_{1}^2\big(\bar{Q}|\mathbf{r}|\big)
        \Big[ f(x_{Bj}, \mathbf{r}) - f(x_{Bj}, \mathbf{0}) \Big]
    \bigg\} + {\rm sub. \; reminder} + {\rm NNEik} \; ,
\end{align}
in the massless case and 
\begin{align}
     F_T(x_{Bj} &,Q^2)\Big|_{{\rm NLO}, \Psi+\overline{\Psi}}^{m}
    = x_{Bj} \sum_{f}
    e_f^2 \,
    \frac{\alpha_s}{2\pi}
    \;{\rm Re} :
    \int_0^1 \frac{dz}{z}
    \int \frac{ d^{2} \mathbf{r} }{\pi} 
    \nonumber
\\
   & \times 
    \Big\{ \Big[ \bar{Q}^2 K_1^2(\bar{Q}|\mathbf{r}|)- m^2 K_1^2(m|\mathbf{r}|)
            + m^2\Big( K_0^2(\bar{Q}|\mathbf{r}|) - K_0^2(m|\mathbf{r}|) \Big)\Big] f(x_{Bj}, \mathbf{r})
            \nonumber\\ 
            & 
        -\,
        2 z(1-z)\,\bar{Q}^2
        K_1^2(\bar{Q}|\mathbf{r}|)\, 
        \Big[ f (x_{Bj}, \mathbf{r}) - f(x_{Bj}, \mathbf{0}) \Big]  
    \Big\} + {\rm sub. \; reminder} + {\rm NNEik}
    \; ,
\end{align}
in the massive case, where "sub. reminder" denotes the finite terms surviving the subtraction of divergences, which are resummation scheme and scale dependent.

%%%%%%%%%%%%%%%%%%%%%%%%%%%%%%%%%%%%%%%%%%%%%%%%%%%%%%%%%%%%%
%%%%%%%%%%%%%%%%%%%%%%%%%%%%%%%%%%%%%%%%%%%%%%%%%%%%%%%%%%%%%
%%%%%%%%%%%%%%%%%%%%%%%%%%%%%%%%%%%%%%%%%%%%%%%%%%%%%%%%%%%%%
\section{NEik DIS cross-section at the next-to-leading order: Gluon background field contribution}
\label{Sec:NLOXsec_gb}
%%%%%%%%%%%%%%%%%%%%%%%%%%%%%%%%%%%%%%%%%%%%%%%%%%%%%%%%%%%%%
%%%%%%%%%%%%%%%%%%%%%%%%%%%%%%%%%%%%%%%%%%%%%%%%%%%%%%%%%%%%%

In addition to the quark and antiquark background contributions, the NEik inclusive DIS cross section receives ${\mathcal O}(\alpha_s)$ 
corrections at NLO  arising from the gluon background field of the target. As discussed previously, NEik corrections in a gluon background field can be classified into two categories that yield complementary yet independent effects. In the first category, the target is described by a gluon background field within the static approximation, in which the background fields are taken to be independent of the minus longitudinal coordinate. In this case, the NEik corrections originate from the finite longitudinal extent of the target and from the interaction of the projectile partons with the transverse components of the background gluon field. These effects are collectively encoded in the decorated Wilson lines defined in Eqs.~\eqref{eq:decW_1}, \eqref{eq:decW_2} and \eqref{eq:decW_3}. The second category of NEik corrections arises from going beyond the static approximation by allowing for minus longitudinal coordinate dependence in the gluon background field. Since these two sources provide independent contributions, we study their effects separately in this section.

Before proceeding with the explicit calculation of the NEik contributions in a gluon background field, it is instructive to discuss the associated power counting in more detail. The only diagram that yields NLO corrections to the quark background contribution to the NEik inclusive DIS cross section in a gluon background field is shown in Fig.~\ref{fig:NLODISNEikgluon}. In this configuration, the virtual photon splits into a quark–antiquark pair outside the medium. The resulting NEik corrections are encoded in the decorated Wilson lines appearing in the before-to-after quark or antiquark propagators. In terms of the QCD coupling $g_s$, this diagram provides the LO contribution in dipole factorization, both at eikonal order and at NEik order, in the presence of a gluon background field.
In the dense target regime, where the gluon background field scales as $A={\cal O}(1/g_s)$ and the corresponding field strength tensor as ${\cal F}={\cal O}(1/g_s)$, this diagram yields a contribution of ${\cal O}(1)$. 

 However, additional care is required when NEik effects associated with quark background fields are included in the analysis of observables. The Yang–Mills equations for the gluon background field imply that the quark background field $\Psi$ also scales with the strong coupling and { can be  at most of order ${\cal O}(1/g_s)$} in the dense regime. With this power counting, one immediately finds that the quark background contribution the LO inclusive DIS, computed in Sec.~\ref{Sec:LOCrossSec}, scales as ${\cal O}(1/g_s^2)$. Consequently, the NEik contribution in a gluon background field shown in Fig.~\ref{fig:NLODISNEikgluon}, which is of ${\cal O}(g_s^0)$, constitutes a relative ${\cal O}(g_s^2)$ correction to the quark background contribution to the LO inclusive DIS cross section.\footnote{{In the dilute regime for the target,  the quark and gluon background fields are instead assumed to both be of order ${\cal O}(g_s^0)$. In that case, the contribution discussed in section \ref{Sec:LOCrossSec} is of order ${\cal O}(g_s^0)$, whereas the ones calculated in the section \ref{Sec:NLOCrossSecQuark} and in the present section are of order ${\cal O}(g_s^2)$, so that they have the same relative suppression by ${\cal O}(g_s^2)$ as allowed in the power counting for the  dense target case.}}

In summary, although the NEik corrections in a gluon background field are leading order in dipole factorization, they are parametrically suppressed by $g_s^2$ relative to the enhanced quark background contribution to the inclusive DIS in the dense regime. In the rest of this section, we provide a detailed analysis of those corrections. 

%%%%%%%%%%%%%%%%%%%%%%%%%%%%%%%%%%%%%%%%%%%%%%%%
\subsection{Contributions from the gluon background field within the static approximation}
\label{Sec:Before-to-after}
%%%%%%%%%%%%%%%%%%%%%%%%%%%%%%%%%%%%%%%%%%%%%%%
\begin{figure}
    \centering
    \includegraphics[width=0.70 \linewidth]{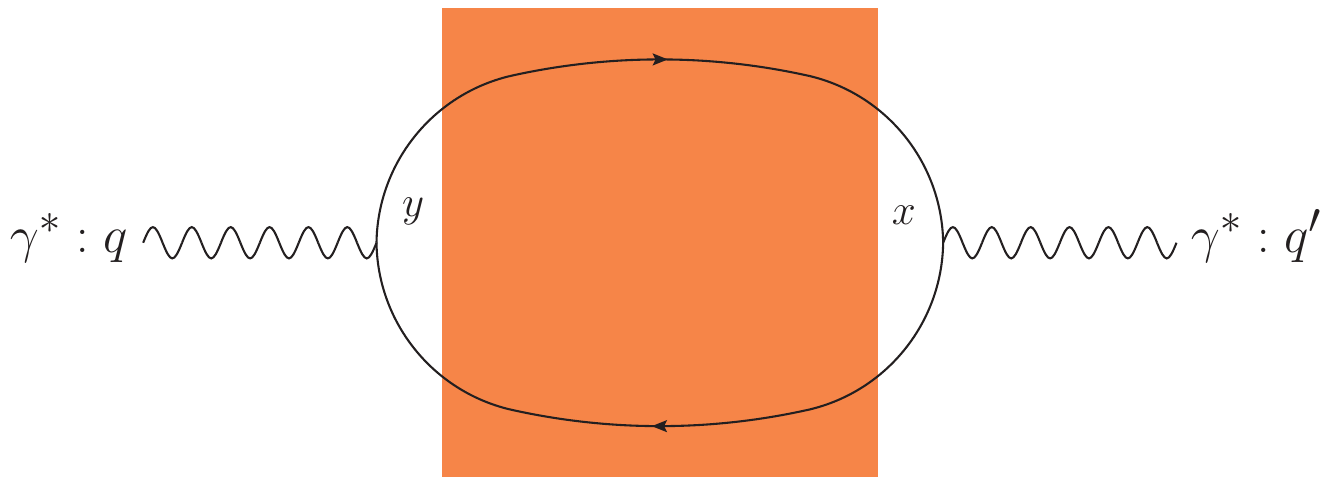} 
    \caption{Next-to-leading order contribution to DIS at the NEik accuracy from the gluon background of the target.}
    \label{fig:NLODISNEikgluon}
\end{figure}

Within the static target field approximation in a pure gluon background, the uncontracted $S$-matrix associated with Fig. \ref{fig:NLODISNEikgluon}, can be written as 
\begin{align}
    [S_{{\rm NLO , \; G}}^{b.a.}]^{\mu \nu} = e_f^2 \; g^2 \int_{-\infty^+}^{-\frac{L}{2}^+} d^D y \int_{ \frac{L}{2}^+}^{\infty^+} d^D x \; e^{i q' \cdot x -i q \cdot y} \; \mathcal{T}: {\rm Tr_{D,c}} \left[  \gamma^{\nu} S_{F,q}^{b.a.} (x, y) \gamma^{\mu}  S_{F, \bar{q}}^{b.a.} (y, x) \right] \; .
    \label{Eq:GluonConBeg}
\end{align}
where the before-to-after quark and antiquark propagators are given in Eqs. \eqref{eq:quark_BA} and \eqref{eq:antiquark_BA} respectively. The fully decorated Wilson lines ${\cal U}_F^\star(\z)$ entering in the before-to-after quark propagators include both eikonal Wilson line and also the NEik decorated Wilson lines. Consequently, the uncontracted $S$-matrix given in Eq. \eqref{Eq:GluonConBeg}, receives contributions from the eikonal dipole-corresponding to selecting  eikonal Wilson lines both in the quark and antiquark propagators-as well as from decorated dipoles, obtained by selecting  eikonal Wilson line in one propagator and a decorated Wilson line in the other one. These operators structures must subsequently be integrated over the longitudinal coordinates $x^+$ and $y^+$. 

We first consider the contribution of the eikonal dipole operator to the uncontracted $S$-matrix element. Since the eikonal dipole is independent of longitudinal coordinates $x^+$ and $y^+$, the corresponding integrals can be performed trivially. For instance, the integration over $x^+$ yields   
\begin{align}
    \int^{\infty^+}_{\frac{L}{2}^+} d x^+ e^{i x^+ (q^- + \check{p}_i - \check{k}_f ) } =  \int^{\infty^+}_{0} d x^+ e^{i x^+ (q^- + \check{p}_i - \check{k}_f ) } - \int_{0}^{\frac{L}{2}^+} d x^+ e^{i x^+ (q^- + \check{p}_i - \check{k}_f ) } \nonumber \\
    \simeq \frac{i}{(q^- + \check{p}_i - \check{k}_f )} -\frac{L}{2}^+ = i \frac{2 k^+ (k^+-q^+)}{q^+\big[\mathbf{k}_f^2 + {\bar Q}^2\big]} -\frac{L}{2}^+ \; .
    \label{Eq:energyDenWithL+}
\end{align}
The first term in Eq.~\eqref{Eq:energyDenWithL+} yields the eikonal contribution to the cross section, while the second term represents an ${\cal O}(L^+)$ correction. However, it is straightforward to observe that the energy denominator is absent in the second term. As a consequence, the integration over one of the transverse momenta produces a factor $\delta^{(2)}(\x_1-\x_2)$ with $\x_1$ and $\x_2$ denoting the transverse coordinates of the eikonal Wilson lines, both for longitudinally and transversely polarized photons. The eikonal dipole operator therefore reduces to the identity, implying that this term corresponds to a vacuum contribution and does not contribute to the $S$-matrix. 

Even though, such terms do not contribute to the NLO corrections to the quark contribution to the NEik inclusive DIS, there are other type of diagrams in which the dipole is formed or annihilated inside the medium,  as illustrated in Fig.~\ref{fig:InsideMedium}. The analogous ${\cal O}(L^+)$ terms arising in these diagrams can likewise be shown to correspond to vacuum contributions and therefore do not contribute to the $S$-matrix. A more detailed discussion on these "inside medium" contributions is presented in Appendix \ref{Sec:AppInsideMedium}. 

We now focus on the NEik contributions induced by the decorated Wilson lines appearing in the before-to-after quark and antiquark propagators. We begin by analyzing the case in which the decoration is located on the antiquark propagator. The integrations over $x^+$ and $y^+$  can be carried out using Eq.~\eqref{Eq:Energy denominator}, and the subsequent integrations over the transverse coordinates $\x$ and $\y$ in the resulting expression can be performed straightforwardly, leading to
\begin{align}
\label{SNLO_BA_1}
[S_{{\rm NLO , G, \Psi}}^{b.a.}]^{\mu \nu} & = 
\frac{2\,  e_f^2 \, g^2}{\pi} 
(2 \pi) \delta (q^+-q'^+) 
\int d^{D-2} \mathbf{x}_1 \int d^{D-2} \mathbf{x}_2   
\int_0^{q^+} d k^{+} \frac{(k^+)^2 (q^+-k^+)^2}{ (q^+)^2} 
\nonumber \\ 
& \times  
\int \frac{d^{D-2} \mathbf{k}_i}{(2 \pi)^{D-2}}
 \int \frac{d^{D-2} \mathbf{k}_f}{(2 \pi)^{D-2}}  
 \frac{e^{ - i \mathbf{x}_{12} \cdot (\mathbf{k}_f - \mathbf{k}_i)}}{[\mathbf{k}_i^2 + \bar{Q}^2][\mathbf{k}_f^2 + \bar{Q}^2]} 
 \nonumber \\ 
 & \times 
 {\rm Tr_{D,c}} 
 \bigg[ 
 \gamma^{\nu} \frac{(\slashed{\check{k}}_f+m)}{2 k^+} \gamma^+ \; 
 \mathcal{U}_F \left( \mathbf{x}_1 \right) 
 \frac{(\slashed{\check{k}}_i+m)}{2 k^+}  
  \gamma^{\mu} \frac{(\slashed{\check{p}}_f+m)}{2 (k^+ - q^+)} \gamma^+ 
 \nonumber \\ 
 & \hspace{1.5cm}
 \times  
% \gamma^{\mu} \frac{(\slashed{\check{p}}_f+m)}{2 (k^+ - q^+)} \gamma^+ 
 \bigg( \overline{\mathcal{U}}_F^{\dagger} ( \mathbf{x}_2, \mathbf{k}_f, \mathbf{k}_i )
  - \frac{[\gamma^l, \gamma^n]}{ 8 ( k^{+} - q^{+} ) } 
  \mathcal{U}_{F,l n}^{(3) \dagger}( \mathbf{x}_2) \bigg) 
  \frac{(\slashed{\check{p}}_i+m)}{2 (k^+ - q^+)}  \bigg]  ,
\end{align}
where we have introduced a short hand notation $\overline{\mathcal{U}}^{\dagger}_F$ that is defined as 
 \begin{align}
    \overline{\mathcal{U}}_F^{\dagger} \left( \mathbf{x}_2, \mathbf{k}_f, \mathbf{k}_i \right) = 
    \frac{ \mathbf{k}_i^j +\mathbf{k}_f^j }{ 4 (k^{+} - q^{+}) } \mathcal{U}_{F,j}^{(1) \dagger} \left( \mathbf{x}_2 \right) + \frac{i}{ 2 (k^{+} - q^{+})} \mathcal{U}_{F}^{(2) \dagger} \left( \mathbf{x}_2  \right) \; .
\end{align}
For a longitudinally polarized photon one uses Eq. \eqref{Eq:LonPolVec} for the longitudinal polarization vector and the $S$-matrix element given in Eq. \eqref{SNLO_BA_1} immediately reduces to 
\begin{align}
\big[S_{{\rm NLO , G, \Psi}}^{b.a.}\big]^{\rm L}_{\rm NEik} & = 
\frac{2\,  e_f^2 \, g^2}{\pi} 
(2 \pi) \delta (q^+-q'^+) 
\int d^{D-2} \mathbf{x}_1 \int d^{D-2} \mathbf{x}_2   
\int_0^{q^+} d k^{+} \frac{(k^+)^2 (q^+-k^+)^2}{ (q^+)^2} 
\nonumber \\ 
& \times  
\int \frac{d^{D-2} \mathbf{k}_i}{(2 \pi)^{D-2}}
 \int \frac{d^{D-2} \mathbf{k}_f}{(2 \pi)^{D-2}}  
 \frac{e^{ - i \mathbf{x}_{12} \cdot (\mathbf{k}_f - \mathbf{k}_i)}}{[\mathbf{k}_i^2 + \bar{Q}^2][\mathbf{k}_f^2 + \bar{Q}^2]} \, 
 \nonumber\\
 & \times 
 2 \; {\rm Tr_{c}}
  \bigg[   \mathcal{U}_F ( \mathbf{x}_1) \bigg(   \frac{ \mathbf{k}_i^j +\mathbf{k}_f^j }{ 4 (k^{+} - q^{+}) } \mathcal{U}_{F,j}^{(1) \dagger} \left( \mathbf{x}_2 \right) + \frac{i}{ 2 (k^{+} - q^{+})} \mathcal{U}_{F}^{(2) \dagger} \left( \mathbf{x}_2  \right) \bigg)\bigg] \; .
 \end{align}
One can readily verify that the contribution from the decorated Wilson line of type $\mathcal{U}_{F,j}^{(1) \dagger} $ vanishes, owing to the fact that 
\begin{align}
    \int \frac{d^{D-2} \mathbf{k}_i}{(2 \pi)^{D-2}} \int \frac{d^{D-2} \mathbf{k}_f}{(2 \pi)^{D-2}}  \frac{e^{ - i \mathbf{x}_{12} \cdot (\mathbf{k}_f - \mathbf{k}_i)}}{[\mathbf{k}_i^2 + \bar{Q}^2][\mathbf{k}_f^2 + \bar{Q}^2]} (\mathbf{k}_i^j +\mathbf{k}_f^j) = 0 \; ,
\end{align}
The remaining contribution arises from the decorated Wilson line of type $\mathcal{U}_{F}^{(2) \dagger}$. For this term, we perform the transverse momentum integrations using Eq. \eqref{Eq:AppTransvMomInt1}. The forward scattering amplitude is then obtained from the corresponding $S$-matrix element via Eq.~\eqref{Eq:S_M_matrix_relation}, after which its imaginary part is taken. Finally, upon evaluating the quantum expectation value of the resulting expression in the target states, one obtains
\begin{align}
 \big \langle p_t'\big|  2 \; {\rm Im } :   \big[\mathcal{M}_{{\rm NLO , \; G, \Psi}}^{b.a.}\big]_{\rm Neik}^{\rm L} \big| p_t \big\rangle  
& = 
- 
\frac{ 2 \, e_f^2 \,  g^2 \, Q^2 }{ \pi q^+ } \int_0^1 dz \; z (1-z)^2 
\int \frac{ d^{2}  \mathbf{x}_1  d^{2} \mathbf{x}_2}{(2\pi)^2}
K_{0}^2 (\bar{Q} |\mathbf{x}_{12}|)
\nonumber \\ 
& \times 
 \; {\rm Re } : (-i)  \big\langle p_t'\big|  {\rm Tr_c}  \big[ \mathcal{U}_{F} (\mathbf{x}_{2}) \mathcal{U}^{ (2) \dagger}_{F} (\mathbf{x}_1) \big ] \big| p_t \big\rangle  \; ,
\end{align}
where we set $D=4$ since the integrals are finite. Using the explicit expression for the decorated Wilson given in Eq. \eqref{eq:decW_2}, one gets 
\begin{align}
&
  \langle p_t'|  2 \; {\rm Im } :  [\mathcal{M}_{{\rm NLO , \; G, \Psi}}^{b.a.}]_{\rm Neik}^{\rm L} | p_t \rangle   =  
  \frac{ 2\,  e_f^2 \, g^2 \, Q^2 }{ \pi q^+ } 
  \int_0^1 dz \; z (1-z)^2 
  \int \frac{ d^{2}  \mathbf{x}_1  d^{2} \mathbf{x}_2}{(2\pi)^2} 
  K_{0}^2 (\bar{Q} |\mathbf{x}_{12}|) 
  \nonumber \\ 
  & \times  
  {\rm Re } : (-i)  
  \int_{-\frac{L}{2}^+}^{\frac{L}{2}^+} d x_1^{+} 
  \int_{-\frac{L}{2}^+}^{\frac{L}{2}^+} d x_2^{+} 
  (x_2^+ - x_1^+) \theta ( x_2^+ - x_1^+) 
  {\rm Tr_c}  
  \bigg[ \mathcal{U}_{F} (\mathbf{x}_{2}) \mathcal{U}_F\Big(-\frac{L}{2}^+, x_1^+ ; \mathbf{x}_1 \Big) 
  \nonumber \\ 
  & \hspace{2cm} 
  \times g_st \cdot \mathcal{F}_j^{\; -} (x_1^+, \mathbf{x}_1  ) \; \mathcal{U}_F\Big( x_1^{+}, x_2^{+} ; \mathbf{x}_1 \Big) g_st \cdot \mathcal{F}_j^{\; -} (x_2^+, \mathbf{x}_1  ) \mathcal{U}_F\Big(x_2^{+}, \frac{L}{2}^+ ; \mathbf{x}_1 \Big) 
  \bigg ]  \; .
  \label{MNLO_GPsi_1}
\end{align}
Upon performing the following translation 
\begin{equation}
    \mathcal{O} (x_2^+, x_1^+ , \mathbf{x}_1, \mathbf{x}_2 ) = e^{i \hat{p} \cdot x_1 }  \mathcal{O} (x_2^+-x_1^+,0^+, \mathbf{0}, \mathbf{x}_{21}) e^{-i \hat{p} \cdot x_1 } \; ,
    \label{Eq:GluonTransl}
\end{equation}
and restoring the normalization of the target states in the resulting expression as described in Eq. \eqref{Eq:DefForwMatrElem}, the NLO correction to the NEik quark contribution to the inclusive DIS in a gluon background for longitudinally polarized photon can be written as 
\begin{align}
\label{sigma_NLOG_Psi}
   \big[\sigma_{{\rm NLO , \; G, \Psi}}^{b.a.}\big]_{\rm Neik}^{\rm L}  =    \frac{ e_f^2 g^2  g_s^2}{ \pi } x_{Bj} \int_0^1 dz \; z (1-z)^2 \int \frac{d^{2} \mathbf{r}}{(2 \pi)} \; K_{0}^2 (\bar{Q} |\mathbf{r}|)  {\rm Re } : G_q (\mathbf{r}) \; ,
\end{align}
where we have used $x_{Bj}\sim Q^2/ {  2 q^+ p_t^-} $ and  the target-averaged color operator in the gluon background $G_q(\r)$ is defined as 
\begin{align}
\label{de:Gq}
 G_q (\mathbf{r}) &= - \frac{i}{g_s^2}  
 \int \frac{d z^{+}}{ \pi} \theta ( z^+ )  z^+ \;  
  \big\langle p_t\big|  {\rm Tr_c}  \Big[  \mathcal{U}_{F} (\mathbf{r})   \mathcal{U}_F\left(-\infty^+, 0^+ ; \mathbf{0} \right) 
  \nonumber \\ 
  &
  \times  \; g_st \cdot \mathcal{F}_j^{\; -} (0^+, \mathbf{0}  ) \; \mathcal{U}_F\left( 0^{+}, z^+ ; \mathbf{0} \right) g_st \cdot \mathcal{F}_j^{\; -} (z^+ , \mathbf{0}  ) \mathcal{U}_F\left( z^+ , \infty^+ ; \mathbf{0} \right) \big] \big| p_t \big\rangle \;  .
\end{align}
Two comments are in order before we proceed with computation of the cross section for transversely polarized photon. 
First, as discussed in detail at the beginning of the section, to maintain a consistent power counting in terms of $g_s$, the non-perturbative color operator $G_q$ is defined with an explicit factor of $1/g_s^2$ in Eq. \eqref{de:Gq}. 
With this normalization, both the non-perturbative quark distribution at LO and the non-perturbative gluon distribution at NLO given by $G_q$ scale as ${\cal O}(1/g_s^2)$. Consequently, an explicit factor of  $g_s^2$ factor appears in Eq. \eqref{sigma_NLOG_Psi}, making the power counting between LO and NLO manifest.  Second, the power counting arguments regarding $x_{Bj}$ follow exactly the same reasoning as in the case of  NLO quark contributions analyzed in Sec.~\ref{Sec:NLOCrossSecQuark}. Therefore up to NNEik corrections the cross section in Eq. \eqref{sigma_NLOG_Psi} can be written as 
\begin{align}
\label{sigma_NLOG_Psi_2}
   \big[\sigma_{{\rm NLO , \; G, \Psi}}^{b.a.}\big]_{\rm NEik}^{\rm L}  =    \frac{ e_f^2 g^2  g_s^2}{ \pi } x_{Bj} \int_0^1 dz \; z (1-z)^2 \int \frac{d^{2} \mathbf{r}}{(2 \pi)} \; K_{0}^2 (\bar{Q} |\mathbf{r}|)  {\rm Re } : G_q (x_{Bj},\mathbf{r}) \; ,
\end{align}
with 
\begin{align}
\label{de:Gq_2}
 G_q (x_{Bj},\mathbf{r}) &= - \frac{i}{g_s^2}  
 \int \frac{d z^{+}}{ \pi} \theta ( z^+ )  z^+  e^{-ix_{Bj}{{p_t^-}} z^+}\;  
  \big\langle p_t\big|  {\rm Tr_c}  \Big[  \mathcal{U}_{F} (\mathbf{r})   \mathcal{U}_F\left(-\infty^+, 0^+ ; \mathbf{0} \right) 
  \nonumber \\ 
  &
  \times  \; g_st \cdot \mathcal{F}_j^{\; -} (0^+, \mathbf{0}  ) \; \mathcal{U}_F\left( 0^{+}, z^+ ; \mathbf{0} \right) g_st \cdot \mathcal{F}_j^{\; -} (z^+ , \mathbf{0}  ) \mathcal{U}_F\left( z^+ , \infty^+ ; \mathbf{0} \right) \big] \big| p_t \big\rangle \;  .
\end{align}
{ 
In particular, let us note that at $\mathbf{r}=0$, the color structure can be simplified, and one finds
\begin{align}
\label{de:Gq_2_new}
 G_q (x_{Bj},\mathbf{0}) &= - \frac{i}{2}  
 \int \frac{d z^{+}}{ \pi} \theta ( z^+ )  z^+  e^{-ix_{Bj}{{p_t^-}} z^+}\;   
  %\nonumber \\ 
  %&
  \langle p_t| \mathcal{F}_j^{\; - a} (0^+, \mathbf{0}  ) \; [ \mathcal{U}_A \left( 0^{+}, z^+ ; \mathbf{0} \right)]^{ab} \mathcal{F}_j^{\; - b} (z^+ , \mathbf{0}  )   | p_t \rangle \;  ,
\end{align}
so that
\begin{align}
\label{de:Gq_2_new_dag}
 G_q (x_{Bj},\mathbf{0})^{\dagger} &=  \frac{i}{2}  
 \int \frac{d z^{+}}{ \pi} \theta ( z^+ )  z^+  e^{ix_{Bj}{{p_t^-}} z^+}\;   
  %\nonumber \\ 
  %&
  \langle p_t| \mathcal{F}_j^{\; - b} (z^+ , \mathbf{0}  ) \; [ \mathcal{U}_A \left( z^{+}, 0^+ ; \mathbf{0} \right)]^{ba} \mathcal{F}_j^{\; - a} (0^+, \mathbf{0}  )   | p_t \rangle \;  ,
  \nonumber\\
  &=  \frac{i}{2}  
 \int \frac{d z^{+}}{ \pi} \theta ( z^+ )  z^+  e^{ix_{Bj}{{p_t^-}} z^+}\;   
  %\nonumber \\ 
  %&
  \langle p_t| \mathcal{F}_j^{\; - b} (0^+ , \mathbf{0}  ) \; [ \mathcal{U}_A \left( 0^{+}, -z^+ ; \mathbf{0} \right)]^{ba} \mathcal{F}_j^{\; - a} (-z^+, \mathbf{0}  )   | p_t \rangle \;  ,
  \nonumber\\ 
    &=  \frac{i}{2}  
 \int \frac{d z^{+}}{ \pi} \theta ( -z^+ )  (-z^+)  e^{-ix_{Bj}{{p_t^-}} z^+}\;   
  %\nonumber \\ 
  %&
  \langle p_t| \mathcal{F}_j^{\; - a} (0^+ , \mathbf{0}  ) \; [ \mathcal{U}_A \left( 0^{+}, z^+ ; \mathbf{0} \right)]^{ab} \mathcal{F}_j^{\; - b} (z^+, \mathbf{0}  )   | p_t \rangle \;  ,
\end{align}
where, in the second step, we have shifted the whole operator by $-z^+$, and in the last step, we have relabeled the color indices and made the change of variable $z^+\mapsto -z^+$. Hence, we find
\begin{align}
    2{\rm Re } : G_q (x_{Bj},\mathbf{0}) 
    =&\,
- \frac{i}{2}  
 \int \frac{d z^{+}}{ \pi}   z^+  e^{-ix_{Bj}{{p_t^-}} z^+}\;   
  %\nonumber \\ 
  %&
  \langle p_t| \mathcal{F}_j^{\; - a} (0^+, \mathbf{0}  ) \; [ \mathcal{U}_A \left( 0^{+}, z^+ ; \mathbf{0} \right)]^{ab} \mathcal{F}_j^{\; - b} (z^+ , \mathbf{0}  )   | p_t \rangle \;  ,
    \nonumber\\
    =&\,
    \frac{\partial}{\partial {\rm x}} \Big[{\rm x} g ({\rm x})\Big] \bigg|_{{\rm x}=x_{Bj}} \; ,
\end{align}
with the background gluon PDF $g ({\rm x})$ defined as 
\begin{align}
    g ({\rm x}) = \frac{1}{{\rm x} p_t^{-}} \int \frac{dz^+}{2 \pi} e^{-i  {\rm x}  z^+ p_t^{-}}   \langle p_t| \mathcal{F}_j^{\; - a} (0^+, \mathbf{0}  ) \; [ \mathcal{U}_A \left( 0^{+}, z^+ ; \mathbf{0} \right)]^{ab} \mathcal{F}_j^{\; - b} (z^+ , \mathbf{0}  )   | p_t \rangle \; .
\end{align}
}

For a transversely polarized photon, we first average over the $(D-2)$ photon polarizations and then sum over photon polarizations in Eq.~\eqref{SNLO_BA_1}. Although the computation of the Dirac trace in $D$ dimensions is slightly more involved, it can be straightforwardly evaluated using {\tt FeynCalc}~\cite{Mertig:1990an,Shtabovenko:2016sxi}, yielding
\begin{align}
\label{eq:MNLOG_T_1}
   &  \langle p_t'| 2 {\rm Im } : [\mathcal{M}_{{\rm NLO , \; G, \Psi}}^{b.a.}]_{\rm Neik}^{\rm T} | p_t \rangle = \frac{ -e_f^2 g^2 }{ \pi q^+ (D-2)} \int_0^1 \frac{dz}{z} \int d^{D-2} \mathbf{x}_1 \int d^{D-2} \mathbf{x}_2  \int \frac{d^{D-2} \mathbf{k}_i}{(2 \pi)^{D-2}} \nonumber \\ 
   & \hspace{0.5cm} 
   \times 
   \int \frac{d^{D-2} \mathbf{k}_f}{(2 \pi)^{D-2}} 
    \frac{e^{ - i \mathbf{x}_{12} \cdot (\mathbf{k}_f - \mathbf{k}_i)}}{[\mathbf{k}_i^2 + \bar{Q}^2][\mathbf{k}_f^2 + \bar{Q}^2]} \Big[ \mathbf{k}_i \cdot \mathbf{k}_f (z^2 + (1-z)^2) + m^2 + \frac{D-4}{2} (\mathbf{k}_i \cdot \mathbf{k}_f + m^2) \Big] 
    \nonumber \\ 
    & \hspace{0.5cm}
    \times 
    {\rm Re} : (-i)  {\rm Tr_c}  \Big[ \mathcal{U}_{F} (\mathbf{x}_{1}) \mathcal{U}^{ (2) \dagger}_{F} (\mathbf{x}_2) \Big ] \; ,
\end{align}
In Eq. \eqref{eq:MNLOG_T_1}, transverse integrations can be performed using Eqs. \eqref{Eq:AppTransvMomInt1} and \eqref{Eq:AppTransvMomInt2}, and one obtains 
\begin{align}
 &
 \langle p_t'| 2 {\rm Im } :  [\mathcal{M}_{{\rm NLO , \; G, \Psi}}^{b.a.}]_{\rm Neik}^{\rm T} | p_t \rangle 
  = \frac{ - e_f^2 g^2 }{ \pi q^+ (D-2)} \int_0^1 \frac{dz}{z} \int \frac{d^{2-2 \epsilon} \mathbf{x}_1 d^{2-2 \epsilon} \mathbf{x}_2 }{ (2 \pi)^{2-2 \epsilon} } \left( \frac{\bar{Q}^2}{\mathbf{x}_{12}^2} \right)^{-\epsilon} 
  \nonumber \\ 
   & \times  
  \Big[ \big(z^2 + (1-z)^2\big) \bar{Q}^2 K_{1-\epsilon}^2 (\bar{Q} |\mathbf{x}_{12}|)  + m^2 K_{\epsilon}^2 (\bar{Q} |\mathbf{x}_{12}|) 
  - 
  \epsilon \Big( \bar{Q}^2 K_{1-\epsilon}^2 (\bar{Q} |\mathbf{x}_{12}|) + m^2 K_{\epsilon}^2 (\bar{Q} |\mathbf{x}_{12}|) \Big) \Big] 
  \nonumber \\ 
  & \times {\rm Re} : (-i) \big\langle p_t'\big|  {\rm Tr_c}  \big[ \mathcal{U}_{F} (\mathbf{x}_{2}) \mathcal{U}^{ (2) \dagger}_{F} (\mathbf{x}_1) \big ] \big| p_t \big\rangle  \; .
\end{align}
Finally, performing the translation in eq.~(\ref{Eq:GluonTransl}), we can write the cross section as 
\begin{align}
    &  \big[\mathcal{\sigma}_{{\rm NLO , \; G, \Psi}}^{b.a.}\big]_{\rm NEik}^{\rm T}  
     =  \frac{  e_f^2 \, g^2 \, g_s^2 }{ {  2 q^+ p_t^-} (D-2) }\,  {\rm Re} : \int_0^1 \frac{dz}{z} \int \frac{d^{2-2 \epsilon} \mathbf{r} }{ (2 \pi)^{2-2 \epsilon} } \left( \frac{\bar{Q}^2}{\mathbf{r}^2} \right)^{-\epsilon}  
     \nonumber \\ 
     &
     \times  \Big[ (z^2 + (1-z)^2) \bar{Q}^2 K_{1-\epsilon}^2 (\bar{Q} |\mathbf{r}|)  + m^2 K_{\epsilon}^2 (\bar{Q} |\mathbf{r}|) - \epsilon ( \bar{Q}^2 K_{1-\epsilon}^2 (\bar{Q} |\mathbf{r}|) + m^2 K_{\epsilon}^2 (\bar{Q} |\mathbf{r}|) ) \bigg] G_q (\mathbf{r}) \; ,
     \label{Eq:Sigma_NLO_Psi_T_GluonBack}
\end{align}
with the gluon distribution $G_q$ defined in Eq.~\eqref{de:Gq}. 

By comparing Eqs.~\eqref{Eq:Sigma_NLO_Psi_T_GluonBack} and~\eqref{Eq:Sigma_NLO_Psi_T}, one finds that the NLO corrections to the quark contribution to the inclusive DIS cross section in a gluon background field are identical to those in a quark background field, up to the replacement of the gluon and quark distributions, which are given by
\begin{equation}
\label{rel_F2G}
    f_q (x_{Bj},\mathbf{r}) \longrightarrow  \frac{2}{D-2} G_q(x_{Bj},\mathbf{r}) \; , \hspace{1cm} C_F\, q_f (x_{Bj}) \longrightarrow  \frac{2}{D-2} T_R\, \frac{\partial}{\partial {\rm x}} \Big[{\rm x} g ({\rm x})\Big] \bigg|_{{\rm x}=x_{Bj}} \; .
\end{equation}
Therefore, the analysis of the divergences and the extraction of the finite terms proceed in exactly the same way, up to the above-mentioned interchange of the quark and gluon distributions.\footnote{Since the same type of divergences are found than in Sec.~\ref{Sec:NLOCrossSecQuark}, this NLO gluon background contribution should also participate to the double log evolution of the LO background quark contribution from Sec.~\ref{Sec:LOCrossSec}.}

Finally, the remaining NLO contribution in a gluon background field arises when the decoration is located on the quark propagator. Following exactly the same steps as before, and after a $z \rightarrow 1-z$ change of variables, one obtains the same cross section as given in Eq.~\eqref{Eq:Sigma_NLO_Psi_T_GluonBack}, up to the replacement of $G_q$ by $G_{\bar q}$, where the operator $G_{\bar q}$ is defined as
\begin{align}
    G_{\bar{q}} (x_{Bj},\mathbf{r}) & = -\frac{i}{g_s^2}  \int \frac{d z^{+}}{ \pi} \theta ( -z^+ )  (-z^+) \, e^{-ix_{Bj}p_t^- z^+}\;   \big\langle p_t\big|  {\rm Tr_c}  \Big[  \mathcal{U}^{\dagger}_{F} (\mathbf{r})  \,  \mathcal{U}_F \left( \infty^+, 0^+ ; \mathbf{0} \right) 
    \nonumber \\ 
  &  \hspace{1cm}
  \times  g_st \cdot \mathcal{F}_j^{\; -} (0^+, \mathbf{0}  ) \; \mathcal{U}_F\left( 0^{+}, z^+ ; \mathbf{0} \right) g_st \cdot \mathcal{F}_j^{\; -} (z^+ , \mathbf{0}  )\,  \mathcal{U}_F\left( z^+ , -\infty^+ ; \mathbf{0} \right) \Big ] \big| p_t \big\rangle \; .
\end{align}
%
%\color{blue} Thus, summing quark and the antiquark contribution amounts to the replacement: ${\rm Re}: G_q (\mathbf{r}) \rightarrow {\rm Re}: G_q (\mathbf{r}) + G_{\bar{q}} (\mathbf{r})$, which vanishes.

%%%%%%%%%%%%%%%%%%%%%%%%%%%%%%%%%%%%%%%%%%%%%%%%%%%%%%%%%%%%%%%%
%%%%%%%%%%%%%%%%%%%%%%%%%%%%%%%%%%%%%%%%%%%%%%%%%%%%%%%%%%%%%%%%
\subsection{Contributions from the gluon background field beyond the static approximation}
%%%%%%%%%%%%%%%%%%%%%%%%%%%%%%%%%%%%%%%%%%%%%%%%%%%%%%%%%%%%%%%
%%%%%%%%%%%%%%%%%%%%%%%%%%%%%%%%%%%%%%%%%%%%%%%%%%%%%%%%%%%%%%%

At NEik order, non-static corrections with $z^- \neq 0$ contribute to physical observables, enabling longitudinal momentum transfer between the projectile and the target during the interaction. This contribution reads\footnote{We use the label b.s., meaning "beyond static", to denote eikonal + non-static contributions. After removing the eikonal part, we will use n.s., meaning "non static".}
\begin{align}
    [S_{{\rm NLO , \; G}}^{\rm b.s.}]^{\mu \nu} = e_f^2 \; g^2 \int_{-\infty^+}^{-\frac{L}{2}^+} d^D y \int_{ \frac{L}{2}^+}^{\infty^+} d^D x \; e^{i q' \cdot x -i q \cdot y} \;  {\rm Tr_{D,c}} \left[  \gamma^{\nu} S_{F,q}^{\rm n.s.} (x, y) \gamma^{\mu}  S_{F, \bar{q}}^{\rm n.s.} (y, x) \right] \; ,
    \label{Eq:GluonConBegNonStat}
\end{align}
where the non-static quark and antiquark propagators are given in Eqs.~(\ref{Eq:NonStatQuarkProp}) and~(\ref{Eq:NonStatAntiQuarkProp}), respectively. Using their explicit expressions, one readily obtains
\begin{align}
\label{SNLO_bs_1}
 \big[S_{{\rm NLO , G}}^{\rm b.s.}\big]^{\mu \nu} 
 & = -e^2_f \, g^2 
 \int^{-\frac{L}{2}^+}_{-\infty^+}d^Dy
 \int^{+\infty^+}_{\frac{L}{2}^+}d^Dx 
 \int_{0^+}^{\infty^+} \frac{d^{D-1}\underline{k_f}}{(2\pi)^{D-1}}\frac{d^{D-1}\underline{k_i}}{(2\pi)^{D-1}}
 \int_{- \infty^+ }^{0^+} \frac{d^{D-1}\underline{p_f}}{(2\pi)^{D-1}}\frac{d^{D-1}\underline{p_i}}{(2\pi)^{D-1}}
 \nonumber\\ 
 & \times 
 e^{ix\cdot(q'+\check{p}_i-\check{k}_f)-iy\cdot(q+\check{p}_f-\check{k}_i)}\; 
 {\rm Tr}_{\rm D}\bigg[\gamma^{\nu}\frac{(\slashed{k}_f+m)}{2k^+_f}\gamma^+\frac{(\slashed{k}_i+m)}{2k^+_i}\gamma^{\mu}\frac{(\slashed{p}_f+m)}{2p^+_f}\gamma^+\frac{(\slashed{p}_i+m)}{2p^+_i}\bigg]
 \nonumber \\ 
 & \times 
 \int d^{D-2}\mathbf{x}_1
 \int d^{D-2}\mathbf{x}_2
 \int dx^-_1
 \int dx^-_2
 e^{ix^-_1(k^+_f-k^+_i)-i\mathbf{x}_1\cdot(\mathbf{k}_f-\mathbf{k}_i)+ix^-_2(p^+_f-p^+_i)-i\mathbf{x}_2\cdot(\mathbf{p}_f-\mathbf{p}_i)}
 \nonumber\\ 
 & \times 
 {\rm Tr}_{\rm c}\left[\mathcal{U}_F(\mathbf{x}_1,x^-_1)\mathcal{U}^{\dagger}_F(\mathbf{x}_2,x^-_2)-1\right] ,
\end{align}
where we kept the eikonal part and thus subtracted the non-interacting contribution. It is convenient to perform the change of variables with unit Jacobian
\begin{equation}
     b^-=\frac{x^-_1+x^-_2}{2}\hspace{0.1cm},\hspace{0.5cm} r^-= x^-_2-x^-_1 \; .
\end{equation}
In this way, the operator structure becomes
\begin{gather}
    \mathcal{U}_F(\mathbf{x}_1,x^-_1)\mathcal{U}^{\dagger}_F(\mathbf{x}_2,x^-_2)  \longrightarrow \mathcal{U}_F\left(\mathbf{x}_1,b^--\frac{r^-}{2}\right)\mathcal{U}^{\dagger}_F\left(\mathbf{x}_2,b^-+\frac{r^-}{2}\right) \; .
\end{gather}
Since we are ultimately interested in the forward matrix element of the operator, we can shift it by $b^-$ and then expand around $r^- = 0$, i.e.
\begin{align}
 &
 \mathcal{U}_F\left(\mathbf{x}_1, -\frac{r^-}{2}\right)\mathcal{U}^{\dagger}_F\left(\mathbf{x}_2, \frac{r^-}{2}\right)  
 \nonumber\\
 & =  
 \mathcal{U}_F\left(\mathbf{x}_1,0^-\right)\mathcal{U}^{\dagger}_F\left(\mathbf{x}_2,0^-\right) 
% \nonumber \\ 
 %& 
 + \frac{r^-}{2} \left[\mathcal{U}_F(\mathbf{x}_1,b^-)\overleftrightarrow{\partial}_{b^-}\mathcal{U}^{\dagger}_F(\mathbf{x}_2,b^-)\right]_{b^-=0} + \mathcal{O} ((r^-)^2)
 \nonumber \\ 
 & = \mathcal{U}_F\left(\mathbf{x}_1,0^-\right)\mathcal{U}^{\dagger}_F\left(\mathbf{x}_2,0^-\right) 
 %\nonumber \\ 
 %& 
 + \frac{r^-}{2} \left[\mathcal{U}_F(\mathbf{x}_1,b^-)\overleftrightarrow{D}^F_{b^-}\mathcal{U}^{\dagger}_F(\mathbf{x}_2,b^-)\right]_{b^-=0} + \mathcal{O} ((r^-)^2) \; .
\end{align}
where $\overleftrightarrow{\partial}_{b^-} = \overrightarrow{\partial}_{b^-} - \overleftarrow{\partial}_{b^-}$ and, in the last equality, we used the gauge condition $A^+=0$. Thanks to translation invariance, it is now clear that the first term of the expansion in $r^-$, with the non-interacting part subtracted, is nothing but the eikonal contribution. Terms $\mathcal{O} ((r^-)^2)$ are  next-to-next-to-eikonal (NNEik) contributions and therefore can be discarded in the accuracy of our calculation. We must therefore concentrate only on the operator containing the covariant derivative in fundamental representation. A useful property of this operator is that it vanishes in the UV limit, i.e.
\begin{equation}
\lim_{\mathbf{x}_1\rightarrow\mathbf{x}_2} {\rm Tr}_{\rm c}\left[\mathcal{U}_F(\mathbf{x}_1,b^-)\overleftrightarrow{D}^F_{b^-}\mathcal{U}^{\dagger}_F(\mathbf{x}_2,b^-)\right]_{b^-=0} =  0,
\label{Eq:UVnonStatic}
\end{equation}
as it is easy to verify by using the property 
\begin{align}
 \partial_{z^-} \mathcal{U}_R \left(\mathbf{z},z^- \right)  
 =-i g_s 
% \nonumber \\ 
 %\times 
 \int_{-\frac{L}{2}^+}^{\frac{L}{2}^+} \!\! d z^{+} \; \mathcal{U}_R\Big(\frac{L}{2}^+, z^{+} ; \mathbf{z}, z^- \Big) T_R \cdot \mathcal{F}^{+ -}(z^+, \mathbf{z}, z^-) \; \mathcal{U}_R \Big(  z^{+}, - \frac{L}{2}^+ ; \mathbf{z}, z^- \Big) . 
 \label{Eq:MinDeriv}
\end{align}
As in previous computations, we first integrate over the  positions of the photon vertices $x$ and $y$, followed by integration over the anti-quark internal momenta, $p_i$ and $p_f$. In the absence of a strict equality between between $k^+_i$ and $k^+_f$, it is convenient to introduce
\begin{gather}
\bar{Q}^2_f=m^2+\frac{k^+_f(q^+-k^+_f)}{(q^+)^2}Q^2 \hspace{1 cm} {\rm and} \hspace{1 cm} \bar{Q}^2_i=m^2+\frac{k^+_i(q^+-k^+_i)}{(q^+)^2}Q^2 \; \; .
\end{gather}
Consequently, Eq. \eqref{SNLO_bs_1} can be written as 
%Then, we obtain
\begin{gather}
\label{SNLO_ns_G_gen}
    [S_{{\rm NLO , G}}^{\; \rm n.s.}]^{\mu \nu} =  \frac{e^2_f g^2}{2} (2 \pi) \delta (q^+ - q'^+) \int d^{D-2}\mathbf{x}_1\int d^{D-2}\mathbf{x}_2 \int_{0^+}^{q^+} \frac{d^{D-1}\underline{k_i}}{(2\pi)^{D-1}} \frac{d^{D-1}\underline{k_f}}{(2\pi)^{D-2}} \delta' (k^+_f-k^+_i) \nonumber \\ \times  \; \frac{4 k_i^+ k_f^+ (q^+-k_i^+ ) (q^+-k_f^+ ) }{(q^+)^2\left[\mathbf{k}^2_f+\bar{Q}^2_f\right]\left[\mathbf{k}^2_i+\bar{Q}^2_i\right]}   e^{-i\mathbf{x}_{12} \cdot(\mathbf{k}_f-\mathbf{k}_i)} {\rm Tr}_{\rm c}\left[\mathcal{U}_F(\mathbf{x}_1,b^-) i\overleftrightarrow{D}^F_{b^-}\mathcal{U}^{\dagger}_F(\mathbf{x}_2,b^-)\right]_{b^-=0} \nonumber \\ 
    \times {\rm Tr}_{\rm D}\left[\gamma^{\nu}\frac{(\slashed{k}_f+m)}{2k^+_f}\gamma^+\frac{(\slashed{k}_i+m)}{2k^+_i}\gamma^{\mu}\frac{(\slashed{p}_f+m)}{2p^+_f}\gamma^+\frac{(\slashed{p}_i+m)}{2p^+_i}\right]  \bigg|_{\substack{p_i^+ = k_f^+ - q^+ \\ p_f^+ = k_i^+ - q^+ \\ \mathbf{k}_i = \mathbf{p}_f \; , \; \mathbf{p}_i = \mathbf{k}_f}} \nonumber \; ,
\end{gather}
where the derivative of the Dirac Delta distribution, $\delta'$, is defined by the action
\begin{equation}
   \int_{-\infty}^{\infty} dx \; \delta'(x-y) f(x) = - \left. \partial_xf(x)\right|_{x=y} \;.
   \label{Eq:DeltaDer}
\end{equation}

In the longitudinally polarized case, we consider $[S_{{\rm NLO , G}}^{\; \rm n.s.}]^L = \frac{Q^2}{(q^+)^2} [S_{{\rm NLO , G}}^{\; \rm n.s.}]^{+ +}$. Thus, the Dirac structure reads
\begin{gather}
\label{trace_ns_1}
     {\rm Tr}_{\rm D}\left[\gamma^{+}\frac{(\slashed{k}_f+m)}{2k^+_f}\gamma^+\frac{(\slashed{k}_i+m)}{2k^+_i}\gamma^{+}\frac{(\slashed{p}_f+m)}{2p^+_f}\gamma^+\frac{(\slashed{p}_i+m)}{2p^+_i}\right]  \bigg|_{\substack{p_i^+ = k_f^+ - q^+ \\ p_f^+ = k_i^+ - q^+ \\ \mathbf{k}_i = \mathbf{p}_f \; , \; \mathbf{p}_i = \mathbf{k}_f}} = 2 \; ,
\end{gather}
Using Eq. \eqref{trace_ns_1}, extracting the forward scattering amplitude from the  $S$-matrix, taking its imaginary part, and evaluating the quantum expectation value of the resulting expression in the target states, one obtains 
\begin{align}
\label{expectation_ns_L_1}
  &\big\langle p_t'\big| 2 {\rm Im } :  \big[\mathcal{M}_{{\rm NLO , \; G}}^{\; \rm n.s.}\big]_{\rm Neik}^{\rm L} \big| p_t \big\rangle 
  =  
  \frac{- 4 \, e^2_f \, g^2 \, Q^2}{(2 \pi)}   
  \int_0^{q^+} \frac{d k_i^+}{q^+} 
  \int_0^{q^+}  \frac{d k_f^+}{q^+} \delta' \bigg( \frac{k^+_f}{q^+}- \frac{k^+_i}{q^+} \bigg)   
  \nonumber \\ 
  & \times 
  \frac{ k_i^+ k_f^+ (q^+-k_i^+ ) (q^+-k_f^+ ) }{(q^+)^4 } 
   \int d^{D-2}\mathbf{x}_1
   \int d^{D-2}\mathbf{x}_2 
   \int \frac{d^{D-2}\mathbf{k_i}}{(2\pi)^{D-2}} \frac{d^{D-2}\mathbf{k_f}}{(2\pi)^{D-2}} 
   \frac{ e^{-i\mathbf{x}_{12} \cdot(\mathbf{k}_f-\mathbf{k}_i)} }{\big[\mathbf{k}^2_f+\bar{Q}^2_f\big]\big[\mathbf{k}^2_i+\bar{Q}^2_i\big]}
    \nonumber \\ 
    & \times 
    {\rm Re} : \big\langle p_t'\big| {\rm Tr}_{\rm c} \big[\mathcal{U}_F(\mathbf{x}_1,b^-) i\overleftrightarrow{D}^F_{b^-}\mathcal{U}^{\dagger}_F(\mathbf{x}_2,b^-) \big]_{b^-=0} \big| p_t \big\rangle \; .
\end{align}
In Eq. \eqref{expectation_ns_L_1}, the transverse integrals over $\k_i$ and $\k_f$ can be performed by using Eq. \eqref{Eq:AppTransvMomInt1}. In the resulting expression, performing the following change of variables 
\begin{equation}
    z = \frac{k_i^+}{q^+} \hspace{0.5 cm}  {\rm and} \hspace{0.5 cm}  \bar{z} = \frac{k_f^+}{q^+} \; ,
\end{equation}
yields 
\begin{align}
&  \langle p_t'| 2 {\rm Im } :  [\mathcal{M}_{{\rm NLO , \; G}}^{\; \rm n.s.}]_{\rm Neik}^{\rm L} | p_t \rangle =  -\frac{ e^2_f g^2 Q^2}{  \pi^2}   \int_0^{1} d z \int_0^{1} d \bar{z} \; \delta' \left( \bar{z} - z \right) z \bar{z} (1-z)  (1-\bar{z}) \nonumber \\ 
&\times  \int \frac{ d^{2} \mathbf{x}_1  d^{2}\mathbf{x}_2}{ 2 \pi } \; K_0 (\bar{Q}_f |\mathbf{x}_{12}|) K_0 (\bar{Q}_i |\mathbf{x}_{12}|)  {\rm Re} : \langle p_t'| {\rm Tr}_{\rm c} \big[\mathcal{U}_F(\mathbf{x}_1,b^-) i\overleftrightarrow{D}^F_{b^-}\mathcal{U}^{\dagger}_F(\mathbf{x}_2,b^-) \big]_{b^-=0} | p_t \rangle \nonumber \; .
\end{align}
By using eq.~(\ref{Eq:DeltaDer}), we can further write
\begin{align}
  \langle p_t'| 2 {\rm Im } :  [\mathcal{M}_{{\rm NLO , \; G}}^{\; \rm n.s.}]_{\rm Neik}^{\rm L} | p_t \rangle & =  
  \frac{  e^2_f \, g^2 \, Q^2}{ \pi^2}   \int_0^{1} d z  \;  z (1-z)  ( 1 - 2 z ) 
   \nonumber \\ 
   & \times  
   \int \frac{ d^{2} \mathbf{x}_1  d^{2}\mathbf{x}_2}{ 2 \pi } \left[ K_0^2 (\bar{Q} |\mathbf{x}_{12}|) - \frac{Q^2 - m^2}{2 \bar{Q}} |\mathbf{x}_{12}| K_0 (\bar{Q} |\mathbf{x}_{12}|) K_1 (\bar{Q} |\mathbf{x}_{12}|) \right] 
   \nonumber \\ 
   & \times {\rm Re} : \langle p_t'| {\rm Tr}_{\rm c} \big[\mathcal{U}_F(\mathbf{x}_1,b^-) i\overleftrightarrow{D}^F_{b^-}\mathcal{U}^{\dagger}_F(\mathbf{x}_2,b^-) \big]_{b^-=0} | p_t \rangle  \nonumber \; .
\end{align}
Since the integrand is of the form $(1-2z) f(z)$, where $f(z)$ is symmetric under the transformation $z \to 1-z$, the integration over $z$ yields zero. Consequently, we can conclude that non-static NEik corrections to the quark background contribution to the inclusive DIS cross section for a longitudinally polarized photon vanishes identically. 

In the transversally polarized case, one considers
\begin{align}
 [S_{{\rm NLO , G}}^{\; \rm n.s.}]^T = \frac{-g_{\perp \mu \nu}}{D-2} [S_{{\rm NLO , G}}^{\; \rm n.s.}]^{\mu \nu}
 \end{align} 
 Consequently, the Dirac trace in Eq. \eqref{SNLO_ns_G_gen} reduces to 
\begin{align}
 & \hspace{-0.5cm}
 {\rm Tr}_{\rm D}\left[\gamma_{\perp \mu}\frac{(\slashed{k}_f+m)}{2k^+_f}\gamma^+\frac{(\slashed{k}_i+m)}{2k^+_i}\gamma_{\perp}^{\mu}\frac{(\slashed{p}_f+m)}{2p^+_f}\gamma^+\frac{(\slashed{p}_i+m)}{2p^+_i}\right]  \bigg| {\substack{p_i^+ = k_f^+ - q^+ \\ p_f^+ = k_i^+ - q^+ \\ \mathbf{k}_i = \mathbf{p}_f \; , \; \mathbf{p}_i = \mathbf{k}_f}} \nonumber \\ 
 & \hspace{4cm}
 = - \frac{ \mathbf{k}_i \cdot \mathbf{k}_f ( q^+(q^+ (D-2) - 2 k_i^+ ) + 2 k_f^+ (2 k_i^+ - q^+) )}{2 k_i^+ k_f^+ (k_i^+ - q^+) (k_f^+ - q^+)} \; .
\end{align}
Proceeding in a manner entirely analogous to the longitudinal case, we obtain 
\begin{align}
& \hspace{-0.5cm}
 \langle p_t'| 2 {\rm Im } :  [\mathcal{M}_{{\rm NLO , \; G}}^{\; \rm n.s.}]_{\rm Neik}^{\rm T} | p_t \rangle =  \frac{  e^2_f g^2 }{ (2\pi)^2}   \int_0^{1} d z  \;  ( 2 z - 1 ) \int \frac{ d^{2} \mathbf{x}_1  d^{2}\mathbf{x}_2}{ 2 \pi } 
\nonumber \\ 
& \hspace{1cm}
\times \left[\left((1-z)^2+z^2+\frac{m^2}{\bar{Q}^2}\right)Q^2|\mathbf{x}_{12}|  \bar{Q} K_{0}(\bar{Q}|\mathbf{x}_{12}|)K_{1}(\bar{Q} |\mathbf{x}_{12}|)+2\bar{Q}^2K_{1}(\bar{Q}|\mathbf{x}_{12}|)^2\right] 
 \nonumber \\ 
 & \hspace{1cm} 
 \times {\rm Re} : \langle p_t'| {\rm Tr}_{\rm c} \big[\mathcal{U}_F(\mathbf{x}_1,b^-) i\overleftrightarrow{D}^F_{b^-}\mathcal{U}^{\dagger}_F(\mathbf{x}_2,b^-) \big]_{b^-=0} | p_t \rangle \; ,
\end{align}
where we set $D=4$, since the result is UV-finite due to Eq.~(\ref{Eq:UVnonStatic}). Once again, the antisymmetric nature of the integrand implies that its contribution vanishes when integrated over a symmetric interval. As a result, the non-static NEik corrections to the quark background contribution to the inclusive DIS cross section for a transversely polarized photon vanish identically, as in the case of longitudinally polarized photon.

\subsection{Total gluon contribution}
\label{sec:total_gluon}

Similarly to the quark background case, we can present the total gluon background contribution to the DIS structure functions at NEik and NLO, in the longitudinal case, as 
\begin{align}
F_L(x_{Bj},Q^2)\Big|_{{\rm NLO, G}}&= 4  \, x_{Bj} Q^2 \sum_{f} e_f^2 \, \, \frac{\alpha_{s}}{2 \pi} \, 
  \nonumber \\ & \times \int \frac{d^{2} \mathbf{r}}{\pi} \int_0^1 dz \; z^2 (1-z) K_{0}^2 (\bar{Q} |\mathbf{r}|) \; {\rm Re} : G(x_{Bj},\mathbf{r}) +{\rm NNEik}  \, , 
\end{align}
where we introduced the total target-averaged color structure $G(x_{Bj},\mathbf{r}) = G_q(x_{Bj},\mathbf{r}) + G_{\bar{q}}(x_{Bj},\mathbf{r}) $.
In the transverse case, we have 
\begin{align}
    F_T(x_{Bj} &,Q^2)\Big|_{{\rm NLO,G}}^{m=0}
    =
    x_{Bj} \sum_{f}
    e_f^2 \frac{\alpha_s}{2 \pi}
    \,{\rm Re} :
    \int_0^1 \frac{dz}{z}
    \int \frac{ d^{2} \mathbf{r} }{ \pi }
    \bigg\{
  \Big[ \bar{Q}^2 K_{1}^2\big(\bar{Q}|\mathbf{r}|\big) - \frac{1}{\mathbf{r}^2} \Big]
        G(x_{Bj},\mathbf{r}) 
        \nonumber \\
    & 
        - 2 z(1-z)\,\bar{Q}^2 K_{1}^2\big(\bar{Q}|\mathbf{r}|\big)
        \Big[ G(x_{Bj},\mathbf{r}) - G(x_{Bj},\mathbf{0}) \Big]
    \bigg\} + {\rm sub. \; reminder} + {\rm NNEik} \; ,
\end{align}
in the massless case and 
\begin{align}
     F_T(x_{Bj} &,Q^2)\Big|_{{\rm NLO,G}}^{m}
    = x_{Bj} \sum_{f}
    e_f^2 \,
    \frac{\alpha_s}{2\pi}
    \;{\rm Re} :
    \int_0^1 \frac{dz}{z}
    \int \frac{ d^{2} \mathbf{r} }{\pi} 
    \nonumber
\\
   & \times 
    \Big\{ \Big[ \bar{Q}^2 K_1^2(\bar{Q}|\mathbf{r}|)- m^2 K_1^2(m|\mathbf{r}|)
            + m^2\Big( K_0^2(\bar{Q}|\mathbf{r}|) - K_0^2(m|\mathbf{r}|) \Big)\Big] G(x_{Bj},\mathbf{r})
            \nonumber\\ 
            & 
        -\,
        2 z(1-z)\,\bar{Q}^2
        K_1^2(\bar{Q}|\mathbf{r}|)\, 
        \Big[ G(x_{Bj},\mathbf{r}) - G(x_{Bj},\mathbf{0}) \Big]  
    \Big\} + {\rm sub. \; reminder} + {\rm NNEik}
    \; ,
\end{align}
in the massive case, where "sub. reminder" denotes the scheme-dependent finite terms surviving the subtraction of divergences.

\section{Summary and outlook}
\label{Sec:sum_out}

In this work, we investigate NEik  contributions to DIS structure functions beyond the leading order in $\alpha_s$. We start our analysis with the LO computation and provide a brief summary of the results originally obtained in \cite{Altinoluk:2025ang}. At LO, the only NEik corrections arise from $t$-channel quark exchanges, which can be formulated in terms of insertions of the quark background field of the target. Consequently, these corrections to the DIS structure functions involve the presence two quark background fields $\Psi$. Therefore, these are NEik corrections and are suppressed by one power of energy compared the eikonal DIS structure functions computed in dipole factorization. One the other hand, in the dense regime for the target, the power counting is such that the quark field $\Psi={\cal O}(1/g_s)$ in QCD coupling. Hence, these corrections are parametrically of ${\cal O}(1/\alpha_s)$ and therefore perturbatively enhanced compared to eikonal DIS structure functions, corresponding to the dipole factorization formula. This relative enhancement by $(1/\alpha_s)$ of the NEik corrections with respect to the Eikonal dipole factorization result is actually valid not only in the dense regime but also in the dilute regime for the target.
At LO, we have shown that NEik quark background contributions to longitudinal structure function vanishes identically while the contribution to the transverse structure function is given in Eq. \eqref{FT_LO_fin}.  

At NLO, the NEik quark background contributions to DIS structure functions receives contributions both from quark and antiquark backgrounds of the target. The NEik and NLO corrections to the NEik quark background contributions to both longitudinal and transverse structure functions are computed in subsection \ref{Sec:QuarkCon}. While the NEik and NLO corrections to the NEik longitudinal structure function are completely finite, the corrections to NEik transverse structure function exhibit rapidity- and/or UV-divergences. We analyzed these divergences and extract the finite contributions. The same analysis is performed for the antiquark contributions and the total result is presented in subsection \ref{sec:total_fermion}.  

At NLO, the NEik DIS structure functions also receive contributions purely from the gluon background of the target. In this work, we classify these corrections according to whether they originate from a static or a non-static gluon background. We find that the corrections beyond the static approximation vanish identically at this order for the DIS structure functions. By contrast, the corrections arising within the static approximation (coming from the finite width of the target or from the transverse components of the background field) are non-vanishing. Furthermore, their structure closely resembles the corresponding quark contributions and can be obtained by replacing the gluon and quark distributions as specified in Eq. \eqref{rel_F2G}.

The results presented in this manuscript constitute the first study of NLO corrections to NEik DIS structure functions. A natural continuation of this work is the computation of NLO corrections to NEik hadron production in single-inclusive deep inelastic scattering (SIDIS). These studies increase the precision of CGC calculations, not only in terms of perturbative corrections but also by systematically including finite-energy effects, which are expected to play a key role in the phenomenological analysis of DIS observables at the future EIC. 

An interesting finding of the present manuscript is that the NLO corrections to the NEik DIS structure functions for a transversely polarized photon include a contribution that exhibits both rapidity and UV divergences, suggesting the emergence of a double-logarithmic evolution. We plan to explore this direction further and to derive this double-logarithmic evolution of the quark operator directly from its operator definition at low $x_{Bj}$.

\acknowledgments
We thank the anonymous referee for comments that helped us improve the quality of this paper. TA and JF are supported in part by the National Science Centre (Poland) under the research Grant No. 2023/50/E/ST2/00133 (SONATA BIS 13). GB is supported in part by the National Science Centre (Poland) under the research Grant No. 2020/38/E/ST2/00122 (SONATA BIS 10). The work of MF is supported by the ULAM fellowship program of NAWA No. BNI/ULM/2024/1/00065 “Color glass condensate effective theory beyond the eikonal approximation”.

\appendix

\section{Useful formulas}
\label{Sec:AppUsefulFormulas}

\paragraph*{Transverse momentum integrals.}
In the main body of the paper, we use extensively the following results for the transverse momentum integrals:
\begin{equation}
    \int \frac{d^{2-2 \epsilon} \mathbf{k}}{(2 \pi)^{2-2 \epsilon}}  \frac{e^{\pm  i \mathbf{k} \cdot \mathbf{r} }}{\mathbf{k}^2 + \bar{Q}^2} = \frac{K_{-\epsilon} (\bar{Q} |\mathbf{r}|) }{(2 \pi)^{1- \epsilon}} \left( \frac{\bar{Q}}{|\mathbf{r}|} \right)^{-\epsilon} \; ,
    \label{Eq:AppTransvMomInt1}
\end{equation}
\begin{equation}
    \int \frac{d^{2-2 \epsilon} \mathbf{k}}{(2 \pi)^{2-2 \epsilon}}  \frac{e^{\pm  i \mathbf{k} \cdot \mathbf{r} } }{\mathbf{k}^2 + \bar{Q}^2} \mathbf{k}  = \pm  \frac{ i \mathbf{r} }{|\mathbf{r}| }\frac{ \bar{Q} K_{1-\epsilon} (\bar{Q} |\mathbf{r}|) }{(2 \pi)^{1- \epsilon}} \left( \frac{\bar{Q}}{|\mathbf{r}|} \right)^{-\epsilon} \; .
    \label{Eq:AppTransvMomInt2}
\end{equation}
\paragraph*{Transverse coordinate integrals.} In the main body of the paper, we use extensively the following results for the transverse coordinate integrals:
\begin{equation}
    \int \frac{ d^{2-2 \epsilon} \mathbf{r} }{(2 \pi)^{2-2 \epsilon}} \frac{1}{ (\mathbf{r}^2)^{1 - 2 \epsilon}} \theta \left( \frac{1}{\Lambda^2} - \mathbf{r}^2 \right) = \frac{\Omega_{2-2 \epsilon}}{(2 \pi)^{2-2 \epsilon}} \int_0^{1/\Lambda} d |\mathbf{r}| \; |\mathbf{r}|^{-1 + 2 \epsilon} = \frac{\Lambda^{-2\epsilon}}{(4 \pi)^{1- \epsilon} \Gamma(1-\epsilon) 2 \epsilon} \; ,
    \label{Eq:AppTransvCoordInt1}
\end{equation}
where $\Omega_{2-2 \epsilon}$ is the solid angle in $2-2 \epsilon$ dimensions.
\begin{gather}
    \int \frac{ d^{2-2 \epsilon} \mathbf{r} }{(2 \pi)^{2-2 \epsilon}} \frac{ \bar{Q}^2  K_{1-\epsilon}^2 (\bar{Q} | \mathbf{r} |)  }{(\mathbf{r}^2)^{-\epsilon}} = \bar{Q}^2 \frac{ \Omega_{2-2 \epsilon} }{ (2 \pi)^{2-2 \epsilon} } \int d |\mathbf{r}| |\mathbf{r}| K_{1-\epsilon}^2 (\bar{Q} | \mathbf{r} |) \nonumber \\ = \frac{2}{(4 \pi)^{1-\epsilon} \Gamma (1-\epsilon) } \int_0^{\infty} \; dx \; x K_{1-\epsilon}^2 (x) = \frac{(1-\epsilon) \Gamma (\epsilon)}{(4 \pi)^{1-\epsilon}}  \; .
    \label{Eq:AppTransvCoordInt2}
\end{gather}
\paragraph*{Color properties.} In the main body of the paper, we use the relation 
\begin{equation}
    \mathcal{U}_A^{ba} (x) t^b = \mathcal{U}_F (x) t^a \mathcal{U}_F^{\dagger} (x) \; ,
    \label{Eq:AppenAdjFun}
\end{equation}
and the color Fierz identity 
\begin{equation}
    t^a_{kl}   t^a_{jn} = \frac{1}{2} \left( \delta_{kn} \delta_{lj} - \frac{1}{N_c} \delta_{kl} \delta_{jn} \right) \; .
    \label{Eq:AppenColorFierz}
\end{equation}

\section{Dirac trace in the quark background field contribution}
\label{Sec:AppDiracTrace}
\noindent In this section, we consider the Dirac trace appearing in eq.~(\ref{Eq:QuarkBackConTransverse}). Removing the trivial denominator, it reads
\begin{gather*}
     {\rm Tr_{D,c}} \left[ \gamma_{\perp}^{\mu} (\slashed{\check{k}}_f+m) \gamma^{+} \mathcal{U}_F\left(\mathbf{x}_2  \right) (\slashed{\check{k}}_i+m)  \gamma_{\perp \mu} (\slashed{\check{p}}_f + m ) \mathcal{U}_F \left( -\frac{L}{2}^+ , x_1^+, \mathbf{x}_1 \right) t^b \gamma^{i} \gamma^+ \gamma^-  \right. \nonumber \\ \left.  \Psi (x_1^+ , \mathbf{x}_1) \left[ \mathcal{U}_A \left(x_1^{+}, x_2^{+}, \mathbf{x}_1 \right) \right]^{ba}  \overline{\Psi} (x_2^+, \mathbf{x}_1) \gamma^- \gamma^+ t^a \gamma^{i}  \mathcal{U}_F\left( x_2^+, \frac{L}{2}^+ , \mathbf{x}_1 \right) (\slashed{\check{p}}_i+m) \right] \bigg| {\substack{p_i^+ = p_f^+ = k^+ - q^+ \\ k_i^+ = k_f^+ = k^+ \\ \mathbf{k}_i = \mathbf{p}_f \; , \; \mathbf{p}_i = \mathbf{k}_f}} \nonumber \; . 
    \end{gather*}
We start by rewriting it as
    \begin{gather*}
    -  {\rm Tr_{D,c}} \left[   \gamma^-  \gamma^{\nu}_{\perp} \gamma^+ (\slashed{\check{p}}_i+m) \gamma_{\perp}^{\mu} (\slashed{\check{k}}_f+m) \gamma^{+}  (\slashed{\check{k}}_i+m)  \gamma_{\perp \mu} (\slashed{\check{p}}_f + m ) \gamma^+ \gamma_{\perp \nu}  \gamma^- \mathcal{O}_q   \right] \bigg| {\substack{p_i^+ = p_f^+ = k^+ - q^+ \\ k_i^+ = k_f^+ = k^+ \\ \mathbf{k}_i = \mathbf{p}_f \; , \; \mathbf{p}_i = \mathbf{k}_f}}  \; ,
\end{gather*}
where $\mathcal{O}_q$ is defined in eq.~(\ref{Eq:Oope}). Now, we first use
\begin{gather}
    \gamma^+ (\slashed{\check{p}}_i+m)  \gamma_{\perp}^{\mu} (\slashed{\check{k}}_f+m) \gamma^{+} \bigg|_ {\substack{p_i^+ = p_f^+ = k^+ - q^+ \\ k_i^+ = k_f^+ = k^+ \\ \mathbf{k}_i = \mathbf{p}_f \; , \; \mathbf{p}_i = \mathbf{k}_f}} = 2 ( (k^+-q^+) \gamma_{\perp}^{\mu} \slashed{k}_{f \perp} + k^+ \slashed{k}_{f \perp} \gamma_{\perp}^{\mu} - m q^+ \gamma_{\perp}^{\mu} ) \gamma^+ \; ,
\end{gather}
and then
\begin{gather}
    \gamma^{+}  (\slashed{\check{k}}_i+m)  \gamma_{\perp \mu} (\slashed{\check{p}}_f + m ) \gamma^+ \bigg|_ {\substack{p_i^+ = p_f^+ = k^+ - q^+ \\ k_i^+ = k_f^+ = k^+ \\ \mathbf{k}_i = \mathbf{p}_f \; , \; \mathbf{p}_i = \mathbf{k}_f}} = 2  ( (k^+-q^+) \slashed{k}_{i \perp} \gamma_{\perp}^{\mu}  + k^+  \gamma_{\perp}^{\mu} \slashed{k}_{i \perp} + m q^+ \gamma_{\perp}^{\mu} ) \gamma^+ \; ,
\end{gather}
to reach the form
\begin{gather}
     8 {\rm Tr_{D,c}} \bigg[ \gamma^{\nu}_{\perp}  ( (k^+-q^+) \gamma_{\perp}^{\mu} \slashed{k}_{f \perp} + k^+ \slashed{k}_{f \perp} \gamma_{\perp}^{\mu} + m q^+ \gamma_{\perp}^{\mu} ) \nonumber \\ \times ( (k^+-q^+) \slashed{k}_{i \perp} \gamma_{\perp}^{\mu}  + k^+  \gamma_{\perp}^{\mu} \slashed{k}_{i \perp} - m q^+ \gamma_{\perp}^{\mu} ) \gamma_{\perp \nu} \gamma^- \mathcal{O}_q    \bigg] \; .
     \label{Eq:DiracTraceQuarkToFin}
\end{gather}
The contraction can be shown to be
\begin{gather}
    \gamma^{\nu}_{\perp}  ( (k^+-q^+) \gamma_{\perp}^{\mu} \slashed{k}_{f \perp} + k^+ \slashed{k}_{f \perp} \gamma_{\perp}^{\mu} + m q^+ \gamma_{\perp}^{\mu} ) \nonumber \\ \times ( (k^+-q^+) \slashed{k}_{i \perp} \gamma_{\perp}^{\mu}  + k^+  \gamma_{\perp}^{\mu} \slashed{k}_{i \perp} - m q^+ \gamma_{\perp}^{\mu} ) \gamma_{\perp \nu} \gamma^- = A m^2 + B m + C \; .
    \label{Eq:Contrac}
\end{gather}
where
\begin{gather}
    A = - (D-2)^2 (q^+)^2 \gamma^- \; , \nonumber \\
    B = q^+ (D-4) ((D-4) q^+ + 2 k^+ ) (\slashed{k}_{f \perp } - \slashed{k}_{i \perp }) \gamma^- \nonumber \\
    C = (D-2) (k_{i \perp} \cdot k_{f \perp}) \gamma^- ((D-2) (q^+)^2 - 4 k^+ q^+ + 4 (k^+)^2) + {\rm antisymmetric \; terms} \; .
    \label{Eq:Coeff}
\end{gather}
In the coefficient $C$, for antisymmetric terms, we mean contributions proportional to $[ \slashed{k}_{f \perp} , \slashed{k}_{i \perp} ]$. These terms will vanish after integration over transverse momenta because
\begin{gather}
    \slashed{k}_{f \perp} \slashed{k}_{i \perp} - \slashed{k}_{i \perp} \slashed{k}_{f \perp}  \xrightarrow{{\rm integration \; over} \; k_{i \perp} \; {\rm and} \; k_{f \perp } } \rm  \slashed{x}_{12} \slashed{x}_{12} - \slashed{x}_{12}\slashed{x}_{12} = 0 \; .
\end{gather}
Using eq.~(\ref{Eq:Contrac}) and~(\ref{Eq:Coeff}) in eq.~(\ref{Eq:DiracTraceQuarkToFin}), the result (\ref{Eq:BigTraceQuarkFin}) is found.

\section{Discussion on the inside-the-medium contributions}
\label{Sec:AppInsideMedium}

\begin{figure}
    \centering    \includegraphics[width=1\linewidth]{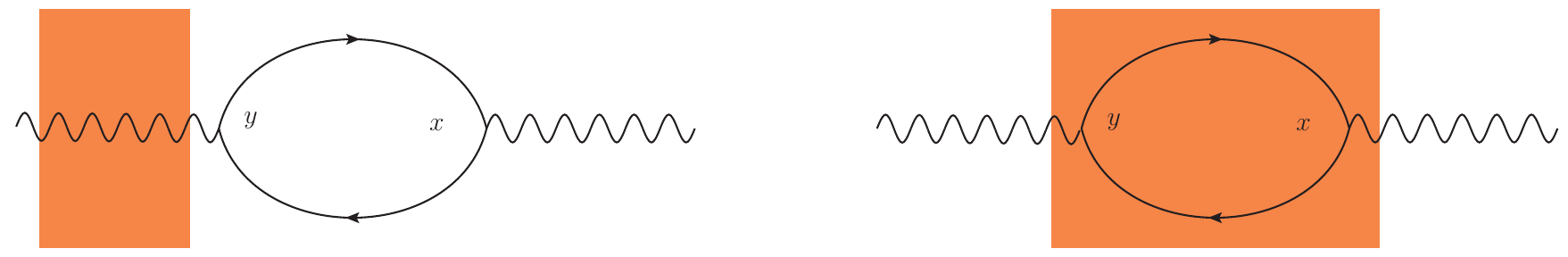}
    (a) \hspace{7.6 cm} (b) \vspace{0.5 cm} \\ 
    \includegraphics[width=1\linewidth]{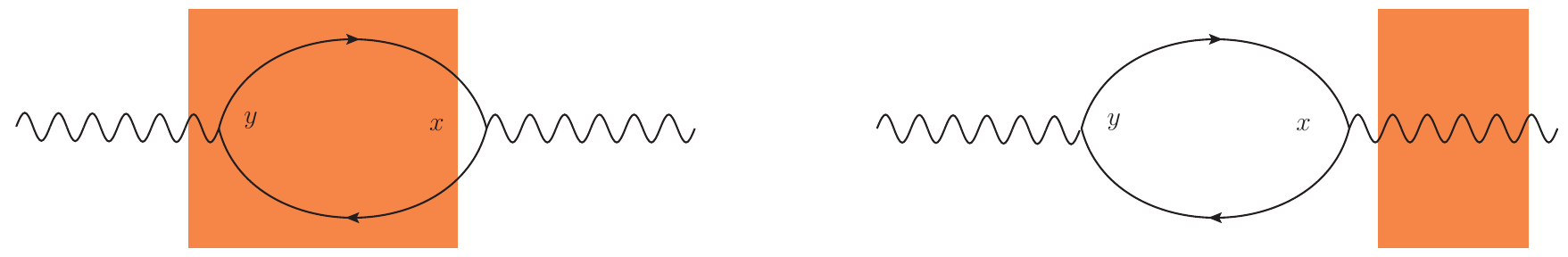}
    (c) \hspace{7.6 cm} (d) \vspace{0.5 cm} \\ 
    \includegraphics[width=1\linewidth]{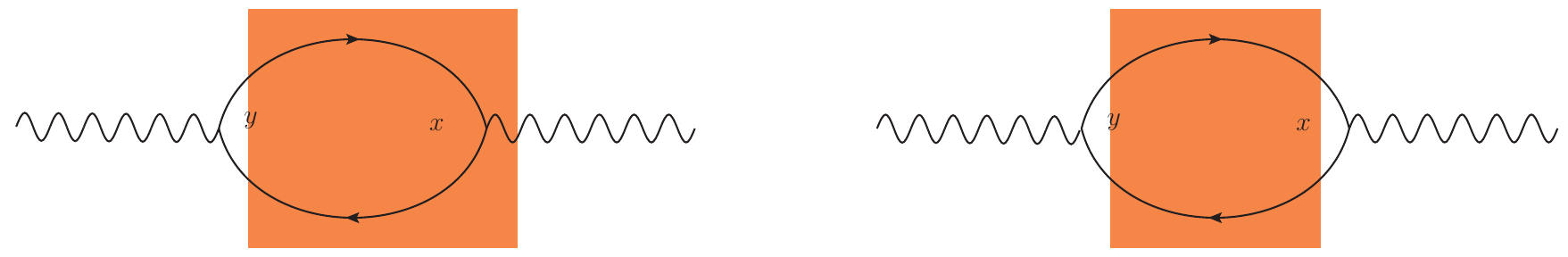}
    (e) \hspace{7.6 cm} (f)
    \caption{All the NEik contributions from the gluon background of the target, considering the $x^+ > y^+$ configuration.}
    \label{fig:InsideMedium}
\end{figure}

In the main text, we did not discuss gluon background contributions in which one or two electromagnetic vertices are inside the medium (see Fig. \ref{fig:InsideMedium}). In this section, we show that these contributions are vacuum ones. First, let's recall that the pure vacuum contribution (photon-to-photon diagram in the absence of a medium) does not contribute to the S-matrix in DIS. This can be easily verified by calculating 
\begin{gather}
    [S_{{\rm NLO , \; G}}^{\rm vacuum}]^{\mu \nu} = - e_f^2 \; g^2 \int d^D y \int d^D x \; e^{i q' \cdot x -i q \cdot y}  {\rm Tr_{D,c}} \left[  \gamma^{\nu} S_{F,q} (x, y) \gamma^{\mu}  S_{F, \bar{q}} (y, x) \right] \nonumber \\ \nonumber = - e_f^2 \; g^2 (2 \pi)^{D-1} \delta ( q^+ - q'^+ ) \delta^{(D-2)} ( \mathbf{q} - \mathbf{q}' ) \int \frac{d^{D-1} \underline{k}}{(2 \pi)^{D-1}} {\rm Tr_{D,c}}  \left[  \gamma^{\nu} ( \check{\slashed{k}} + m) \gamma^{\mu} ( \check{\slashed{p}} + m ) \right] \nonumber \\ \times \theta(k^+ ) \theta (q^+-k^+) \int_{- \infty}^{\infty} \hspace{-0.3 cm} d y^+ \; e^{-i y^+ \left( q^- - \frac{\mathbf{k}^2 + m^2}{2 k^+} - \frac{ (\mathbf{k}-\mathbf{q})^2 + m^2 }{2 (q^+ - k^+)} \right)} \int_{-\infty}^{\infty} \hspace{-0.3 cm} d x^+ \; e^{i x^+ \left( q^- - \frac{\mathbf{k}^2 + m^2}{2 k^+} - \frac{ (\mathbf{k}-\mathbf{q})^2 + m^2 }{2 (q^+ - k^+)} \right) }  .
    \label{Eq:Vacuum}
\end{gather}
Integrating over $x^+$ or $y^+$, in eq.~(\ref{Eq:Vacuum}), yields a Dirac delta function that is incompatible with the kinematics of the DIS, i.e., it enforces $q^2 = - Q^2 > 0$. \\

In our framework, this vacuum contribution appears separated into several terms. Assuming the order $x^+ > y^+$, one of these vacuum contributions is \\
\begin{gather}
    [S_{{\rm NLO , \; G}}^{\rm vacuum}]_{\rm Eik}^{\mu \nu} \bigg |_{x^+ > y^+} = - e_f^2 \; g^2 (2 \pi)^{D-1} \delta ( q^+ - q'^+ ) \delta^{(D-2)} ( \mathbf{q} - \mathbf{q}' ) \nonumber \\ \times \int \frac{d^{D-1} \underline{k}}{(2 \pi)^{D-1}} {\rm Tr_{D,c}}  \left[  \gamma^{\nu} ( \check{\slashed{k}} + m) \gamma^{\mu} ( \check{\slashed{p}} + m ) \right] \theta(k^+ ) \theta (q^+-k^+) \nonumber \\ \times  \int_{- \infty}^{-\frac{L}{2}^+} \hspace{-0.3 cm} d y^+ \; e^{-i y^+ \left( q^- - \frac{\mathbf{k}^2 + m^2}{2 k^+} - \frac{ (\mathbf{k}-\mathbf{q})^2 + m^2 }{2 (q^+ - k^+)} \right)} \int_{\frac{L}{2}^+}^{\infty} \hspace{-0.1 cm} d x^+ \; e^{i x^+ \left( q^- - \frac{\mathbf{k}^2 + m^2}{2 k^+} - \frac{ (\mathbf{k}-\mathbf{q})^2 + m^2 }{2 (q^+ - k^+)} \right) }  .
\end{gather}
Setting $L^+ = 0$, one obtains the vacuum that is usually subtracted from the eikonal calculation. Keeping also linear terms in the $L^+ \rightarrow 0$ expansion (i.e. NEik accuracy), we find the vacuum contributions discussed in Sec. \ref{Sec:Before-to-after}. Indeed, we have shown that in these contributions the Wilson lines combine to give the unity in a trivial way, which is equivalent to setting the Wilson lines to 1 from the beginning of the calculation. Since setting the Wilson lines to 1 in the before-to-after propagator reduces this latter to the free one, the correspondence with the vacuum is immediate. In any case, at the NEik accuracy, this discussion is not sufficient, since other regions of the two-dimensional domain in $x^+$ and $y^+$ can contribute (see Fig.~\ref{fig:InsideMedium}). As we will show in the remaining part of this appendix, all these contributions are vacuum terms. The only caveat is that, in all other vacuum terms, one or two coordinates are in the medium. In this case, to see the equivalence with the vacuum, one has to expand around $L^+ \rightarrow0$, i.e.  
\begin{gather}
    \int_{-\frac{L}{2}^+}^{\frac{L}{2}^+} d x^+ e^{ix^+ q^-} ... \simeq L^+ + \mathcal{O} ((L^+)^2) \; .
\end{gather}
That said, all that remains to be shown is that Wilson lines can be set to 1 from the start, which is what we are going to show next. Without loss of generality, we can choose $x^+ > y^+$; then the whole domain of integration can be split as 
\begin{gather*}
    \int \theta(x^+-y^+) dx^+ dy^+ \; ... \;  = \int^{\infty}_{\frac{L}{2}^+} dy^+\int^{\infty}_{-\infty} dx^+ \theta(x^+-y^+) \; ... \;  \nonumber \\  + \int_{-\frac{L}{2}^+}^{\frac{L}{2}^+} dy^+\int^{\infty}_{-\infty} dx^+ \theta(x^+-y^+) \; ... \; + \int_{-\infty}^{-\frac{L}{2}^+} dy^+\int^{\infty}_{-\infty} dx^+ \theta(x^+-y^+) \; ... \; \; \; .
\end{gather*}
The contribution from the first region ($ L^+/2 <y^+< \infty$) is non-zero only if $ L^+/2 <x^+< \infty$, which correspond to the diagram (a) in Fig. \ref{fig:InsideMedium}. This contribution is trivially a vacuum one, because the photon cannot interact with the dense QCD medium directly. In the second region ($ -L^+/2 < y^+< L^+/2$), there are to possible contributions: the inside-to-inside quark propagation ($ -L^+/2 < x^+< L^+/2$), corresponding to the diagram (b), and the inside-to-after quark propagation ($ x^+ > L^+/2$), corresponding to the diagram (c). In the third and last region there are three possible contributions: the before-to-before quark propagation ($ -L^+/2 < x^+< -\infty$), corresponding to the diagram (d) and being a trivial vacuum one, the before-to-inside quark propagation ($ -L^+/2 < x^+ < L^+/2$), corresponding to the diagram (e), and the before-to-after, corresponding to the diagram (f). \\

The diagram (f) is the one computed in section \ref{Sec:NLOXsec_gb} (Fig. \ref{fig:NLODISNEikgluon}) and it has been already discussed. Diagrams (a) and (d) are trivial vacuum contributions, and (c) and (e) are specular. Thus, it remains to show that (b) and (e) are pure vacuum contributions. 

\paragraph{Before-to-inside diagram.} This contribution reads
\begin{gather}
    [S_{{\rm NLO , \; G}}^{\rm b.i.}]^{\mu \nu} = e_f^2 \; g^2 \int_{- \infty}^{- \frac{L}{2}^+ } d^D y \int_{- \frac{L}{2}^+ }^{ \frac{L}{2}^+ } d^D x \; e^{i q' \cdot x -i q \cdot y}  {\rm Tr_{D,c}} \left[  \gamma^{\nu} S_{F,q}^{\rm b.i.} (x, y) \gamma^{\mu}  S_{F, \bar{q}}^{\rm b.i.} (y, x) \right] \; ,
\end{gather}
where
\begin{align*} 
S_{F,q}^{b.i.} (x, y)
& =\int \frac{d^{D-1} \underline{k}}{(2 \pi)^{D-1}}  \frac{\theta\left(k^{+}\right)}{2 k^+}  e^{i y \cdot \check{k}}  e^{-i x^- k^+ } e^{i \mathbf{x} \cdot \mathbf{k} } 
\nn  \\ & \hspace{4.5cm}
\times 
\left[ 1 - \frac{\gamma^+ \gamma^i}{2 k^+} i \overrightarrow{D_{\mathbf{x}^i}^F} \right] (\slashed{\check{k}}+m) \mathcal{U}_F\left( x^+,y^+, \mathbf{x} \right)   \; ,
\end{align*}
and
\begin{align*} 
S_{F,\bar{q}}^{b.i.} (y, x) &= (-1) \int \frac{d^{D-1} \underline{p}}{(2 \pi)^{D-1}}  \frac{\theta\left(-p^{+}\right)}{2 p^+}  e^{-i y \cdot \check{p}} (\slashed{\check{p}}+m)  
\nn \\
&  \hspace{4cm}
\times \mathcal{U}^{\dagger}_F\left( x^+, y^+, \mathbf{x} \right) \left[ 1 - \frac{\gamma^+ \gamma^i}{2 p^+} i \overleftarrow{D_{\mathbf{x}^i}^F} \right] e^{i x^- p^+ } e^{-i \mathbf{x} \cdot \mathbf{p} } .
\end{align*}
In the longitudinal case $\mu = \nu = +$, all terms containing the covariant derivative in the fundamental representation vanish. In the surviving term one has
\begin{gather}
    {\rm Tr_c } \left[ \mathcal{U}_F\left( x^+,y^+, \mathbf{x} \right) \mathcal{U}^{\dagger}_F\left( x^+, y^+, \mathbf{x} \right) \right] = N_c \; ,
    \label{Eq:VacuumColorStruct}
\end{gather}
which is identical to the result that one obtains by setting all Wilson lines to the unit from the beginning (i.e., in the absence of the classical gluon background field). The transverse case is slightly less immediate. The structure of the trace this time is
\begin{gather}
{\rm Tr_{D,c}} \left[  \gamma^{\mu}_{\perp} \left( 1 - \frac{\gamma^+ \gamma^i}{2 k^+} i \overrightarrow{D_{\mathbf{x}^i}^F} \right) (\slashed{\check{k}}+m) \mathcal{U}_F\left( x^+,y^+, \mathbf{x} \right) \gamma_{\perp \mu} (\slashed{\check{p}}+m)   \mathcal{U}^{\dagger}_F\left( x^+, y^+, \mathbf{x} \right) \left( 1 - \frac{\gamma^+ \gamma^i}{2 p^+} i \overleftarrow{D_{\mathbf{x}^i}^F} \right) \right] \; ,
\end{gather} 
where $D^F$ is understood as acting only on the nearest Wilson line. The term with no covariant derivatives is equivalent to the vacuum contribution as in the longitudinal case. The term with two covariant derivatives vanishes because $(\gamma^+)^2=0$. Finally, all terms contaning a partial derivative or a gluon field insertion can be shown to be proportional to ${\rm Tr_c} [ t^a ] = 0$, by using Eq.~(\ref{Eq:AppenAdjFun}). Therefore, also in the transverse case, this diagram is a pure vacuum contribution. The proof for the diagram (c) is identical. \\

\paragraph{Inside-to-inside diagram.} This contribution reads
\begin{gather}
    [S_{{\rm NLO , \; G}}^{\rm i.i.}]^{\mu \nu} = e_f^2 \; g^2 \int_{-\frac{L}{2}^+}^{\frac{L}{2}^+} d^D y \int_{-\frac{L}{2}^+}^{\frac{L}{2}^+} d^D x \; e^{i q' \cdot x -i q \cdot y} \; {\rm Tr_{D,c}} \left[  \gamma^{\nu} S_{F,q}^{\rm i.i.} (x, y) \gamma^{\mu}  S_{F, \bar{q}}^{\rm i.i.} (y, x) \right] \; .
    \label{Eq:InsideToInside}
\end{gather}
As mentioned above, we want to show that the additional contributions from these diagrams can at most produce vacuum contributions that, within NEik accuracy, reconstruct the entire vacuum. The inside-to-inside diagram can have a NEik contribution only if one of the two integrations over $x^+$ and $y^+$ is removed by a Dirac Delta function. In all other cases, the integration over the $+$-coordinates leads to the following power counting\footnote{Note that the presence of the Wilson lines and transverse covariant derivatives of the latter cannot decrease the powers of $L^+$.}
\begin{gather}
    \int_{-\frac{L}{2}^+}^{\frac{L}{2}^+} d x^+ e^{ix^+ q^-}  \int_{-\frac{L}{2}^+}^{\frac{L}{2}^+} d y^+ e^{iy^+ q^-} ... \simeq (L^+)^2 + \mathcal{O} ((L^+)^3) \; ,
\end{gather}
which corresponds to a NNEik contribution and can therefore be safely neglected. \\

However, one needs to be careful with instantaneous contributions in the inside-to-inside propagators~\eqref{Eq:Ins-Ins-Quark}, which contain $\delta (x^+-y^+)$. If one picks the instantaneous term in both propagators, the Dirac structure vanishes in both the longitudinal and transverse cases. In the longitudinal case, the fact that $\mu=\nu=+$ immediately nullifies the contributions with the  instantaneous term in a single propagator as well. The only possible NEik contribution arises in the case of  a transverse photon when picking the instantaneous term in only one of the two propagators.\footnote{Note that we employ the standard convention $ \delta (x^+-y^+)\theta (x^+-y^+) = \delta (x^+-y^+)/2$} Nevertheless, this contribution is a vacuum one. Let us consider, for instance, the contribution with the instantaneous term in one propagator and the term in $\theta(x^+-y^+)$ in the other one. We then obtain a term containing the following Dirac structure
\begin{gather*}
 {\rm Tr}_c  \left[  \gamma^{\nu}_{\perp} S_{F,q}^{\rm i.i.} (x, y)   \gamma_{\perp}^{\mu}  \gamma^+ \right] \delta (x^+-y^+) = - {\rm Tr}_c  \left[  \gamma^{\nu}_{\perp} S_{F,q}^{\rm i.i.} (x, y)  \gamma^+ \gamma_{\perp}^{\mu}  \right] \delta (x^+-y^+) \;
\end{gather*}
where\footnote{Remember that we are considering $x^+ > y^+$.}
\begin{align}
S_{F,q}^{\text { i.i. }}  (x, y)  \gamma^+ & = \theta\left(x^{+}-y^{+}\right) \int \frac{d k^{+}}{2 \pi} \frac{\theta\left(k^{+}\right)}{2 k^{+}} e^{-i k^{+}\left(x^{-}-y^{-}\right)}  
 \left[k^{+} \gamma^{-}+m+i \gamma^i \vec{D}_{\mathbf{x}^i}^F\right] \gamma^{+} \nonumber \\ 
& \times \int d^{D-2} \mathbf{z} \; \delta^{(D-2)}(\mathbf{x}-\mathbf{z}) \delta^{(D-2)} (\mathbf{z}-\mathbf{y})  \mathcal{U}_F\left(x^{+}, y^{+} ; \mathbf{z} \right)   \; ,
\end{align}
If this term did not contain the covariant derivative, which can generate a non-trivial effect from the Wilson line in the limit $L^+ \rightarrow 0$ (see Eq.~(\ref{Eq:MinDeriv})), we could immediately use $\delta (x^+-y^+)$ to make the Wilson line $\mathcal{U}_F\left(x^{+}, y^{+} ; \mathbf{z} \right)$ collapse to the identity. However, the term containing the covariant derivative vanishes due to the Dirac trace. From this we can conclude that the terms of order $L^+$ from this diagram are all vacuum contributions, which completes our proof.

\newpage
\bibliographystyle{apsrev}
\bibliography{mybib_New}

\end{document}